\documentclass[a4paper,12pt]{article}
\usepackage[dvipdfmx]{graphicx}
\usepackage{amsfonts,amsthm,amsmath,amssymb,graphicx,enumerate}
\numberwithin{equation}{section}
\usepackage{color}

\begin{document}

\newtheorem{definition}{Definition}[section]
\newcommand{\be}{\begin{equation}}
\newcommand{\ee}{\end{equation}}
\newcommand{\bea}{\begin{eqnarray}}
\newcommand{\eea}{\end{eqnarray}}
\newcommand{\LE}{\left[}
\newcommand{\R}{\right]}
\newcommand{\nn}{\nonumber}
\newcommand{\Tr}{\text{Tr}}
\newcommand{\N}{\mathcal{N}}
\newcommand{\G}{\Gamma}
\newcommand{\vf}{\varphi}
\newcommand{\LL}{\mathcal{L}}
\newcommand{\Op}{\mathcal{O}}
\newcommand{\HH}{\mathcal{H}}
\newcommand{\arctanh}{\text{arctanh}}
\newcommand{\up}{\uparrow}
\newcommand{\down}{\downarrow}
\newcommand{\ket}[1]{\left| #1 \right>}
\newcommand{\bra}[1]{\left< #1 \right|}
\newcommand{\ketbra}[1]{\left|#1\right>\left<#1\right|}
\newcommand{\rd}{\partial}
\newcommand{\de}{\partial}
\newcommand{\ba}{\begin{eqnarray}}
\newcommand{\ea}{\end{eqnarray}}
\newcommand{\db}{\bar{\partial}}
\newcommand{\we}{\wedge}
\newcommand{\ca}{\mathcal}
\newcommand{\lr}{\leftrightarrow}
\newcommand{\f}{\frac}
\newcommand{\s}{\sqrt}
\newcommand{\vp}{\varphi}
\newcommand{\hvp}{\hat{\varphi}}
\newcommand{\tvp}{\tilde{\varphi}}
\newcommand{\tp}{\tilde{\phi}}
\newcommand{\ti}{\tilde}
\newcommand{\ap}{\alpha}
\newcommand{\pr}{\propto}
\newcommand{\mb}{\mathbf}
\newcommand{\ddd}{\cdot\cdot\cdot}
\newcommand{\no}{\nonumber \\}
\newcommand{\la}{\langle}
\newcommand{\lb}{\rangle}
\newcommand{\ep}{\epsilon}
 \def\we{\wedge}
 \def\lr{\leftrightarrow}
 \def\f {\frac}
 \def\ti{\tilde}
 \def\ap{\alpha}
 \def\pr{\propto}
 \def\mb{\mathbf}
 \def\ddd{\cdot\cdot\cdot}
 \def\no{\nonumber \\}
 \def\la{\langle}
 \def\lb{\rangle}
 \def\ep{\epsilon}
\newcommand{\mcl}{\mathcal}
 \def\g{\gamma}
\def\tr{\text{tr}}

\begin{titlepage}
\thispagestyle{empty}

\renewcommand{\thefootnote}{\fnsymbol{footnote}}

\begin{flushright}
OCU-PHYS 492
\\
USTC-ICTS-18-22 
\end{flushright}
\bigskip

\begin{center}
\noindent{\large \textbf{Dynamics of logarithmic negativity and mutual information in smooth quenches}}
\\
\vspace{1cm}

Hiroyuki Fujita\footnote{{\small{\,h-fujita@issp.u-tokyo.ac.jp}}},
Mitsuhiro Nishida\footnote{{\small{\,mnishida@gist.ac.kr}}}, 
Masahiro Nozaki\footnote{{\small{\,masahiro.nozaki@riken.jp}}}, 
and 
Yuji Sugimoto\footnote{{\small{sugimoto@sci.osaka-cu.ac.jp}}}

\vspace{0.8cm}

\footnotesize
{\it $^{*}$ Institute for Solid State Physics, The University of Tokyo, Kashiwa, Chiba 277-8581, Japan}

\vskip.1cm

{\it $^{\dagger}$ School  of  Physics  and  Chemistry,  Gwangju  Institute of  Science  and  Technology, Gwangju  61005,  Korea}

\vskip.1cm

{\it $^{\ddagger}$Kadanoff Center for Theoretical Physics, University of Chicago, Chicago, IL 60637, USA \\}

\vskip.1cm

{\it $^{\ddagger}$iTHEMS Program, RIKEN, Wako, Saitama 351-0198, Japan \\}

\vskip.1cm

{\it $^{\S}$ Osaka City University Advanced Mathematical Institute (OCAMI),  3-3-138, Sugimoto, Sumiyoshi-ku, Osaka, 558-8585, Japan \\}

\vskip.1cm

{\it $^{\S}$ Interdisciplinary Center for Theoretical Study, University of Science and Technology of China, Hefei, Anhui 230026, China \\}
\normalsize

\vspace{0.8cm}

\end{center}
\begin{abstract}
In this paper, we study how quantum correlation between subsystems changes in time by investigating time evolution of mutual information and logarithmic negativity in two protocols of mass quench. Hamiltonian in both protocols is for $2$-dimensional free scalar theory with time-dependent mass: the mass in one case decreases monotonically and vanishes asymptotically (ECP), and that in the other decreases monotonically before $t=0$, but increases monotonically afterward, and becomes constant asymptotically (CCP). We study the time evolution of the quantum correlations under those protocols in two different limits of the mass quench; fast limit and slow limit depending on the speed with which the mass is changed. We obtain the following two results: (1) For the ECP, we find that the time evolution of logarithmic negativity is, when the distance between the two subsystems is large enough, well-interpreted in terms of the propagation of relativistic particles created at a time determined by the limit of the quench we take. On the other hand, the evolution of mutual information in the ECP depends not only on the relativistic particles but also on slowly-moving particles. (2) For the CCP, both logarithmic negativity and mutual information oscillate in time after the quench. When the subsystems are well-separated, the oscillation of the quantum correlations in the fast limit is suppressed, and the time evolution looks similar to that under the ECP in the fast limit.
\end{abstract}
\end{titlepage} 

\tableofcontents

\section{Introduction and summary}
Understanding how quantum states thermalize is one of research themes on the frontier of theoretical physics. 
Thermodynamics well-describes properties of systems with macroscopic degrees of freedom, and thermodynamic systems time-evolve irreversibly. On the other hand, quantum field theories (QFTs)  well-describes properties of systems with microscopic degrees of freedom. The states in QFTs  time-evolve unitarily, and a pure state never thermalizes. Therefore, many researchers have been studying how thermalization is interpreted in terms of microscopic degrees of freedom.
Since the area of black hole is subject to thermodynamic laws, the black hole is a thermal object \cite{Hawking:1971tu,Bekenstein:1972tm,Bardeen:1973gs,Bekenstein:1973ur,Hawking:1974rv,Hawking:1974sw}.
In terms of holography, thermalization is closely related to the dynamics of black hole such as its formation and evaporation \cite{Balasubramanian:2010ce,Balasubramanian:2011ur}. Therefore, we expect a microscopic description of thermalization to help us to study (quantum) dynamics of black holes. 
Thus, understanding thermalization in terms of microscopic degrees of freedom is a challenging but worthwhile problem. 

Thermalization of a whole system of QFTs  is unknown, but that of a subsystem can be studied by looking at its dynamics.
When we focus on the local dynamics in the subsystem $A$ owing to quantum entanglement of a state, the degrees of freedom in the outside of $A$ play as a thermal bath. Therefore, late-time expectation values of local observables, $n$-point functions and entanglement entropy, will be approximated by those of a thermal state. Authors in \cite{Calabrese:2005in} have studied time evolution of entanglement entropy in a $2$-dimensional dynamical system, where a gapped theory changes suddenly to a CFT (a global quench). They have found that the lattice-independent term in a certain scaling limit increases linearly in time and is proportional to the subsystem size at late time. Authors in \cite{AbajoArrastia:2010yt} have numerically studied the time evolution of entanglement entropy for a global quenched state in holographic theories, which have gravity duals. 
The authors in \cite{Liu:2013qca,Hartman:2013qma,Asplund:2015eha,Liu:2013iza} have proposed holographic models, which analytically describe its linear growth in time in holographic theories. In those articles, researchers have been studying the evolution of entanglement entropy when Hamiltonian changes globally and suddenly. 

Also, authors in \cite{Calabrese:2007mtj} have studied the time evolution of entanglement entropy for excited states in local quenches, where a Hamiltonian changes locally and suddenly. Since the late-time 
entropy for locally-quenched states is not approximated by thermal entropy, the subsystem in the local quenches does not thermalize. Various holographic models describing local quenched states are proposed \cite{Nozaki:2013wia,Ugajin:2013xxa,Astaneh:2014fga}.  

\begin{figure}[htbp]
 \begin{minipage}{0.5\hsize}
  \begin{center}
   \includegraphics[width=70mm]{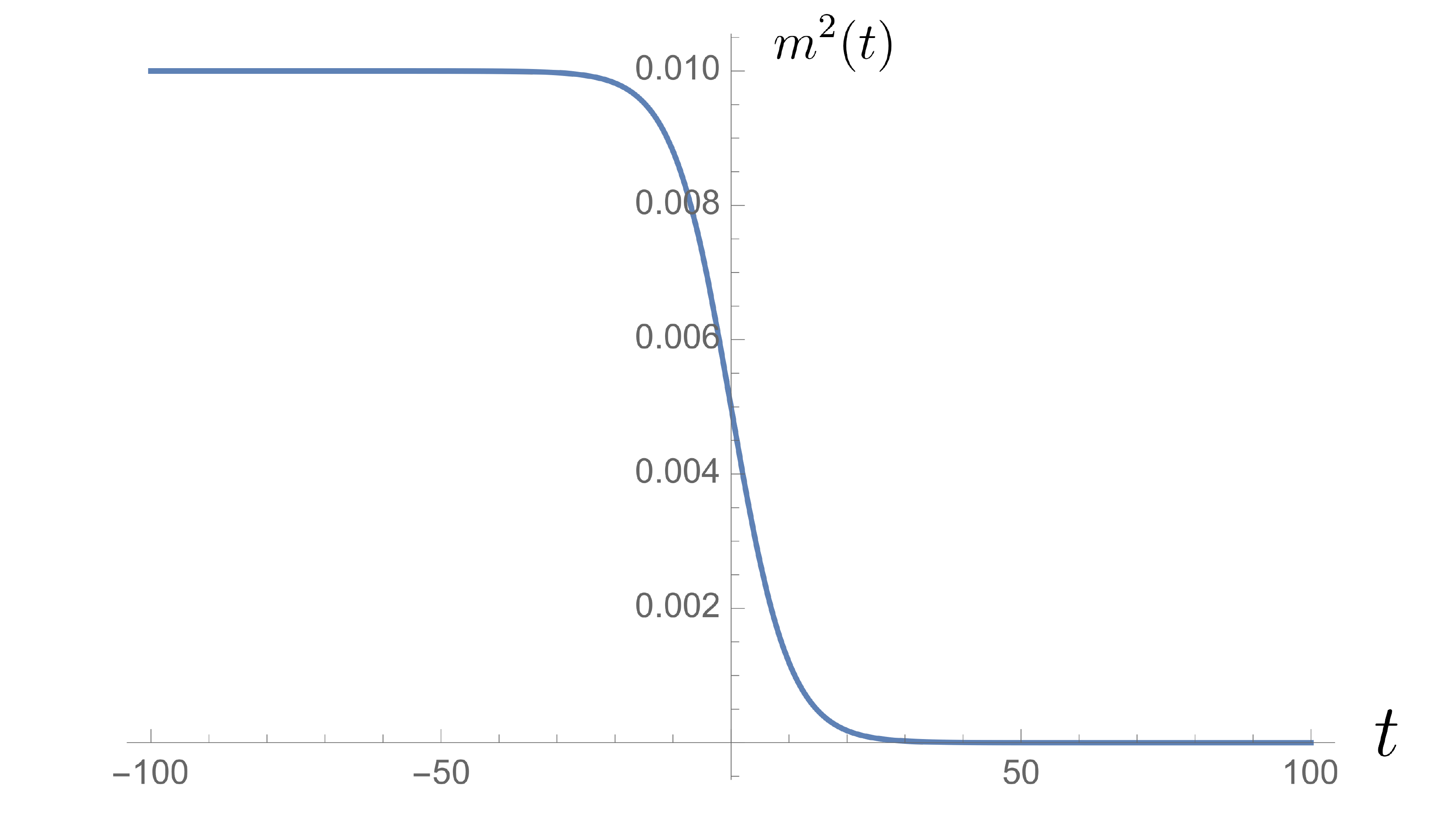}
  \end{center}
  \label{fig:one}
 \end{minipage}
 \begin{minipage}{0.5\hsize}
  \begin{center}
 \includegraphics[width=70mm]{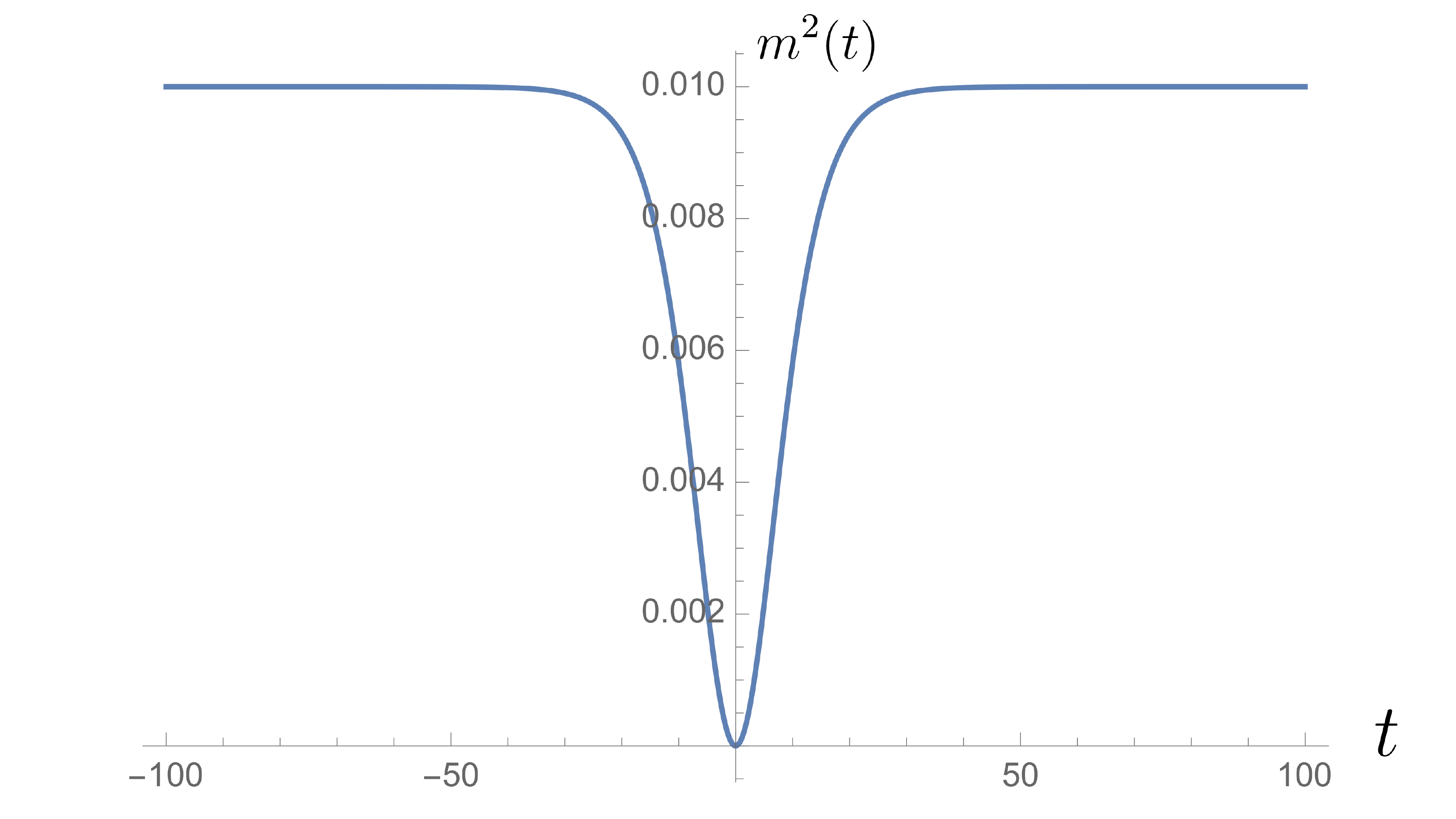}
  \end{center}
  \label{fig:two}
 \end{minipage}
 \caption{Time evolution of two time dependent masses $m^2(t)$ with the parameters $\xi$ and $\delta t$, which are dimensionless papameters, $\xi= \f{\tilde{\xi}}{a}$ and $\delta t=\f{\delta \tilde{t}}{a}$. Here, $a$ is a lattice spacing. The left panel shows the time evolution of $m^2(t)=\f{1}{2\xi^2}\left[1- \tanh{\left(\f{t}{\delta t}\right)}\right]$ with $\xi=\delta t =10$. The right panel shows the evolution of $m^2(t)=\f{1}{\xi^2}\tanh^2{\left(\f{t}{\delta t}\right)}$ with $\xi=\delta t =10$.}\label{quench}
\end{figure}

The time evolution of entanglement entropy for the excited states by the quenches shows thermalization depends on detail of quenches. 
Since three of authors in this paper has been interested in the time evolution of entanglement entropy in more general setups, we have studied the dynamics of quantum entanglement when a time-dependent Hamiltonian changes smoothly in time \cite{Caputa:2017ixa, Nishida:2017hqd}. The time-dependent Hamiltonian is for a $2$-dimensional free scalar theory with two kinds of time-dependent mass \cite{kzp} (See Figure \ref{quench}). If the time-dependent mass is initially constant, it starts to decrease around $t=0$, and it asymptotically vanishes, then the quench is called End-Critical-Protocol (ECP). 
If the time-dependent mass is initially constant, it starts to decrease before $t=0$, it vanishes at $t=0$, it starts to increase after $t=0$, and it asymptotically becomes constant, then the protocol is Cis-Critical-Protocol (CCP). In those smooth protocols, scaling behaviors of observables in QFTs are studied, and their time-evolution is compared to the one in the sudden quenches \cite{Das:2014jna, Das:2015jka}.
The two protocols in this paper have two parameters, which are an initial correlation length $\tilde{\xi}$ and a quench rate $\delta \tilde{t}$. 
We have studied the time evolution of entanglement entropy in the protocols in two limits, $\f{\delta \tilde{t}}{\tilde{\xi}} \gg 1 $ and  $\f{\delta \tilde{t}}{\tilde{\xi}} \ll 1$, which are first considered in \cite{Das:2016lla}.
 The time evolution of entanglement entropy in both limits for ECP shows that the late-time entropy is proportional to the subsystem size because quasi-particles are created at a characteristic time, which is determined by the limit. 
 On the other hand, the entropy in both limits for CCP oscillates in time, and we have found the entropy for the quenched state in the fast limit $\f{\delta \tilde{t}}{\tilde{\xi}}\ll 1$ is proportional to the subsystem size, but we could not find whether the entropy in the slow limit $\f{\delta \tilde{t}}{\tilde{\xi}}\gg 1$ is proportional to the size.

In this paper, we study the time evolution of mutual information $I_{A,B}$ and logarithmic negativity $\mathcal{E}$ in order to  take a  more closely look at how the dynamics of quantum entanglement depends on these protocols. These quantum quantities, $I_{A,B}$ and $\mathcal{E}$, measure quantum correlation between subsystems $A$ and $B$ with several distances between them, and those in 2D CFTs depend on the detail of CFTs because they depend on $n$-point functions where $n>2$.

\subsection*{Summary}
We study the dynamics of quantum entanglement by measuring the time evolution of logarithmic negativity $\mathcal{E}$ and mutual information $I_{A,B}$ in two protocols ECP and CCP. 
We define two dimensionless parameters $\xi $ and $\delta t$ by $\xi =\f{\tilde{\xi}}{a}$ and $\delta t = \f{\delta \tilde{t}}{a}$. Here, $a$ is a lattice spacing. 
The summary of main results is: \\

{\bf ECP}

\begin{itemize}
\item[1.] The evolution of $\mathcal{E}$ and $I_{A,B}$ is interpreted in terms of the relativistic propagation of quasi-particles. In the fast limit where $ \delta t/\xi \ll 1$,  the quasi-particles are created around  $t=0$. In the slow limit where $\delta t/\xi  \gg 1$, the particles are created at $t \sim t_{\text{kz}} =\delta t$, when adiabaticity breaks down.

\item[2.]As we expect from quasi-particle picture, slopes of $\mathcal{E}$ and $I_{A,B}$ for $l_a, l_b \gg \xi$ at early time are independent of the subsystem sizes of $A$ and $B$ when the distance between them is much larger than typical scales, which are determined by these limits.

\item[3.] The late-time $I_{A, B}$ is affected by modes propagating slowly for the disjoint interval where the distance between $A$ and $B$ is much larger than the characteristic length.
However, the quasi-particles which slowly propagates does not affect $\mathcal{E}$ for the disjoint interval.
\end{itemize}

{\bf CCP}

\begin{itemize}

\item[4.] 
If the distance $d$ between $A$ and $B$ is smaller  than $\xi$, then $I_{A,B}$ and $\mathcal{E}$ in the fast CCP oscillate in time. Their late-time period is given by $\pi \xi$. The amplitude of oscillation for the large distance $d \gg \xi$ is smaller than that for $d \ll \xi$. The time evolution of $I_{A,B}$ and $\mathcal{E}$ in this limit is similar to that in the fast ECP. The evolution of $I_{A,B}$ and $\mathcal{E}$ for the subsystems with $d \gg \xi$ is expected to be independent of the late-time protocol qualitatively.

\item[5.] 
The measures $I_{A,B}$ and $\mathcal{E}$ in the slow CCP oscillate with the period, which approaches $\pi\xi$ at late time.  As the distance $d$ between the subsystems increases, the time, when $I_{A,B}$ and $\mathcal{E}$ in the slow CCP start to increase, becomes later.
\end{itemize}

\subsection*{Organization}
In section 1, we have explained why we study the time evolution of logarithmic negativity and mutual information, and we have summarized what we find in this project. 
In section 2,  we will define $\mathcal{E}$ and $I_{A,B}$, and we will explain how to compute them in quantum field theories. In section 3, we will explain the details of two protocols of quench where we study the time evolution of $\mathcal{E}$ and $I_{A,B}$. In the following section, we will explain what we obtain in this project. Finally, we would like to discuss our results and comments on a few of future directions.
\section{Logarithmic negativity and mutual information}

In this section, we define quantum measures for states, logarithmic negativity and mutual information, and explain their properties briefly. After that, we explain how to compute them in path-integral formalism and in terms of correlation function, replica trick and correlator method.   Roughly speaking, the quantum measures in the replica trick are given by a kind of free energy in a partition function on a replicated geometry \cite{RP1, RP2}. The correlator method is applicable to computation of quantum measures for a gaussian state in free field theories. It is a powerful tool to compute them numerically \cite{CM1, CM2, CM3, CM4, CM5,CM6}. 
\subsection{Definition of logarithmic negativity and mutual information}
\begin{figure}[htbp]
 \begin{center}
  \includegraphics[width=80mm]{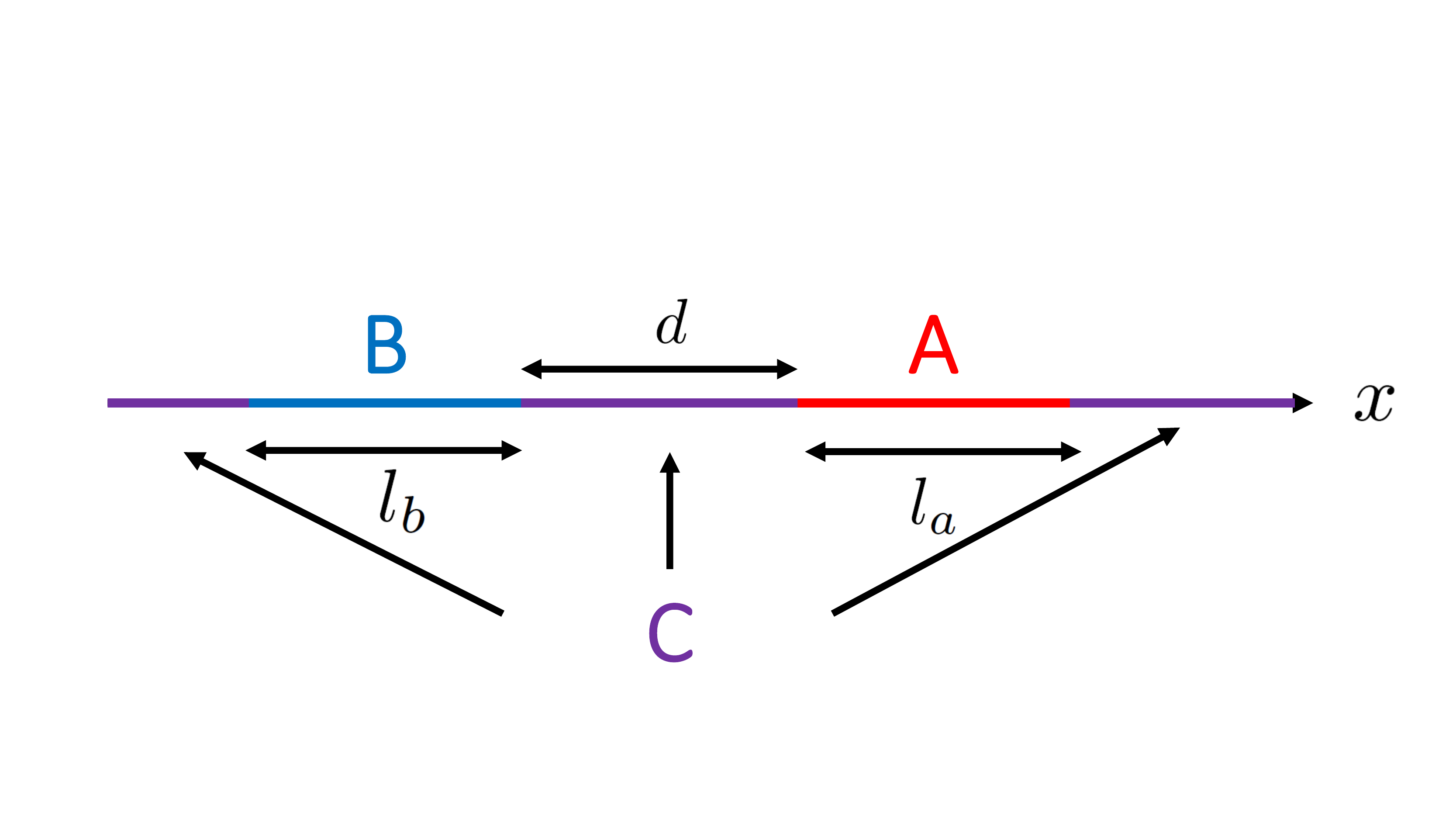}
 \end{center}
 \caption{Geometrical configuration of our subsystems.}
 \label{FG1}
\end{figure}

When the total space is $A \cup B \cup C$, logarithmic negativity and mutual information $\mathcal{E}$ and $I_{A, B}$ can measure non-local correlation which comes from quantum entanglement between subsystems, $A$ and $B$, as defined in Figure \ref{FG1}. Let us explain the definition of $\mathcal{E}$ and $I_{A, B}$, for which Hilbert space on a time slice is $\mathcal{H}=\mathcal{H}_A+\mathcal{H}_B+\mathcal{H}_C$.  Here, subsystem sizes of $A$ and B are $l_a$ and $l_b$, and the distance between $A$ and $B$ is $d$.  For simplicity, we consider $(1+1)$ dimensional quantum field theories. 

\subsubsection*{Definition of mutual information between $A$ and $B$}
Mutual information $I_{A, B}$ is defined by a combination of von Neumann entropies for  reduced density matrix $\rho_\alpha$,
\be
\begin{split}
I_{A,B} = S_A + S_B -S_{A \cup B}, ~~
S_{\alpha} = - \tr_{\alpha} \rho_{\alpha}  \log{\rho_{\alpha} },
\label{def MI}
\end{split}
\ee
where $\rho_{\alpha} $ is the reduced density matrix which is defined by tracing out the degrees of freedom outside $\alpha$:
\be
\rho_{\alpha}  =\tr_{\bar{{\alpha} }}\rho.
\ee
Entanglement entropy $S_{\alpha}$ is given by R$\acute{e}$nyi entanglement entropy $S^{(n)}_{\alpha}$ in von Neumann limit,
\be
S_{\alpha}=\lim_{n\rightarrow 1}S^{(n)}_{\alpha}=\lim_{n\rightarrow 1}\f{1}{1-n}\log{\left[\tr_{\alpha}\left(\rho_{\alpha}^n\right)\right]}
\ee
Mutual information is non-negative quantity and has a upper bound due to Araki-Lieb inequality $S_{A\cup B} \ge \left|S_A-S_B\right| $ \cite{Araki-Lieb},
\be
I_{A, B} \le 2 \text{Min}[S_A, S_B]
\ee
Tripartite information, a linear combination of mutual information, is expected to be a useful tool to characterize conformal field theories (CFTs), which has gravity dual \cite{TMI1}. The tripartite information in holographic CFTs is a non-positive quantity, and  the information in the strongly chaotic systems captures the scrambling properties of holographic CFT\cite{TMI2, TOMIM}.
\subsubsection*{Definition of logarithmic negativity}
Both $I_{A, B}$ and $\mathcal{E}$ can capture the non-local correlation between $A$ and $B$. However, only $\mathcal{E}$ can measure ``pure" quantum entanglement \cite{LN1, LN2}. The quantum measure, which can measure pure quantum entanglement, should be an ``entanglement monotone''. That is, it should decease monotonically under a local manipulation which is called local measurement and classical communication (LOCC) \cite{text1,text2,Horodecki:2009zz}.  Entanglement entropy for a pure state is an entanglement monotone because the entropy decreases monotonically under LOCC. However, the entropy for a mixed state 
is not. Logarithmic negativity is an entanglement monotone even for a mixed state. Let us explain how to define $\mathcal{E}$. The definition of logarithmic negativity $\mathcal{E}$ is 
\be
\mathcal{E} \equiv \log{||\rho_{A\cup B}^{T_B}||}=\log{\tr|\rho_{A\cup B}^{T_B}|},
\ee
where $||\rho_{A\cup B}^{T_B}||\equiv \sum_i\left|\lambda_i\right|$ is the trace norm of $\rho_{A\cup B}^{T_B}$, and $\lambda_i$ are eigenvalues of $\rho_{A\cup B}^{T_B}$.  The operation $\cdot^{T_B}$ is called by {\it partial transpose}.  This operation transposes the matrix component as follows,
\be
\begin{split}
&\rho = \sum_{a,b,\alpha,\beta, \mathcal{A}, \mathcal{B} }\bra{a, \alpha, \mathcal{A}}\rho \ket{b, \beta, \mathcal{B}}\ket{a}\bra{b}_A \otimes \ket{\alpha}\bra{\beta}_B \otimes \ket{\mathcal{A}}\bra{\mathcal{B}}_C \\
&\rightarrow \rho^{T_B} = \sum_{a,b,\alpha,\beta, \mathcal{A}, \mathcal{B} }\bra{a, \alpha, \mathcal{A}}\rho \ket{b, \beta, \mathcal{B}}\ket{a}\bra{b}_A \otimes \ket{\beta}\bra{\alpha}_B \otimes \ket{\mathcal{A}}\bra{\mathcal{B}}_C.
\end{split}
\ee
Thus, the partially-transposed reduced density matrix $\rho_{A\cup B}^{T_B}$ is given by
\be
\rho_{A\cup B}^{T_B}=\sum_{a,b,\alpha,\beta, \mathcal{A}}\bra{a, \alpha, \mathcal{A}}\rho \ket{b, \beta, \mathcal{A}}\ket{a}\bra{b}_A \otimes \ket{\beta}\bra{\alpha}_B .
\ee
\subsubsection{Replica trick}

We review the replica trick for logarithmic negativity $\mathcal{E}$ based on Refs.~\cite{Calabrese:2012ew, Calabrese:2012nk, CM4}. Instead of $\tr|\rho_{A\cup B}^{T_B}|$, consider $\tr(\rho_{A\cup B}^{T_B})^n$ for an integer $n$. The $\lambda_i$-dependence of $\tr(\rho_{A\cup B}^{T_B})^n$ for even $n_e$ is different from that for odd $n_o$ as follows,
\begin{align}
\tr(\rho_{A\cup B}^{T_B})^{n_e}=\sum_{\lambda_i>0}|\lambda_i|^{n_e}+\sum_{\lambda_i<0}|\lambda_i|^{n_e}, \label{trne}\\
\tr(\rho_{A\cup B}^{T_B})^{n_o}=\sum_{\lambda_i>0}|\lambda_i|^{n_o}-\sum_{\lambda_i<0}|\lambda_i|^{n_o}.
\end{align}

Since the trace norm $\tr|\rho_{A\cup B}^{T_B}|$ is given by taking the analytic continuation of the even sequence at $n_e \rightarrow 1$, the negativity $\mathcal{E}$ in the replica trick is given by
\begin{align}
\mathcal{E}=\lim_{n_e\to1}\log\tr(\rho_{A\cup B}^{T_B})^{n_e}.
\end{align}

\begin{figure}[htbp]
 \begin{center}
  \includegraphics[width=60mm]{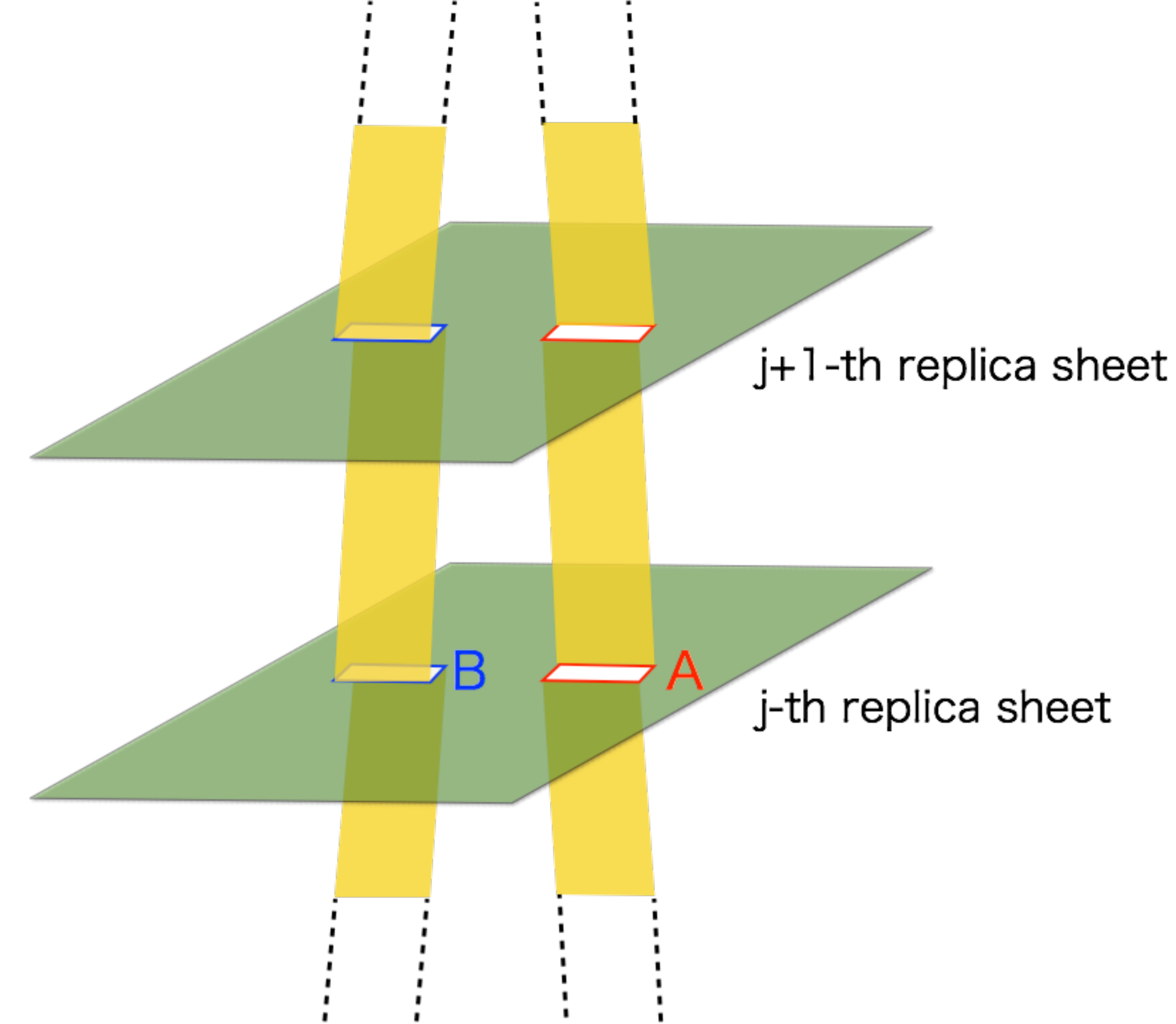}
 \end{center}
 \caption{Replica sheets for the path integral representation of $\tr(\rho_{A\cup B}^{T_B})^{n_e}$. The replica sheets are sewn together cyclically. }
 \label{rs}
\end{figure}

We 
are able to compute $\mathcal{E}$ in 2D CFT by using twist fields. For example, consider the entire space as $A\cup B$ and $A=[u_1, u_2]$. In this system, $\mathcal{E}$ can be expressed as
\begin{align}
\mathcal{E}=\lim_{n_e\to1}\log\langle\mathcal{T}_{n_e}^2(u_1)\bar{\mathcal{T}}_{n_e}^2(u_2)\rangle,
\end{align}
where $\mathcal{T}^2_n$ is the twist field that connects the $j$-th to $(j+2)$-th replica fields.
 For this configuration,  $\mathcal{E}$ corresponds to R\'{e}nyi entanglement entropy with $n=1/2$ \cite{Vidal:2002zz}.  As another example, consider two disjoint intervals $A=[u_1, u_2]$ and $B=[u_3, u_4]$. 
 Logarithmic negativity $\mathcal{E}$  for this configuration is 
\begin{align}
\mathcal{E}=\lim_{n_e\to1}\log\langle\mathcal{T}_{n_e}(u_1)\bar{\mathcal{T}}_{n_e}(u_2)\bar{\mathcal{T}}_{n_e}(u_3)\mathcal{T}_{n_e}(u_4)\rangle, \label{lnAB}
\end{align}
where $\mathcal{T}_n$ is the twist field that connects the $j$-th sheet to $(j+1)$-th one. 
It is well known that the conformal dimensions of $\mathcal{T}_n$ and $\bar{\mathcal{T}}_n$ in 2D CFT with central charge $c$ are \cite{RP1, RP2}
\begin{align}
\Delta_n=\frac{c}{12}\left(n-\frac{1}{n}\right).
\end{align}
Because of the partial transpose $T_B$, the connection between the replica sheets at $B$ is different from that at $A$ as shown in Figure \ref{rs}. This partial transpose $T_B$ has the effect to exchange the twist operators at $u_3$ and $u_4$  compared to the ones in the correlation function $\langle\mathcal{T}_{n_e}(u_1)\bar{\mathcal{T}}_{n_e}(u_2)\mathcal{T}_{n_e}(u_3)\bar{\mathcal{T}}_{n_e}(u_4)\rangle$ for $\tr(\rho_{A\cup B})^{n_e}$.

We consider the time evolution of $\mathcal{E}$ between the two disjoint intervals in 2D CFT with central charge $c$ in a sudden quench, where a mass gap suddenly vanishes. After the gap vanishes, the ground state for an initial Hamiltonian time-evolves under a CFT Hamiltonian.
When we compute $\mathcal{E}$ in the replica trick, $\mathcal{E}$  in Euclidean space is given by the 4-point function of the twist fields on the boundaries of two strips.  
Real time-evolution of $\mathcal{E}$ is given by performing an analytic continuation $\tau=\tau_0+it$, where $\tau$ is the imaginary time of the twist fields, and $t$ is the real time.

We introduce a parameter $\tau_0$ as width of the two strips and interpret $1/\tau_0$ as the energy scale of the initial ground state of the massive theory. In the limit of $t\gg \tau_0$ and $|u_j-u_i|\gg\tau_0$, the time evolution of $\mathcal{E}$ is \cite{CM4}
\begin{align}
\mathcal{E}=\begin{cases}
0& 0\le t<\frac{|u_3-u_2|}{2},\\
\frac{\pi c}{8\tau_0}\left(t-\frac{|u_3-u_2|}{2}\right)& \frac{|u_3-u_2|}{2}<t<\frac{|u_3-u_1|}{2},\\
\frac{\pi c}{8\tau_0}\left(\frac{|u_2-u_1|}{2}\right)& \frac{|u_3-u_1|}{2}<t<\frac{|u_4-u_2|}{2},\\
\frac{\pi c}{8\tau_0}\left(\frac{|u_4-u_1|}{2}-t\right)& \frac{|u_4-u_2|}{2}<t<\frac{|u_4-u_1|}{2},\\
0&\frac{|u_4-u_1|}{2}<t,
\end{cases}\label{2dCFT1}
\end{align}
where we assumed $|u_2-u_1|<|u_4-u_3|$. We note that mutual information $I_{A, B}$ shows the same behavior  up to normalization (for example, see \cite{CM4}).

\subsubsection{Setup and correlator method}
Here, we explain the detail of the dynamical system, a 2D free scalar theory with time dependent masses. We also explain how to compute $I_{A,B}$ and $\mathcal{E}$ by correlation functions.

\subsubsection{Setup}
The excited state in our setup time-evolves with a time-dependent Hamiltonian from the ground state for the initial Hamiltonian. 
In order to study the time evolution of quantum correlation between subsystem $A$ and $B$, 
we define the change of mutual information and logarithmic negativity $\Delta I_{A,B}$ and $\Delta\mathcal{E}$ by
\be
\begin{split}
&\Delta I_{A,B}(t) = I_{A,B}(t)- I_{A,B}(t=-\infty),  \\
&\Delta \mathcal{E}(t)=\mathcal{E}(t)- \mathcal{E}(t=-\infty), \\
\end{split}
\ee
where $I_{A,B}(t)$ and $\mathcal{E}(t)$ are measured at $t$, and  $I_{A,B}(t=-\infty)$ and $\mathcal{E}(t=-\infty)$ are measured at $t=-\infty$. 
We compute $\Delta \mathcal{E}$ and $\Delta I_{A,B}$ with the time-dependent Hamiltonian, which is  for $(1+1)$ dimensional free boson theory with time-dependent mass $\tilde{m}(t)$,
\begin{equation}
H(t)= \frac{1}{2}\int dx \left[\pi(x)\pi(x)+\left(\nabla \phi(x)\right)^2+\tilde{m}(t)^2\phi(x)^2\right],
\end{equation}
where the time-dependent mass $\tilde{m}(t)$ is dimensionful and changes continuously in time.
The time-dependent mass $\tilde{m}(t)$ in early time limit $t \rightarrow -\infty$ approaches a constant mass asymptotically. 
Therefore, we approximate $I_{A,B}(t=-\infty)$ and $\mathcal{E}(t=-\infty)$ by mutual information and logarithmic negativity for the ground state of the free scalar theory with a static mass $\tilde{m}$.

We compute $I_{A,B}(t)$ and $\mathcal{E}(t)$ by a method, correlator method, where  these quantities for gaussian states are given by  the summations of eigenvalue-dependent entropies and logarithmic negativities $s_A(\gamma)$ and $\epsilon(\chi)$ \cite{CM1,CM2,CM3,CM4,CM5,CM6}.
The eigenvalues $\gamma$ and $\chi$ are for matrices whose components are correlators in the subregions.
Since the dimension of these matrices depends on the degrees of freedom in the subsystems, it becomes finite in a discretized system. After discretizing the system, we are able to diagonalize the correlator matrices, and we are able to compute  $\Delta I_{A,B}$ and $\Delta\mathcal{E}$.

We discretize the system by putting it on a circle with the circumference $L=N \epsilon$ and the lattice spacing $\epsilon$.
After replacing $\int dx \rightarrow \epsilon \sum_{n=0}^{N-1}$ and $\phi \rightarrow q_n, \pi \rightarrow p_n/\epsilon, \tilde{m}(t) \rightarrow m(t)/\epsilon$ as well as $\nabla \phi \rightarrow (q_{n+1}-q_n)/\epsilon$, the discrete Hamiltonian on the circle is,
\begin{equation}
H(t)=\frac{1}{2\epsilon}\sum_{n=0}^{N-1}\left[p_{n}^2+(2+ m(t)^2)q_{n}^2-2 q_{n+1}q_{n}\right],
\end{equation}
where we assume that $N=2\mathcal{M}+1$ ($\mathcal{M} \in {\bf Z}$). We choose $\epsilon$ as a unit and define the dimensionless parameters $t, l, \delta t$ and $\xi$ 
\begin{align}
t =\f{\tilde{t}}{\epsilon} ,~~l = \f{\tilde{l}}{\epsilon},~~ \delta t  =\f{\delta \tilde{t}}{\epsilon},~~ \xi= \f{\tilde{\xi}}{\epsilon}.
\end{align}
where $\tilde{t}, \tilde{l}, \delta \tilde{t}$ and $\tilde{\xi}$ are dimensionful.

The variables $q_n$ and $p_n$ are subject to a periodic boundary condition:
\begin{equation}
p_0=p_{N-1}, ~q_0=q_{N-1}.
\end{equation}
Also, they have a canonical commutation relation:
\begin{equation}
[q_a, p_b]= i \delta_{ab}, ~~[q_a, q_b]= [p_a, p_b]=0.
\end{equation}
Discrete Fourier transforms are given by,
\begin{equation}
\begin{split}
q_n=\frac{1}{\sqrt{N}}\sum_{\kappa=-\frac{N-1}{2}}^{\frac{N-1}{2}}e^{i \frac{2\pi n \kappa}{N}}\tilde{q}_{\kappa}, \\
p_n=\frac{1}{\sqrt{N}}\sum_{\kappa=-\frac{N-1}{2}}^{\frac{N-1}{2}}e^{i \frac{2\pi n \kappa}{N}}\tilde{p}_{\kappa}, \\
\end{split}
\end{equation}
where $\tilde{q}^{\dagger}_l=\tilde{q}_{-l}$ and $\tilde{p}^{\dagger}_l=\tilde{p}_{-l}$ because $p_{\kappa}$ and $q_{\kappa}$ are real.
The Hamiltonian in the momentum space is
\begin{equation}
H(t)=\frac{1}{2\epsilon}\sum_{\kappa=-\frac{N-1}{2}}^{\frac{N-1}{2}}\left[\tilde{p}_{\kappa} \tilde{p}^{\dagger} _{\kappa}+ \left(4\sin^2{\left(\frac{\pi {\kappa}}{N}\right)}+ m^2(t)\right)\tilde{q}_{\kappa}\tilde{q}^{\dagger}_{\kappa}\right],
\end{equation}
where $\tilde{p}$ and $\tilde{q}$ are subject to the commutation relations, $\left[\tilde{q}_a, \tilde{p}_b \right]=i\delta_{a, -b}$ and  $\left[\tilde{p}_a, \tilde{p}_b\right]=\left[\tilde{q}_a, \tilde{q}_b\right]=0$.
The variables $\tilde{p}$ and $\tilde{q}$ written in terms of a function $f_{{\kappa}}$, are given by
\begin{equation}\label{fs}
\begin{split}
&\tilde{q}_{\kappa} =f_{\kappa}(t) a_{\kappa} + f_{-\kappa}^*(t) a^{\dagger}_{-\kappa}, \\
&\tilde{p}_{\kappa}=\dot{f}_{\kappa}(t) a_{\kappa} + \dot{f}_{-\kappa}^*(t) a^{\dagger}_{-\kappa},
\end{split}
\end{equation}
where $\left[a_k, a_l\right]=\left[a^{\dagger}_k, a^{\dagger}_l\right]=0$ and $\left[a_k, a^{\dagger}_l\right]=\delta_{k,l}$. The function $f_{\kappa}(t)$ has a property $f_{\kappa}(t)=f_{-{\kappa}}(t)$. Then, $f_{\kappa}(t)$ satisfies the following Wronskian condition:
\begin{equation}
\begin{split}
&\left[\tilde{q}_L, \tilde{p}_{-l}\right]=\left( f_l(t)\dot{f}^*_l(t)-\dot{f}_l(t)f^*_l(t)\right)\delta_{l,L}=i \delta_{l ,L}. \end{split}
\end{equation}

After taking a thermodynamic limit where $N,L\rightarrow \infty$, and $\epsilon$ is finite, $f_\kappa(t)$ is determined by the following equation of motion:
\begin{equation}
\frac{d^2 f_{\kappa}(t)}{dt^2}+\left[4 \sin^2{\left(\frac{{\kappa}}{2}\right)}+ m^2(t)\right]f_{\kappa}(t)=0.
\end{equation}

The coordinates and its conjugate momenta in the thermodynamic limit are given by
\begin{equation}
\begin{split}
q_n=X_n =\int^{\pi}_{-\pi} \frac{d{\kappa}}{\sqrt{2\pi}}\tilde{q}_{\kappa} e^{i {\kappa} n}, \\
p_n=P_n= \int^{\pi}_{-\pi} \frac{d{\kappa}}{\sqrt{2\pi}}\tilde{p}_{\kappa} e^{i {\kappa} n}, \\
\end{split}
\end{equation} 
where $\tilde{p}$ and  $\tilde{q}$ are defined in (\ref{fs}).
\subsubsection{Correlator method}

Here, let us explain the correlator method where entanglement entropy and logarithmic negativity are related to eigenvalues of matrices whose components are given by correlation functions in the subsystem (see, for example, \cite{CM6} for entanglement entropy and \cite{CM1} for logarithmic negativity).
If the given subsystem is $\mathcal{V}$, we can define correlation functions $X_{ab}(t), P_{ab}(t)$ and $D_{ab}(t)$ by
\begin{equation}\label{spect}
\begin{split}
&X_{ab}(t)=\left\langle X_a (t)X_b(t) \right \rangle =\int^{\pi}_{-\pi} \frac{d\kappa}{2\pi}X_{\kappa}\cos{\left(\kappa\left|a-b\right|\right)} =\int^{\pi}_{-\pi} \frac{d\kappa}{2\pi}\left|f_{\kappa}(t)\right|^2\cos{\left(\kappa\left|a-b\right|\right)}, \\
&P_{ab}(t)=\left\langle P_a (t)P_b(t) \right \rangle = \int^{\pi}_{-\pi} \frac{d\kappa}{2\pi}P_\kappa\cos{\left(\kappa\left|a-b\right|\right)} = \int^{\pi}_{-\pi} \frac{d\kappa}{2\pi}\left|\dot{f}_{\kappa}(t)\right|^2\cos{\left(\kappa\left|a-b\right|\right)}, \\
&D_{ab}(t)=\frac{1}{2}\left\langle \left\{X_a (t), P_b(t) \right\} \right \rangle=\int^{\pi}_{-\pi} \frac{d\kappa}{2\pi}D_{\kappa}\cos{\left(\kappa\left|a-b\right|\right)} =\int^{\pi}_{-\pi} \frac{d\kappa}{2\pi}Re\left[\dot{f}^*_{\kappa}(t)f_{\kappa}(t)\right]\cos{\left(\kappa\left|a-b\right|\right)}, \\
\end{split}
\end{equation}
where local operators, $X_a(t)$ and $P_a(t)$, are in $\mathcal{V}$.  If the size of $\mathcal{V}$ is $L$, we can define two $2L \times 2L $ matrices, $\Gamma$ and $J$, by 
\begin{equation}
\begin{split}
&J=\begin{bmatrix}
0 & I_{L \times L} \\
-I_{L \times L} & 0 \\
\end{bmatrix},
\Gamma=\begin{bmatrix}
X_{ab}(t) & D_{ab}(t) \\
D_{ab}(t) & P_{ab}(t) \\
\end{bmatrix}. \\
\end{split}
\end{equation}
Entanglement entropy $S_A$ is given by the summation of eigenvalue-dependent entropies  $s_A(\gamma_{\kappa})$, 
\begin{equation}
S_A=\sum_{\kappa=1}^l s_A(\gamma_{\kappa})=\sum_{\kappa=1}^{l}\left[\left(\gamma_{\kappa}+\frac{1}{2}\right)\log{\left(\gamma_{\kappa}+\frac{1}{2}\right)}-\left(\gamma_{\kappa}-\frac{1}{2}\right)\log{\left(\gamma_{\kappa}-\frac{1}{2}\right)}\right],
\end{equation}
where  $\gamma_\kappa$ are positive eigenvalues of a $2L \times 2L$ matrix $\mathcal{M}=iJ\Gamma$.

In the correlator method, a given partial transposed reduced density matrix $\rho_{A\cup B}^{T_B}$ is 
\be
\begin{split}
\Gamma_{A\cup B}^{T_B} = \begin{pmatrix}
{\bf 1} & {\bf 0} \\
{\bf 0} & {\bf Y}_{B} \\
\end{pmatrix} \cdot \Gamma \cdot 
\begin{pmatrix}
{\bf 1} & {\bf 0} \\
{\bf 0} & {\bf Y}_{B} \\
\end{pmatrix}
=\mathcal{Y}\cdot \Gamma \cdot \mathcal{Y},
\end{split}
\ee
where $\mathcal{Y}$ is a $2L \times 2L$ matrix, and the size of $A\cup B$ is $L$. ${\bf Y}_B$ is a diagonal $L \times L$ matrix which transforms $P_b$ in $B$ into $-P_b$ and $+P_a$ otherwise.
The trace norm of $\rho_{A\cup B}^{T_B} $ in correlator method is
\be
|| \rho_{A \cup B}^{T_B}||= \prod^{L}_{a=1}\left[\left|\chi_a+\f{1}{2}\right|-\left|\chi_a-\f{1}{2}\right|\right]^{-1}= \prod^{L}_{a=1}\text{max}\left[1, \f{1}{2\chi_a}\right]
\ee
where $\chi_a$ are positive eigenvalues of $i J \cdot \Gamma^{T_B}$. The index $a$ runs from $1$ to $L$. Therefore, the logarithmic negativity $\mathcal{E}$ is given by
\be
\mathcal{E}=\log{\left(\prod^{L}_{a=1}\text{max}\left[1, \f{1}{2\chi_a}\right]\right)}.
\ee

\section{Smooth quench}
Here, we explain the protocols which trigger the time evolution of $\Delta I_{{A,B}}$ and $\Delta \mathcal{E}$. The protocols in this paper have the time-dependent masses, one of which is end-critical-protocol (ECP), and the other is Cis-critical-protocol(CCP).
\subsection{End-critical-protocol (ECP) and Cis-critical-protocol (CCP)}

The time-dependent masses in this paper depend on two parameters $\xi$ and $\delta t$. Here, $\xi$ and $\delta t$ are an initial correlation length and quench rate. The time-dependent mass $m(t)$ in ECP is,
\be \label{ecp}
m^2(t)=\f{1}{2\xi^2}\left[1-\tanh{\left(\f{t}{\delta t}\right)}\right],
\ee
which is constant approximately at early time $t \ll -\delta t$, decreases after $t=-\delta t$, and vanishes at late time $\delta t \gg t$.

The mass in CCP is, 
\be \label{ccp}
m^2(t)=\f{1}{\xi^2}\ \tanh^2{\left(\f{t}{\delta t}\right)},
\ee
which is constant approximately in the time region $t\ll -\delta t$, decreases around $t=-\delta t$, vanishes around $t=0$, increases after $t=0$, and approaches a constant values asymptotically. The dynamics of quantum entanglement is governed by the Hamiltonian with the time-dependent masses in (\ref{ecp}) and (\ref{ccp}). 

As in \cite{Das:2014hqa}, the function $f_{\kappa}(t)$ for the time-dependent masses in (\ref{ecp}) and (\ref{ccp}) is analytically computable because the equation of motion for $f_{\kappa}(t)$ is given by
\be 
\frac{d^2 f_{\kappa}(t)}{dt^2}+\left[4 \sin^2{\left(\frac{\kappa}{2}\right)}+\epsilon^2 m^2(t)\right]f_{\kappa}(t)=0.\label{efk}
\ee
The solution of equation of motion (\ref{efk}) for the mass in ECP $f_{\kappa}(t)$  is given  by
\be
\begin{split}
f_{\kappa}(t)=&\f{1}{\sqrt{2\sqrt{\f{A+B}{\delta t}}}}\left(\frac{1+\tanh{\left(t/\delta t\right)}}{2}\right)^{-\alpha}\left(\frac{1-\tanh{\left(t/\delta t\right)}}{2}\right)^{\beta}\\
&\times~_2F_1\left(-\alpha+\beta,1-\alpha+\beta,1-2\alpha,\frac{1+\tanh{\left(t/\delta t\right)}}{2}\right), \\
\end{split}
\ee
where the parameters $A,B, \alpha$ and $\beta$ are given by
\be
\begin{split}
&A=4\delta t^2 \sin^2{\left(\frac{\kappa}{2}\right)}+\f{\delta t^2 \epsilon^2}{2\xi^2}, ~~B=-\f{ \delta t^2 \epsilon}{\xi^2}, \\
&\alpha=i \f{\sqrt{A-B}}{2}, \beta=\f{\sqrt{A+B}}{2}.
\end{split}
\ee
On the other hand, the solution of equation of motion with the mass for CCP is given by
\be
\begin{split}
f_{\kappa}(t)=&C \cosh^{2\alpha}{\left(\f{t}{\delta t}\right)} ~_2F_1\left(a, b, \frac{1}{2}, -\sinh^2{\left(\f{t}{\delta t}\right)} \right) \\
&+D \cosh^{2\alpha}{\left(\f{t}{\delta t}\right)} \sinh{\left(\f{t}{\delta t}\right)}~_2F_1\left(\frac{1}{2}+a, \frac{1}{2}+b,\frac{3}{2}, -\sinh^2{\left(\f{t}{\delta t}\right)} \right)
\end{split}
\ee
where parameters $a,b,C,D,$ and $\alpha$ are 
\be
\begin{split}
&a=\alpha-i \sqrt{A+\frac{B}{4}}, b=\alpha+i \sqrt{A+\frac{B}{4}},C=\f{2^{i \omega_0 \delta t}}{\sqrt{2\omega_0}}\frac{E_{3/2}'}{E_{1/2}E_{3/2}'-E_{3/2}E_{1/2}'}, \\
&D=\f{2^{i \omega_0 \delta t}}{\sqrt{2\omega_0}}\frac{E_{1/2}'}{E_{1/2}E_{3/2}'-E_{3/2}E_{1/2}'}, \alpha=\frac{1+\sqrt{1-4B^2}}{4},\\
&E_{1/2}=\frac{\Gamma(1/2)\Gamma(b-a)}{\Gamma(b)\Gamma(1/2-a)}, E_{3/2}=\f{\Gamma(3/2)\Gamma(b-a)}{\Gamma(b+1/2)\Gamma(1-a)}, E'_c=E(a\leftrightarrow b),\\
&A=\delta t^2 \sin^2{\left(\frac{\kappa}{2}\right)}, B=\epsilon^2 \delta t^2 m_0^2,\omega_0=\sqrt{4\sin^2{\left(\frac{\kappa}{2}\right)}+\epsilon \left(m_0^2\right)}.
\end{split}
\ee

Since the protocols in (\ref{ecp}) and (\ref{ccp}) have two parameters $\xi$ and $\delta t$, the protocols change slowly or fast by changing the parameters. We study the time evolution of $\Delta \mathcal{E}$ and $\Delta I_{A,B}$ in two limits, which will be explained later.

\subsubsection{Kibble-Zurek mechanism}\label{KZ mech}
Let us explain criteria for the adiabaticity. Suppose an ansatz of $f_{\kappa}(t)$ in  (\ref{efk}) is
\begin{align}
f_{\kappa}(t)=\frac{1}{\sqrt{2W_{\kappa}(t)}}e^{-i\int^tW_{\kappa}(t') dt'}.\label{fkt}
\end{align}
By substituting (\ref{fkt}) into (\ref{efk}), $W_{\kappa}(t)$ has to be subject to
\begin{align}
W^2_{\kappa}(t)=\omega^2_{\kappa}(t)-\left[\frac{\ddot{W_{\kappa}}}{2W_{\kappa}}-\frac{3}{4}\left(\frac{\dot{W_{\kappa}}}{W_{\kappa}}\right)^2\right], \;\; \omega_{\kappa}^2(t)=4 \sin^2{\left(\frac{ \kappa}{2}\right)}+m^2(t).\label{AE}
\end{align}
When $\left|\frac{\ddot{W_{\kappa}}}{2W_{\kappa}}-\frac{3}{4}\left(\frac{\dot{W_{\kappa}}}{W_{\kappa}}\right)^2\right|$ in (\ref{AE}) is much smaller than $\omega^2_{\kappa}(t)$, the adiabatic approximation $W_{\kappa}(t)\sim \omega_{\kappa}(t)$ is valid.
Then, the condition where we can use the adiabatic approximation is (see, for example, \cite{Dabrowski:2014ica})
\begin{align}
\left|\frac{\ddot{\omega_{\kappa}}}{2\omega_{\kappa}^3}-\frac{3}{4}\left(\frac{\dot{\omega_{\kappa}}}{\omega_{\kappa}^2}\right)^2\right|\ll1.\label{CAE}
\end{align}
We focus on the LHS at $\kappa=0$ in (\ref{CAE}) because it is the largest for any $\kappa$.
 We also assume $|\frac{\ddot{m}}{m^3}|\lessapprox\left(\frac{\dot{m}}{m^2}\right)^2$. Thus, we obtain criteria for the adiabaticity
\begin{align}
 C_L(t)=\left|\frac{1}{m^2(t)}\times \frac{d m(t)}{dt}\right|\ll1.\label{LC}
 \end{align}
If $C_L(t)$ satisfies the condition (\ref{LC}), we can examine the time evolution of systems by using the adiabatic approximation. The Kibble-Zurek time $t_{\text{kz}}$ is defined by
\begin{align}
 C_L(t_{\text{kz}})\simeq1,
 \end{align}
and it is the time scale at which the adiabaticity breaks down.

In this paper, we consider two limits of the quenches, one of which is a fast limit:
\begin{align}
\xi\gg\delta t,
\end{align}
and $\xi$ is 
the initial correlation length, and the other is a slow limit:
\begin{align}
\xi\ll\delta t,
\end{align}

The adiabaticity is broken at late time in the slow ECP limit and is broken only near $t=0$ in the slow CCP limit.  In the slow ECP and CCP, the Kibble-Zurek time $t_{\text{kz}}$ is (see, for example, \cite{Nishida:2017hqd})
\begin{align}
&t_{\text{kz}}\sim \delta t \log[\delta t/\xi],\;\;\;\xi_{\text{kz}}=\frac{1}{m(t_{\text{kz}})}\sim\frac{1}{\delta t}\;\;\;(\textrm{ECP}),
\label{ECPkz}\\
&t_{\text{kz}}\sim (\delta t\xi)^{\frac{1}{2}},\;\;\;\xi_{\text{kz}}=\frac{1}{m(-t_{\text{kz}})}\sim(\delta t\xi)^{\frac{1}{2}}\;\;\;(\textrm{CCP}),
\label{CCPkz}
\end{align}
where we also defined an length scale $\xi_{\text{kz}}$ by the inverse mass at the Kibble-Zurek time. In the slow quenches, 
$t_{\text{kz}}$ and $\xi_{\text{kz}} $ are the effective time and length scales, respectively.
 We also define an energy scale $E_{\text{kz}}=1/\xi_{\text{kz}}$ in the slow ECP limit. The Kibble-Zurek mechanism \cite{Kibble:1976sj, Zurek:1985qw} conjectures that the dynamics in the slow limit is frozen when the adiabaticity is broken. For example, the effective correlation length in the slow CCP, which is determined at $t\simeq -t_{\text{kz}}$, is almost time-independent from $t\simeq-t_{\text{kz}}$ to $t\simeq t_{\text{kz}}$.

\section{Time evolution of logarithmic negativity and mutual information}

\subsection{Time evolution in ECP}

In the beginning, we calculate $\Delta \mathcal{E}$ and $\Delta I_{A,B}$ numerically in the ECP\footnote{\label{f5}In order to compute the integrals in (\ref{spect}), we replace $\int^\pi_{-\pi}d\kappa g(\kappa)$ with a summation of $g(\kappa_i)$ numerically, where $\kappa_i=-\pi+\varepsilon j$, $j$ is an integer, and $\varepsilon$ is a small parameter for our numerical computations. We compute $\Delta \mathcal{E}$ and $\Delta I_{A,B}$ in the ECP with $\varepsilon=0.0001, 0.00001, 0.000001$. We find no significant difference between them with the different values of $\varepsilon$.   }. Here, we consider the ECP in two limits: the fast limit and the slow limit. We study how $\Delta \mathcal{E}$ and  $\Delta I_{A,B}$ time-evolve in these limits. The similar setup to the fast limit was studied in \cite{CM4}, where the authors have studied time evolution of $\Delta I_{A,B}$ and $\Delta \mathcal{E}$ in a global quench. The quantities $\Delta I_{A,B}$ and $\Delta \mathcal{E}$ in the fast limit change in time in the similar manner to those in the global quench.

In section \ref{sec_ecpf}, we show the time evolution of $\Delta \mathcal{E}$ and  $\Delta I_{A,B}$ in the fast limit, which are computed numerically, and how their time-evolution depends on $\xi$ and $\delta t$. After that, we interpret their time-evolution in terms of our toy model where relativistic propagation of local objects contributes to quantum correlation between the subsystems. In section \ref{sec_ecps}, we show the time evolution of $\Delta \mathcal{E}$ and  $\Delta I_{A,B}$ in the slow limit and how they depend on the parameters. Again, we interpret their evolution in terms of the relativistic propagation of ``particles" created 
at the Kibble-Zurek time $t_{\text{kz}}$.

\subsubsection{Fast limit}\label{sec_ecpf}

\begin{figure}[htbp]
\begin{tabular}{c}
 \begin{minipage}{0.15\hsize}
\begin{center}
\includegraphics[width=45mm]{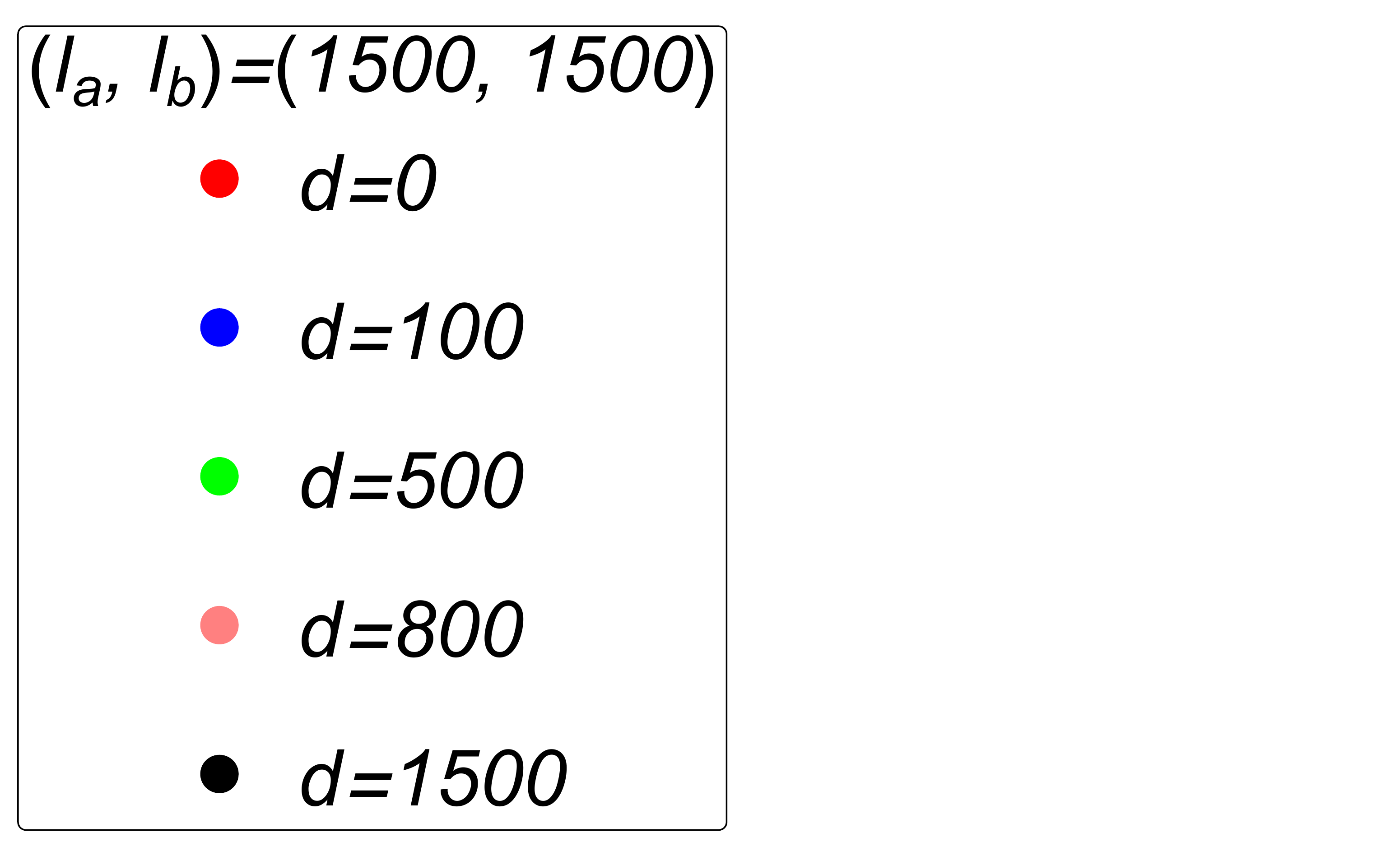}
 \end{center}
 \end{minipage}
 \begin{minipage}{0.4\hsize}
  \begin{center}
   \includegraphics[width=55mm]{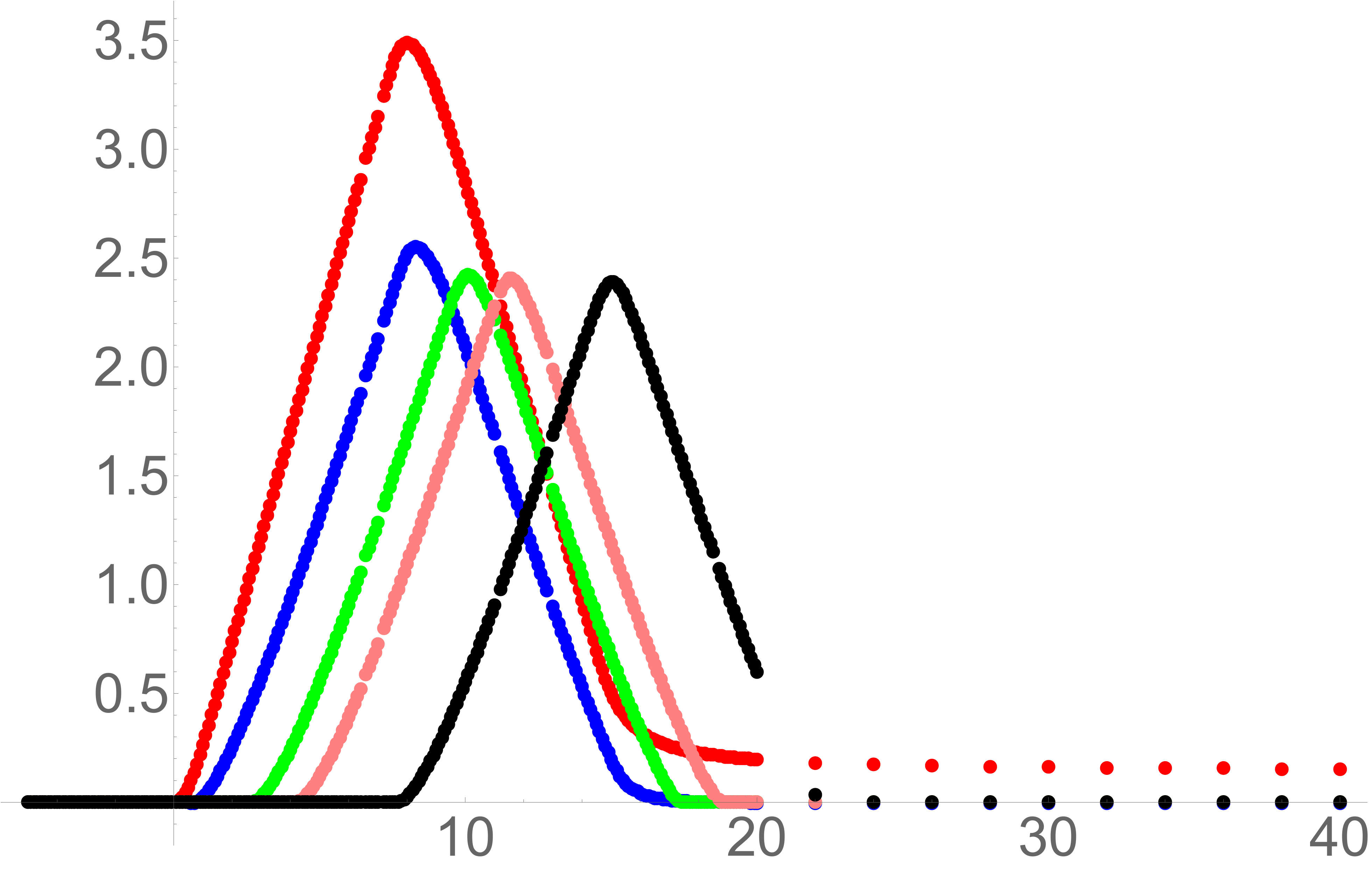}
  \end{center}
 \end{minipage}
   \put(-10,-40){$t/\xi$}
   \put(-160,60){$\Delta \mathcal{E}$}
   \put(-180,65){(a)}
 \begin{minipage}{0.4\hsize}
 \begin{center}
  \includegraphics[width=55mm]{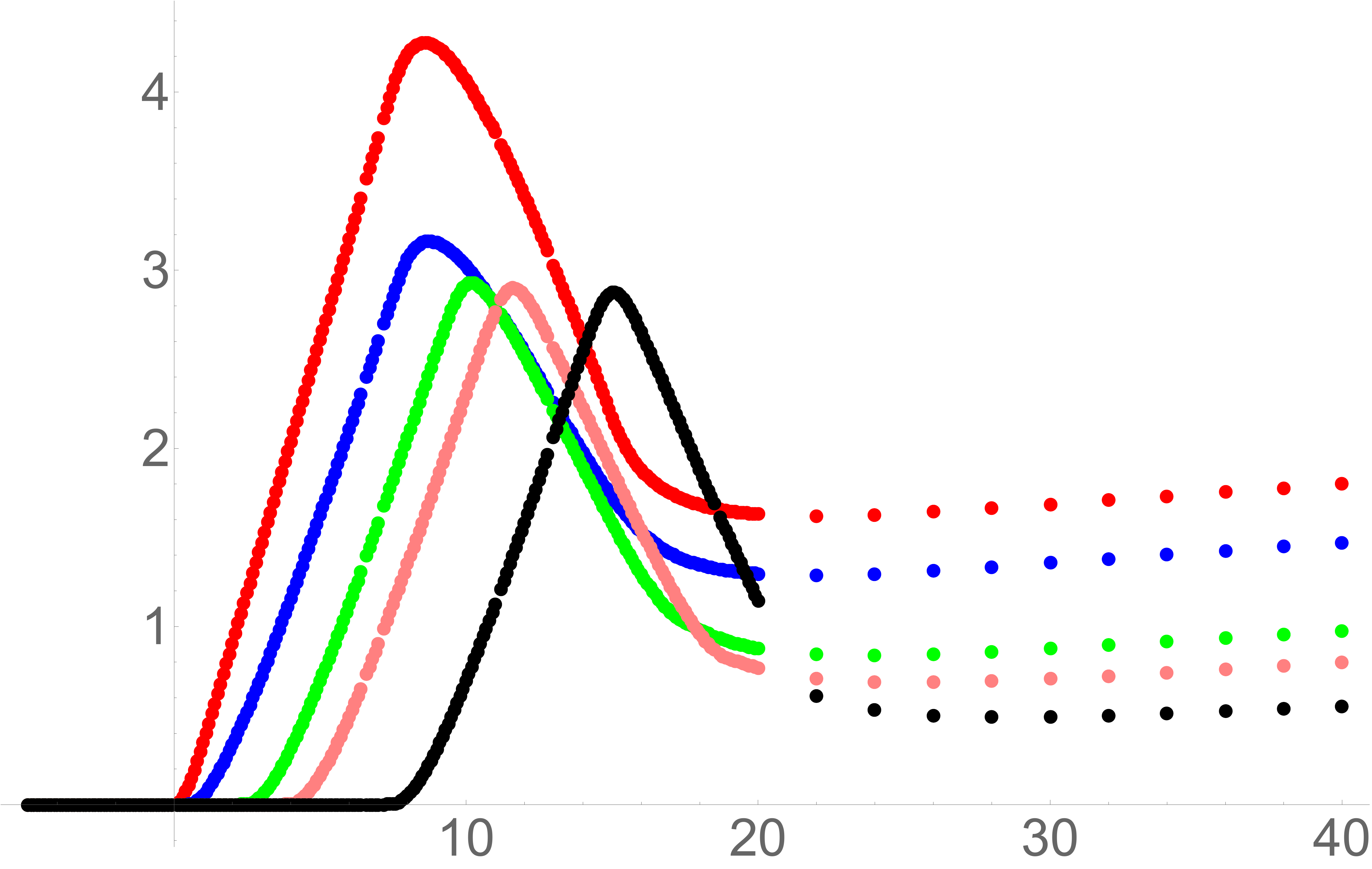}
 \end{center}
  \end{minipage}
     \put(-10,-40){$t/\xi$}
   \put(-160,60){$\Delta I_{A,B}$}
   \put(-180,65){(b)}
\vspace{3mm} \\ 
     \begin{minipage}{0.15\hsize}
  \begin{center}
   \includegraphics[width=45mm]{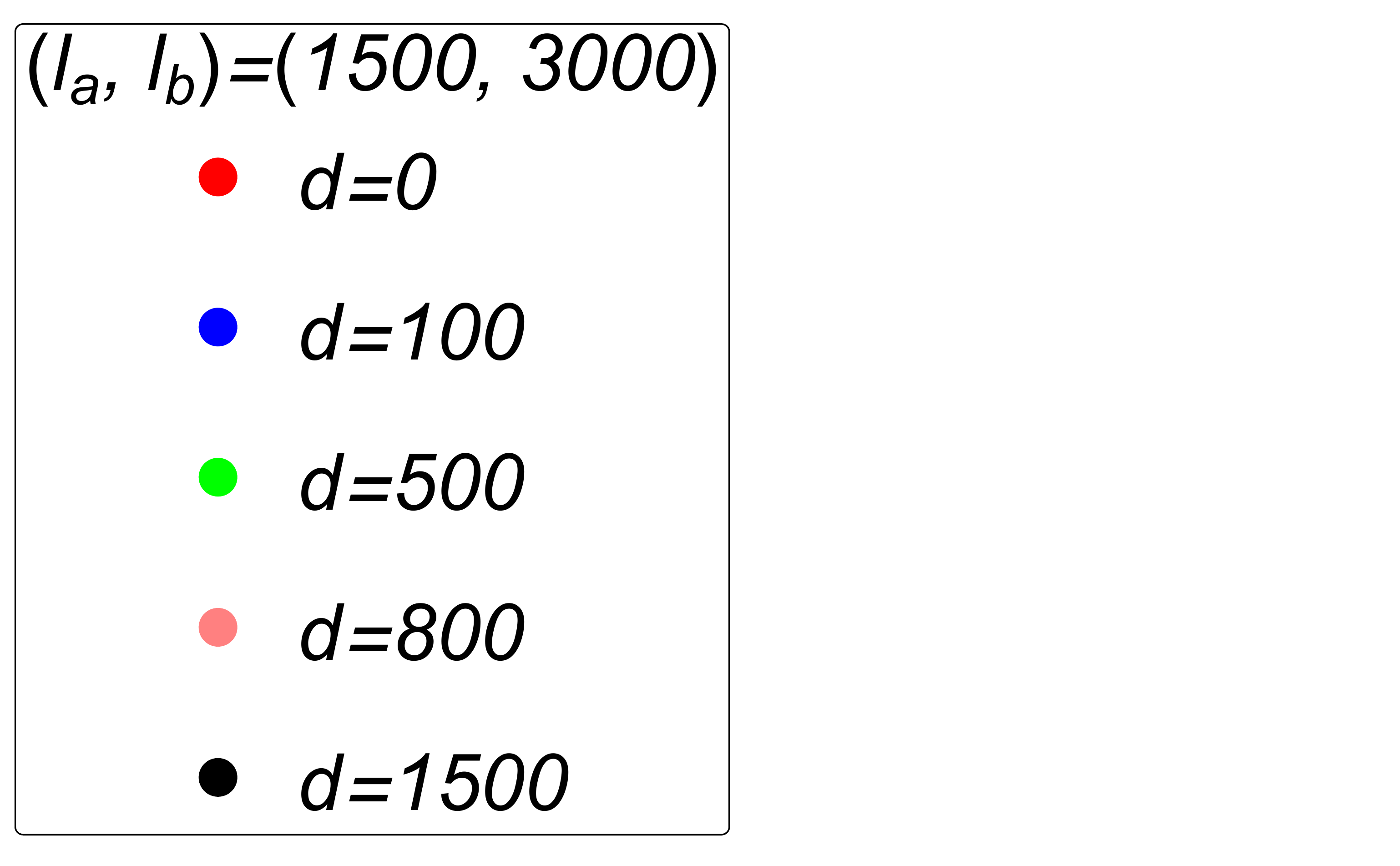}
  \end{center}
 \end{minipage}
 \begin{minipage}{0.4\hsize}
  \begin{center}
   \includegraphics[width=55mm]{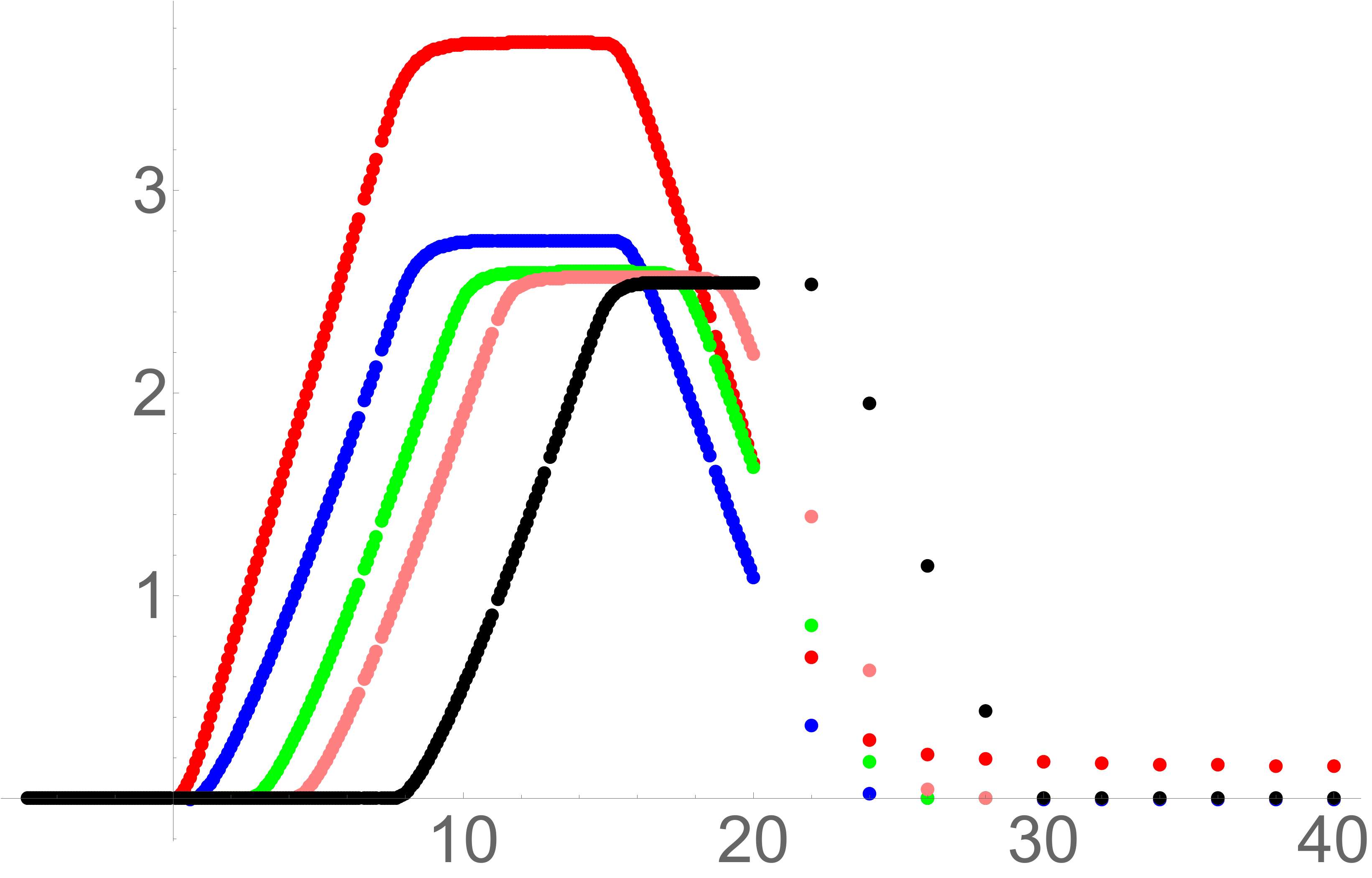}
  \end{center}
 \end{minipage}
   \put(-10,-40){$t/\xi$}
   \put(-160,60){$\Delta \mathcal{E}$}
      \put(-180,65){(c)}
 \begin{minipage}{0.4\hsize}
 \begin{center}
  \includegraphics[width=55mm]{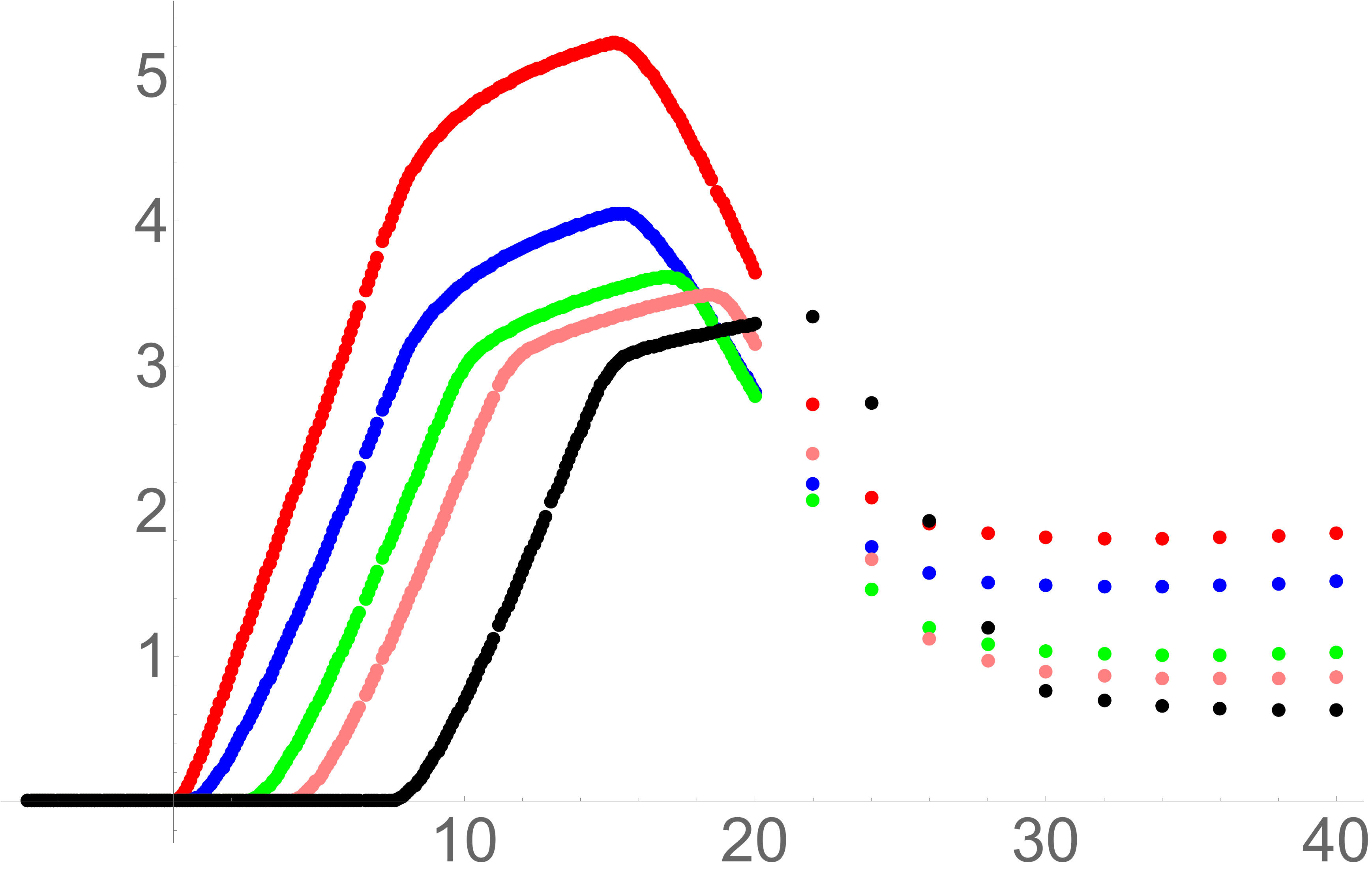}
 \end{center}
  \end{minipage}
     \put(-10,-40){$t/\xi$}
   \put(-160,60){$\Delta I_{A,B}$}
      \put(-180,65){(d)}
 \vspace{3mm} \\ 
     \begin{minipage}{0.15\hsize}
  \begin{center}
   \includegraphics[width=45mm]{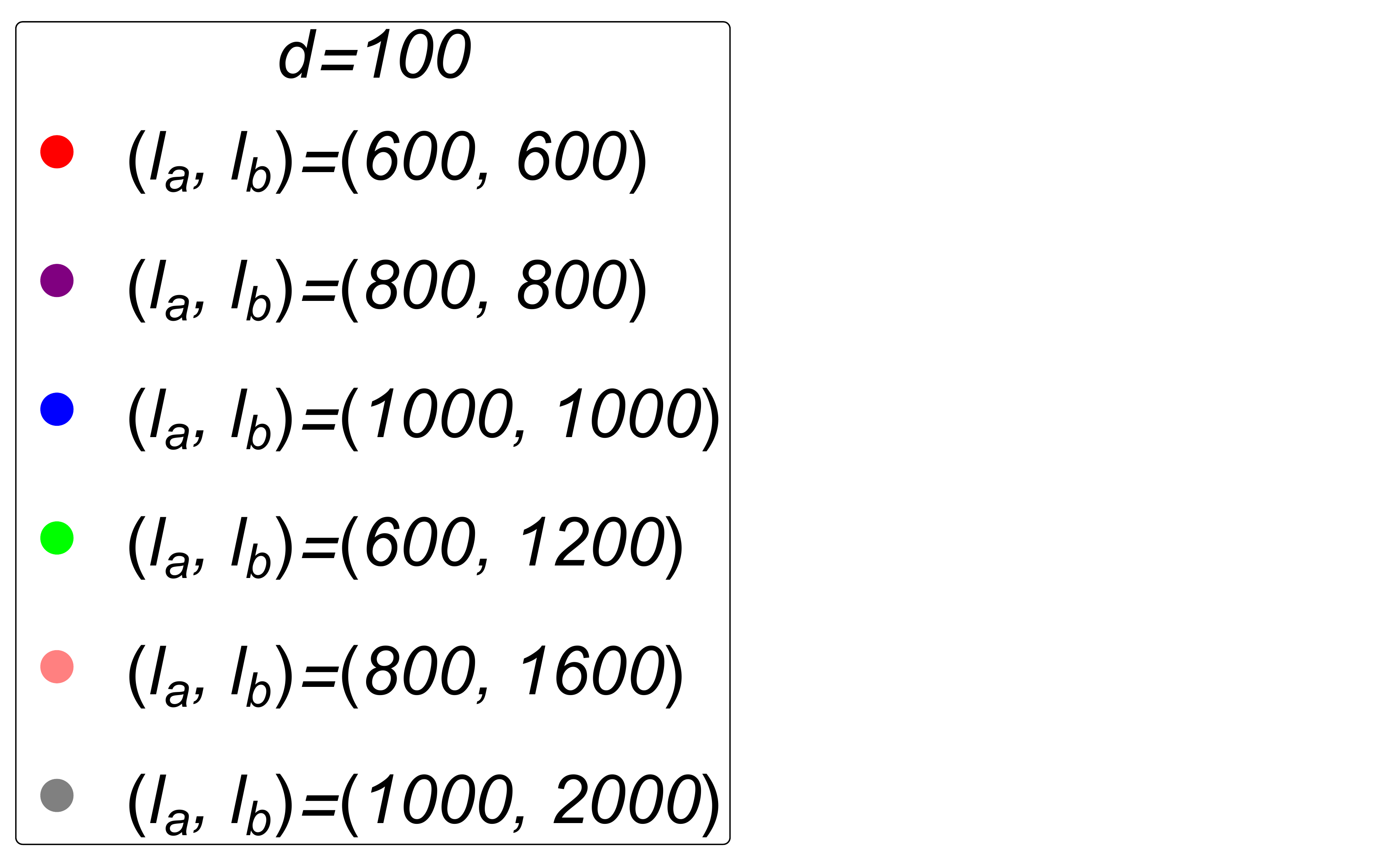}
  \end{center}
 \end{minipage}
 \begin{minipage}{0.4\hsize}
  \begin{center}
   \includegraphics[width=55mm]{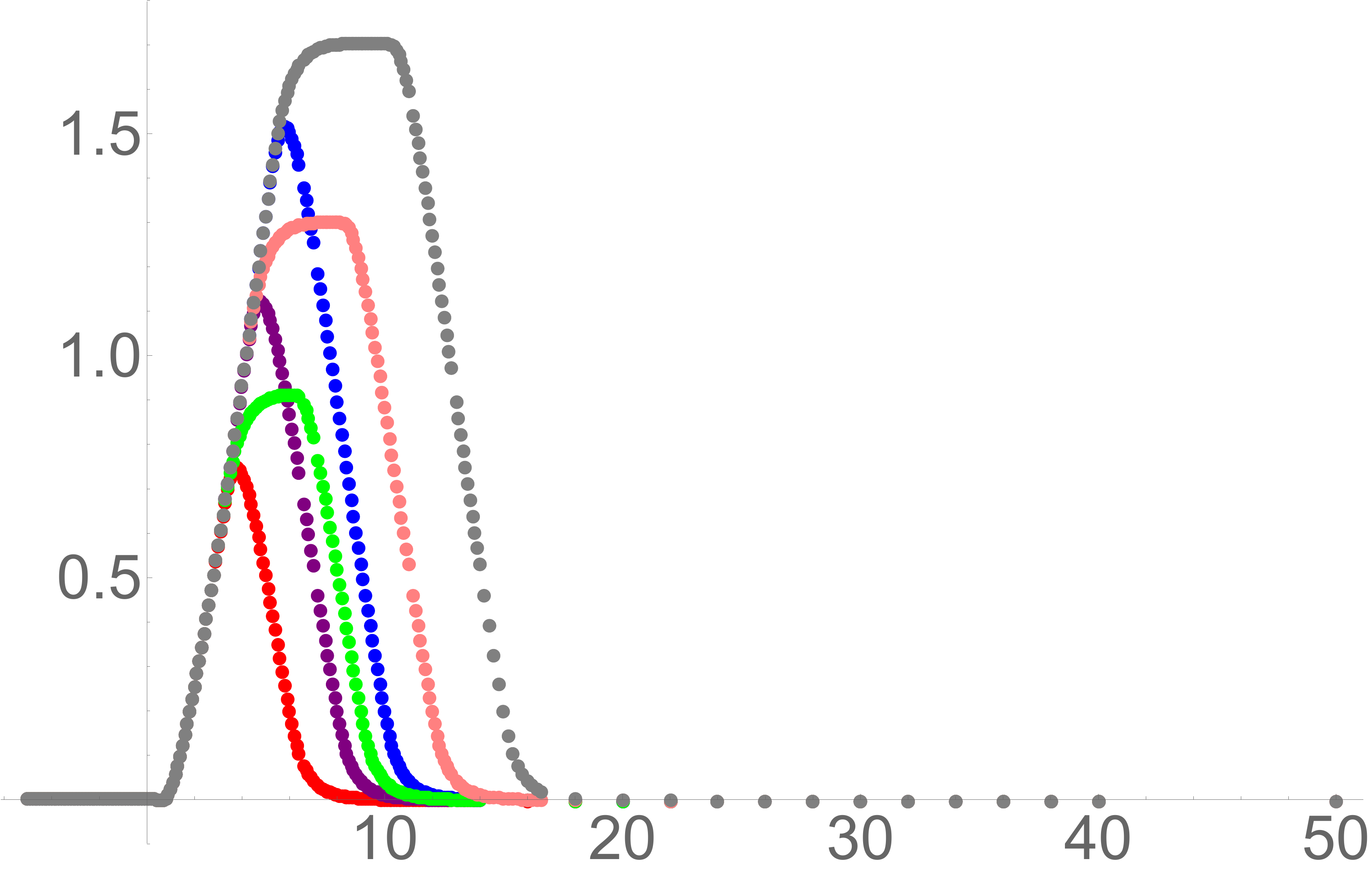}
  \end{center}
 \end{minipage}
   \put(-10,-40){$t/\xi$}
   \put(-160,60){$\Delta \mathcal{E}$}
      \put(-180,65){(e)}
 \begin{minipage}{0.4\hsize}
 \begin{center}
  \includegraphics[width=55mm]{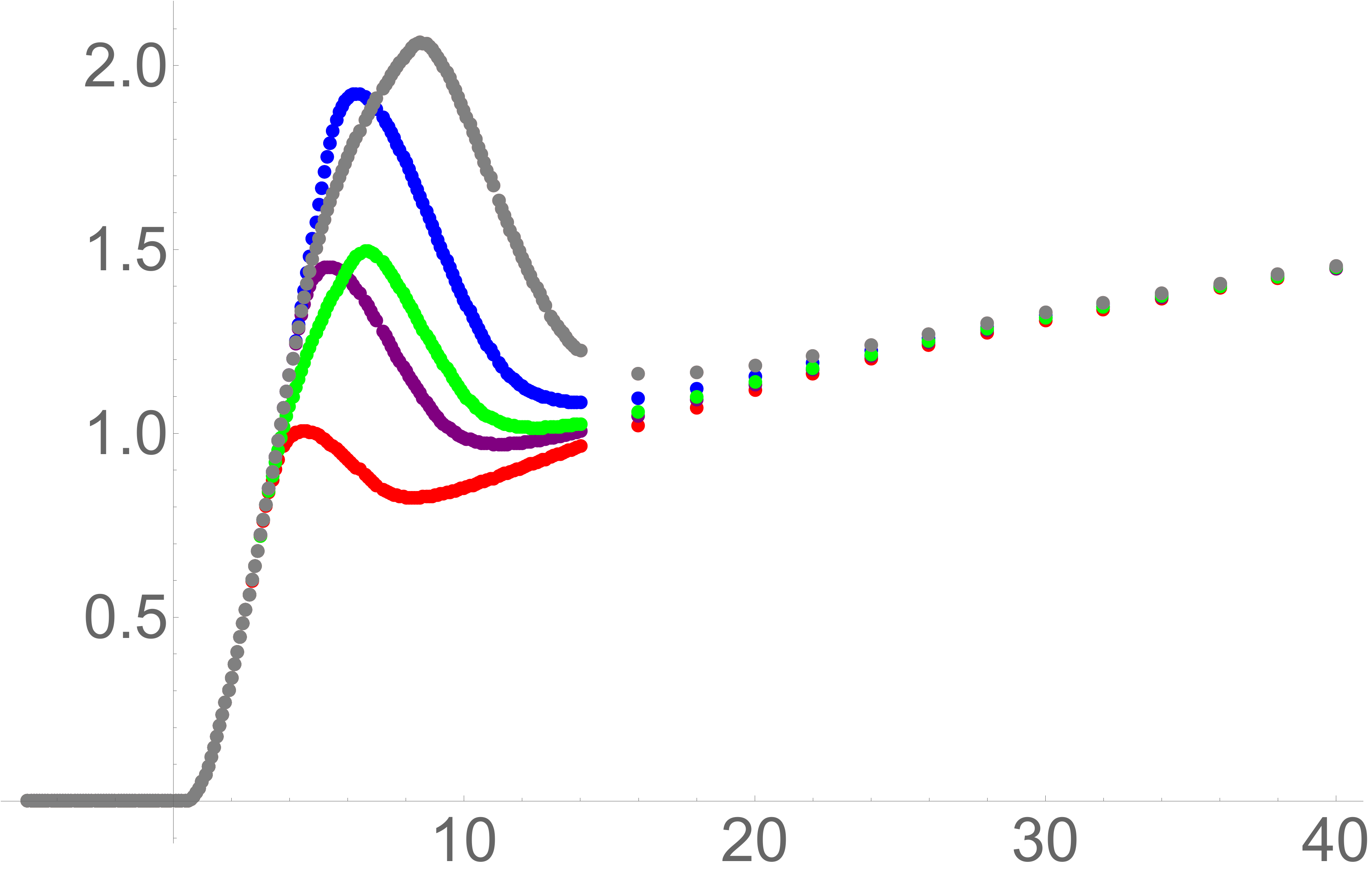}
 \end{center}
  \end{minipage}
     \put(-10,-40){$t/\xi$}
   \put(-160,60){$\Delta I_{A,B}$}
      \put(-180,65){(f)}
  \vspace{3mm} \\      
     \begin{minipage}{0.15\hsize}
  \begin{center}
   \includegraphics[width=45mm]{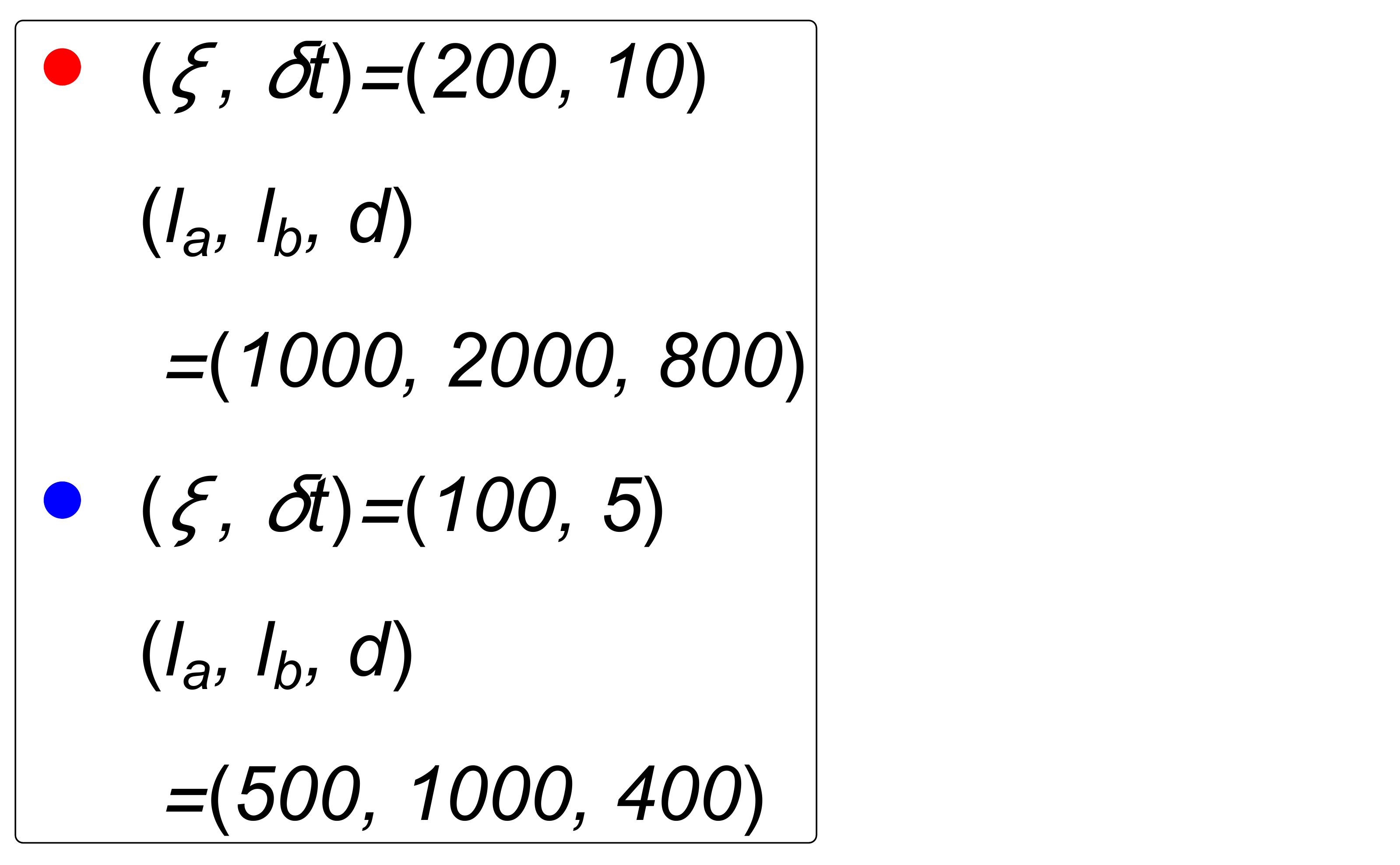}
  \end{center}
 \end{minipage}
 \begin{minipage}{0.4\hsize}
  \begin{center}
   \includegraphics[width=55mm]{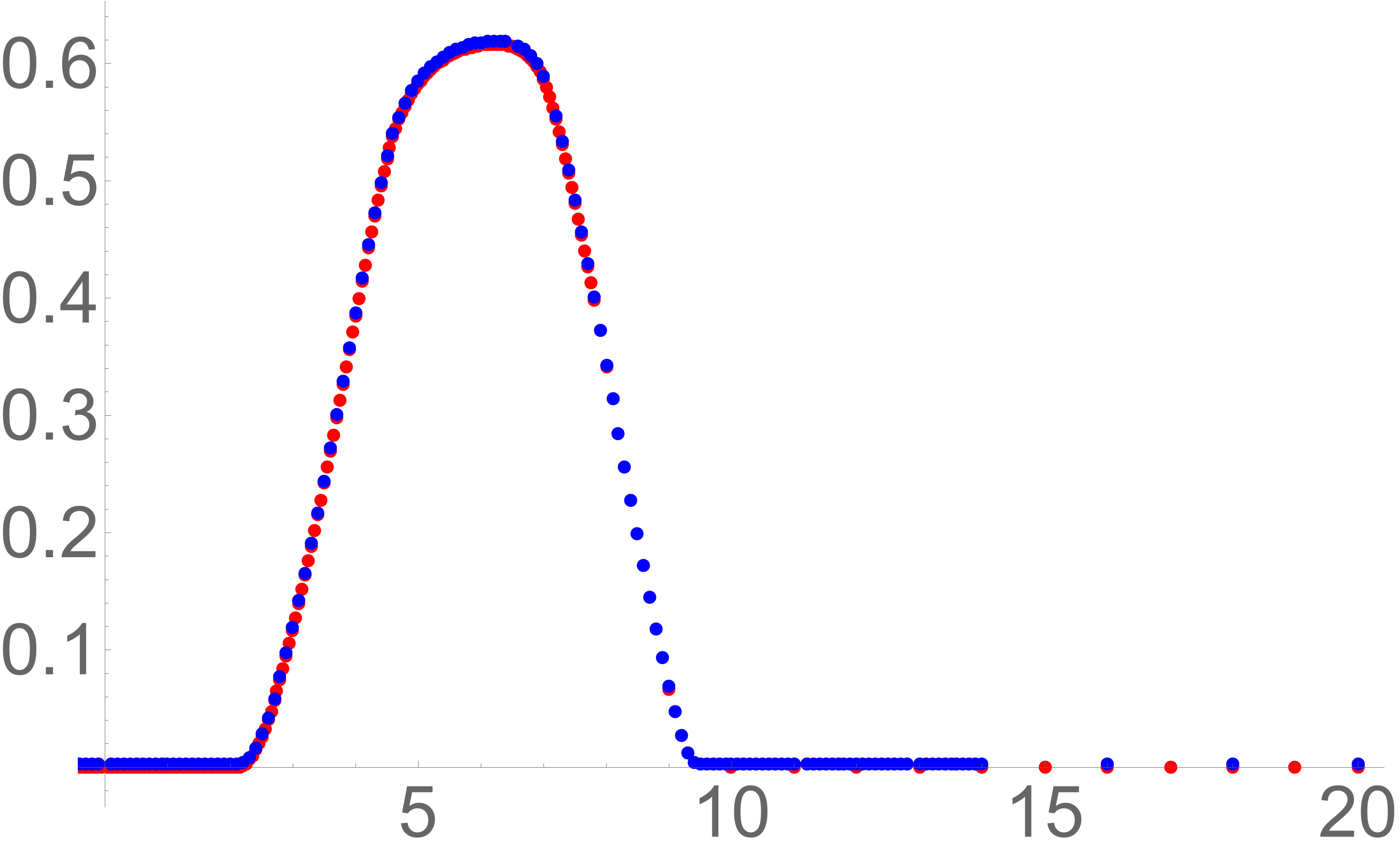}
  \end{center}
 \end{minipage}
   \put(-10,-40){$t/\xi$}
   \put(-170,55){$\Delta \mathcal{E}$}
      \put(-190,65){(g)}
 \begin{minipage}{0.4\hsize}
 \begin{center}
  \includegraphics[width=55mm]{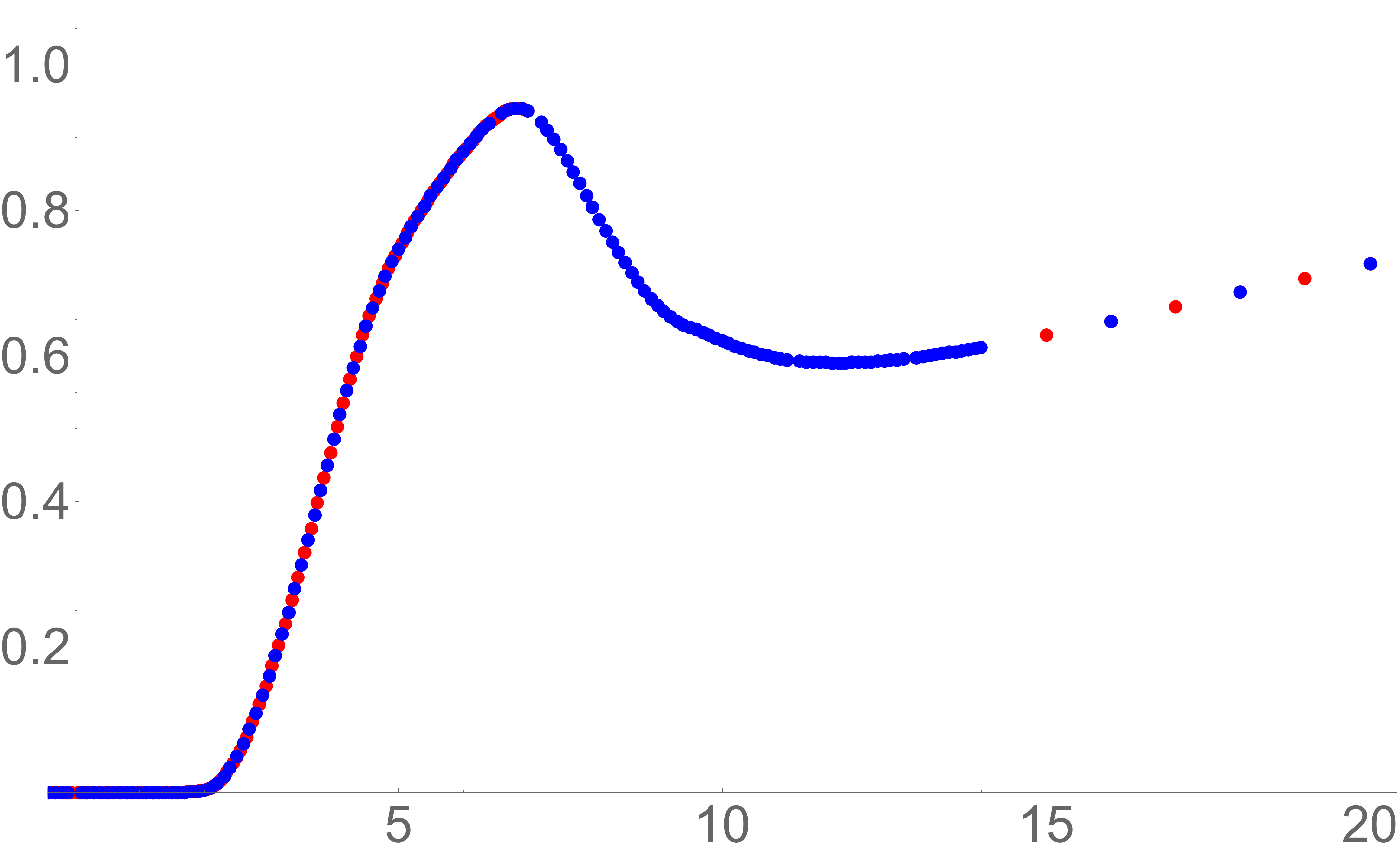}
 \end{center}
  \end{minipage}
     \put(-10,-40){$t/\xi$}
   \put(-170,55){$\Delta I_{A,B}$}
      \put(-190,65){(h)}
        \end{tabular}
  \caption{Time evolution of $\Delta \mathcal{E}$ and $\Delta I_{A,B}$ in the fast ECP. The horizontal axis is labeled with $t/\xi$. The left and right  panels show the time evolution of $\Delta \mathcal{E}$ and $\Delta I_{A,B}$, respectively. The panels (a), (b), (c), and (d) show how $\Delta \mathcal{E}$ and $\Delta I_{A,B}$ depend on the distance between two intervals. We take the subsystem sizes to  $(l_a, l_b)=(1500, 1500)$ in the panels (a) and (b), and  $(l_a, l_b)=(1500, 3000)$ in the panels (c) and (d). The panels (e) and (f) show the subsystem size dependence with $d=100$. The scaling law can be seen from panels (g) and (h), where we take two pairs of parameters for the fast ECP to $(\xi, \delta t)=(200, 10)$ and $(\xi, \delta t)=(100, 5).$   }
  \label{ecpf}
\end{figure}
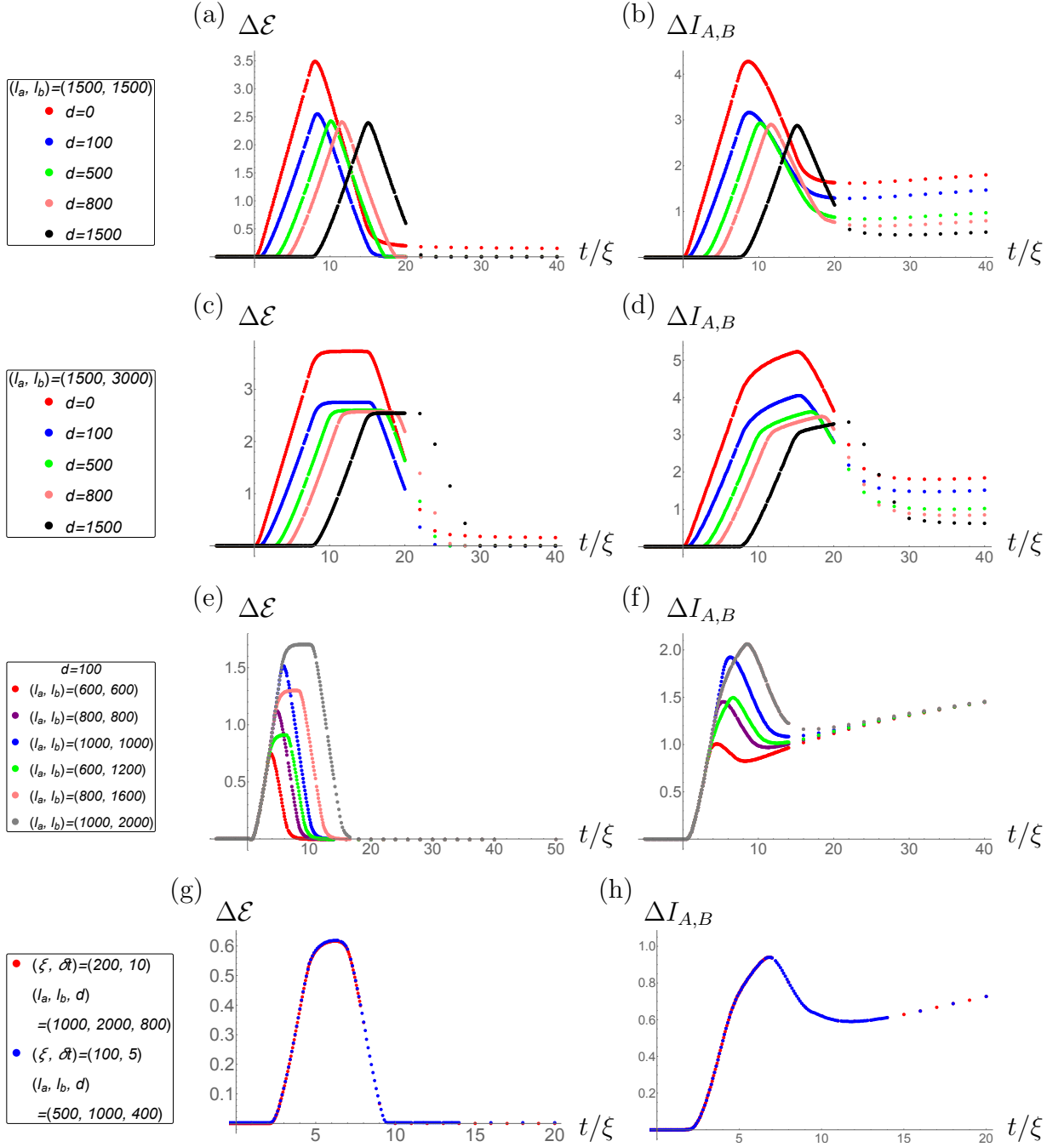

The time evolution of $\Delta \mathcal{E}(t)$ and $\Delta I_{A,B}(t)$ is shown in Figure \ref{ecpf}. Without loss of generality, we can assume $l_b\geq l_a$. The panels (a) and (b) of Figure \ref{ecpf} show how $\Delta \mathcal{E}(t)$ and $\Delta I_{A,B}(t)$  for the protocol in (\ref{ecp}) with $(\xi, \delta t)=(100, 5)$ depend on the distance between two subsystems whose sizes are the same.  The panels (c) and (d) of Figure \ref{ecpf} are similar to the panels (a) and (b), but the subsystem sizes are different. Here, we take the subsystem size $l_b$ to be twice as large as the subsystem size $l_a$, $l_b=2l_a$. We plot $\Delta \mathcal{E}(t)$ and $\Delta I_{A,B}(t)$ with $d=0, 10, 100, 500, 800$, and $1500$ in the panels (a), (b), (c), and (d) of Figure \ref{ecpf}. The panels (e) and (f) of Figure \ref{ecpf} show how $\Delta \mathcal{E}$ and $\Delta I_{A,B}$ depend on the subsystem sizes with fixed distance. We show those for $d=100$. In the panels (g) and (h) of Figure \ref{ecpf}, we plot $\Delta \mathcal{E}$ and $\Delta I_{A,B}$ for ECP with two pair of parameters, $(\xi, \delta t)=(200,10)$ and $(\xi, \delta t)=(100,5)$, to see the scaling law.

We summarize properties of $\Delta \mathcal{E}$ and $\Delta I_{A,B}$ with $2l_a=l_b$. Note that $\Delta \mathcal{E}$ and $\Delta I_{A,B}$ with $l_a=l_b$ have  properties similar to those with $2l_a=l_b$.
\vspace{5mm}

\begin{enumerate}[(1)]
\item Time $t_s$ when $\Delta \mathcal{E}$ and $\Delta I_{A,B}$ begin to increase depends on $d$:
\be \label{ts1}
t_s \sim 0 ~~ \text{for } d \ll \xi,~~~~ t_s \sim \f{d}{2}~~ \text{for } d \gg \xi .
\ee 
It is independent of the subsystem sizes.

\item The quantities $\Delta \mathcal{E}$ and $\Delta I_{A,B}$  monotonically increase in the window $t_s \lessapprox t  \lessapprox t_s+\f{l_a}{2}$.
In this window, $\Delta \mathcal{E}$ for $l_a, l_b \gg \xi$  is  fitted by a following linear function in $t$, $F_f(t,\xi)$, and $\Delta I_{A,B}$ for $l_a, l_b \gg \xi$ is fitted by $G_f(t, \xi)$:
\begin{subequations}
\begin{align}
&F_f(t,\xi) \sim a_1 \f{t}{\xi}+a_2,
\\
&G_f(t,\xi) \sim b_1 \f{t}{\xi}+b_2.
\end{align}
\end{subequations}
Since the slopes of $\Delta \mathcal{E}$ and $\Delta I_{A,B}$ in time do not depend on $l_a$ and $l_b$ but depend on $d$ as one can see in the panels (e) and (f), it is enough to evaluate the coefficients for various $d$ with a pair of subsystem sizes $(l_a, l_b)$. Here we provide the numerical results of $a_1$ and $a_2$ with $(l_a, l_b)=(1500,1500)$ in Table \ref{LF}. 
 By a computation in another protocol wiith $(\xi, \delta t)=(200, 5)$, we find that $a_1$ appears to be independent of $\xi$ and $\delta t$ for $d=0$, and its value is $a_1\sim 0.5$.
\item The slopes of $\Delta \mathcal{E}$ and $\Delta I_{A,B}$ change at $t=t_M$, which depends on $l_a$ as follows:
\be
\begin{split}
t_{M} > \f{l_a+d}{2}~ \text{for } d \ll \xi, ~~
t_M \sim \f{l_a+d}{2} ~\text{for } d \gg \xi.
\end{split}
\ee
\item We define width of the plateau $w$ as a time interval between $t_M$ and a time when the slopes of $\Delta \mathcal{E}$ and $\Delta I_{A,B}$ change from a positive or zero to a negative. Then, $w$  for $d \gg \xi$ is given by $w \sim \f{l_b - l_a}{2}$, and its magnitude is independent of $d$ for $d\gg \xi$.
\item  The quantities $\Delta \mathcal{E}$ and $\Delta I_{A,B}$ in the window $t_s+\f{l_a}{2}+w \lessapprox t \lessapprox t_s+l_a+w$ monotonically decrease.
\item  The quantities $\Delta \mathcal{E}$ and $\Delta I_{A,B}$ for $l_a,l_b,d, \xi \gg 1$ obey the scaling law,
\begin{subequations}
\begin{align}
&\Delta \mathcal{E} = 
\Delta \mathcal{E} \biggl(\frac{l_a}{\xi},\hspace{0.3mm} \frac{l_b}{\xi},\hspace{0.3mm} \frac{d}{\xi},\hspace{0.3mm} \frac{t}{\xi},\hspace{0.3mm}\omega \biggr),
\\
&\Delta I_{A,B} = 
\Delta I_{A,B} \biggl(\frac{l_a}{\xi},\hspace{0.3mm} \frac{l_b}{\xi},\hspace{0.3mm} \frac{d}{\xi},\hspace{0.3mm} \frac{t}{\xi},\hspace{0.3mm}\omega \biggr).
\end{align}
\end{subequations}
\end{enumerate}

The properties (1), (2), (3), (4), and (5) can be seen from the panels (a), (b), (c), and (d) of Figure \ref{ecpf} . We find the last property (6) from the panels (g) and (h) of Figure \ref{ecpf}.

In the following, we list the properties of $\Delta \mathcal{E}$ and $\Delta I_{A,B}$ in the late time. The item (LN) denotes the property of $\Delta \mathcal{E}$, whereas, (MI) denotes the property of  $\Delta I_{A,B}$. 

\begin{enumerate}[(LN1)]
\item The logarithmic negativity $\Delta \mathcal{E}$ for $d \ll \xi$ approaches non-zero constant in the window $t \gg  l_b$.  On the other hand, $\Delta \mathcal{E}$ for $d \gg \xi$ vanishes in the window $t \gg  l_b+d$.
\end{enumerate}

\begin{enumerate}[(M${\rm I}$1)]
\item  The mutual information $\Delta I_{A,B}$ for $d \ll \xi $ decreases monotonically  in the window $t_M < t < \f{l_a+l_b}{2}$, and  increases  logarithmically after $t \sim \frac{l_a+l_b}{2}$. On the other hand, $\Delta I_{A,B}$ for $d \gg \xi $ decreases monotonically in the window $t_M < t < l_a+\f{d}{2}$, and increases after $t \sim \f{l_a + l_b +d}{2}$  as $\Delta I_{A,B}$ for $d \ll \xi $ does.
\item The magnitude of $\Delta I_{A,B}$ in the window, $t\gg \f{l_a+l_b+d}{2}$, depends on the parameters, $l_a, l_b$ and $d$.
However, how $\Delta I_{A,B}$ time-evolves appears to be independent of the parameters. That is, we can fit to $\Delta I_{A,B}$ logarithmic function $t$, $1/2 \log t + c$, where $c$ is a constant which depends on the subsystem size. \end{enumerate}

Interestingly, $\Delta \mathcal{E}$ for the long distance is perfectly consistent with the analytic computation in CFT; around $t\sim d$, $\Delta \mathcal{E}$ increases linearly, and reaches the maximum value at $t\sim d + \f{l_a}{2}$. The maximum value continues to $t\sim d+ \f{l_a}{2}+w$. After that, $\Delta \mathcal{E}$ behaves as linear function with negative slope. Finally, $\Delta \mathcal{E}$ vanishes in the window $t\gg l_b + d$. These behavior agrees with \cite{CM4}.

\begin{table}[htb]
  \begin{center}
    \caption{A fitting function $F_f(t,\xi)$ for $\Delta \mathcal{E}$. \label{LF} }
    \begin{tabular}{|c|c|c|c|c|c|c|c|} \hline
      $(\xi, \delta t)$& $(L_a,L_b, d)$ & Fitting range & $F_f(t, \xi)$ \\ \hline \hline
      $(100, 5)$ & $(1500,1500, 0)$ & $(3\le \f{t}{\xi} \le \f{15}{2})$ &$0.482 \f{t}{\xi}-0.226$ \\
      $(100, 5)$ & $(1500,1500, 100)$ & $(5\le \f{t}{\xi} \le \f{83}{10})$ &$0.397 \f{t}{\xi}-0.666$ \\
      $(100, 5)$ & $(1500,1500, 500)$ & $(7\le \f{t}{\xi} \le \f{19}{2})$ &$0.403 \f{t}{\xi}-1.53$ \\
       $(100, 5)$ & $(1500,1500, 800)$ & $(7\le \f{t}{\xi} \le10)$ &$0.389 \f{t}{\xi}-2.02$ \\
       $(100, 5)$ & $(1500,1500, 1500)$ & $(11\le \f{t}{\xi} \le14)$ &$0.397 \f{t}{\xi}-3.47$ \\ \hline
      $(200, 5)$ & $(1500,1500, 0)$ & $(2\le \f{t}{\xi} \le \f{7}{2})$ &$0.490 \f{t}{\xi}-0.229$ \\
      $(200, 5)$ & $(1500,1500, 100)$ & $(\f{5}{2} \le \f{t}{\xi} \le 4)$ &$0.370 \f{t}{\xi}-0.397$ \\
      $(200, 5)$ & $(1500,1500, 500)$ & $(3\le \f{t}{\xi} \le \f{9}{2})$ &$0.341 \f{t}{\xi}-0.702$ \\
       $(200, 5)$ & $(1500,1500, 800)$ & $(4\le \f{t}{\xi} \le5)$ &$0.341 \f{t}{\xi}-0.960$ \\
       $(200, 5)$ & $(1500,1500, 1500)$ & $(\f{11}{2} \le \f{t}{\xi} \le \f{13}{2})$ &$0.330 \f{t}{\xi}-0.149$ \\ 
\hline
\end{tabular}
\end{center}
\end{table}
\begin{table}[htb]  
  \begin{center}
    \caption{A fitting function $G_f(t,\xi)$ for $\Delta I_{A,B}$. \label{MF}}
    \begin{tabular}{|c|c|c|c|c|c|c|c|} \hline
      $(\xi, \delta t)$& $(L_a,L_b, d)$ & Fitting range & $G_f(t, \xi)$ \\ \hline  \hline
      $(100, 5)$ & $(1500, 1500, 0)$ & $(3 \le \f{t}{\xi} \le \f{15}{2})$ & $0.568 \f{t}{\xi} -0.231$ \\
      $(100, 5)$ & $(1500, 1500, 100)$ & $(4 \le \f{t}{\xi} \le 8)$ & $0.484 \f{t}{\xi} -0.792$ \\
      $(100, 5)$ & $(1500, 1500, 500)$ & $(5 \le \f{t}{\xi} \le 9)$ & $0.467 \f{t}{\xi} -1.67$ \\
      $(100, 5)$ & $(1500, 1500, 800)$ & $(7 \le \f{t}{\xi} \le \f{21}{2})$ & $0.473 \f{t}{\xi}-2.43$ \\
      $(100, 5)$ & $(1500, 1500, 1500)$ & $(10 \le \f{t}{\xi} \le 14)$  & $0.467\f{t}{\xi}-4.01$ \\ \hline
      $(200, 5)$ & $(1500, 1500, 0)$ & $(2 \le \f{t}{\xi} \le \f{75}{20})$ & $0.568 \f{t}{\xi} -0.229$ \\
      $(200, 5)$ & $(1500, 1500, 100)$ & $(\f{5}{2} \le \f{t}{\xi} \le \f{85}{20})$ & $0.429 \f{t}{\xi} -0.409$ \\
      $(200, 5)$ & $(1500, 1500, 500)$ & $(3 \le \f{t}{\xi} \le 5)$ & $0.394 \f{t}{\xi} -0.777$ \\
      $(200, 5)$ & $(1500, 1500, 800)$ & $(\f{18}{5} \le \f{t}{\xi} \le \f{21}{4})$ & $0.397 \f{t}{\xi}-1.08$ \\
      $(200, 5)$ & $(1500, 1500, 1500)$ & $(\f{11}{2} \le \f{t}{\xi} \le 7)$  & $0.395\f{t}{\xi}-1.77$ \\ \hline
          \end{tabular}
  \end{center}
\end{table}

\subsubsection{Slow limit}\label{sec_ecps}
\begin{figure}[htbp]
\begin{tabular}{c}
 \begin{minipage}{0.14\hsize}
\begin{center}
\includegraphics[width=45mm]{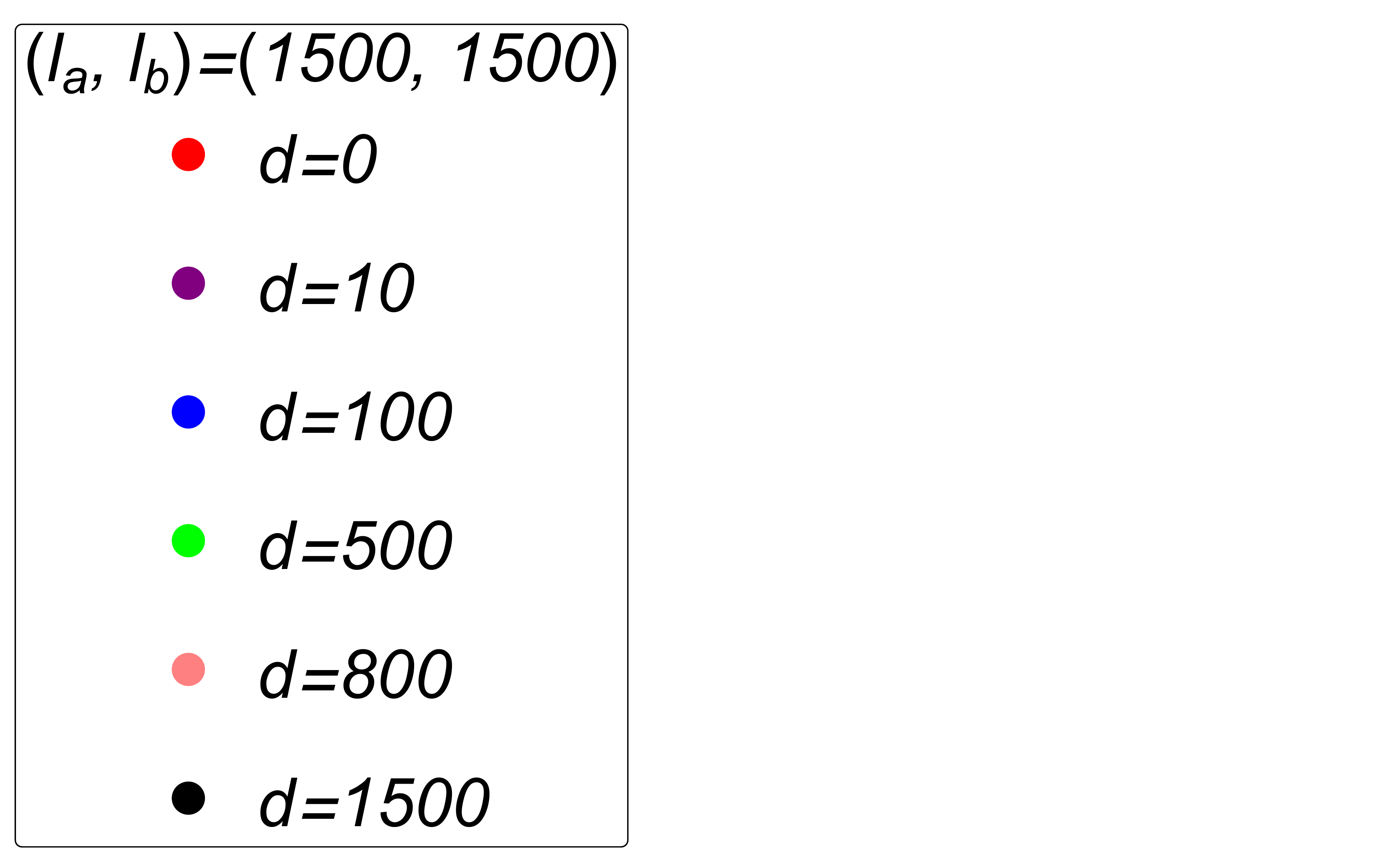}
 \end{center}
 \end{minipage}
 \begin{minipage}{0.4\hsize}
  \begin{center}
   \includegraphics[width=50mm]{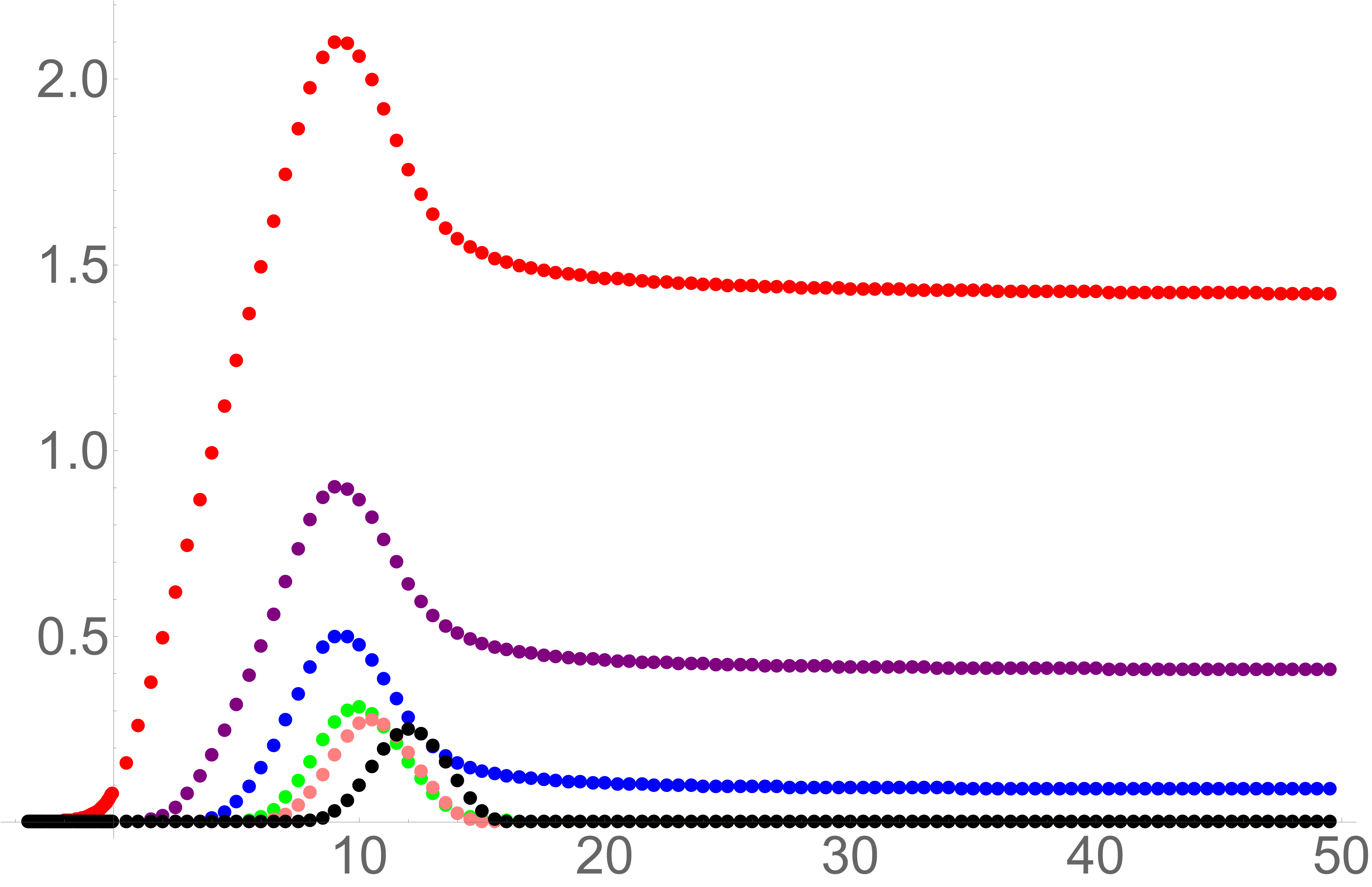}
  \end{center}
 \end{minipage}
   \put(-18,-38){$t \cdot E_{\text{kz}}$}
   \put(-160,52){$\Delta \mathcal{E}$}
   \put(-180,65){(a)}
   \hspace{5mm}
 \begin{minipage}{0.4\hsize}
 \begin{center}
  \includegraphics[width=50mm]{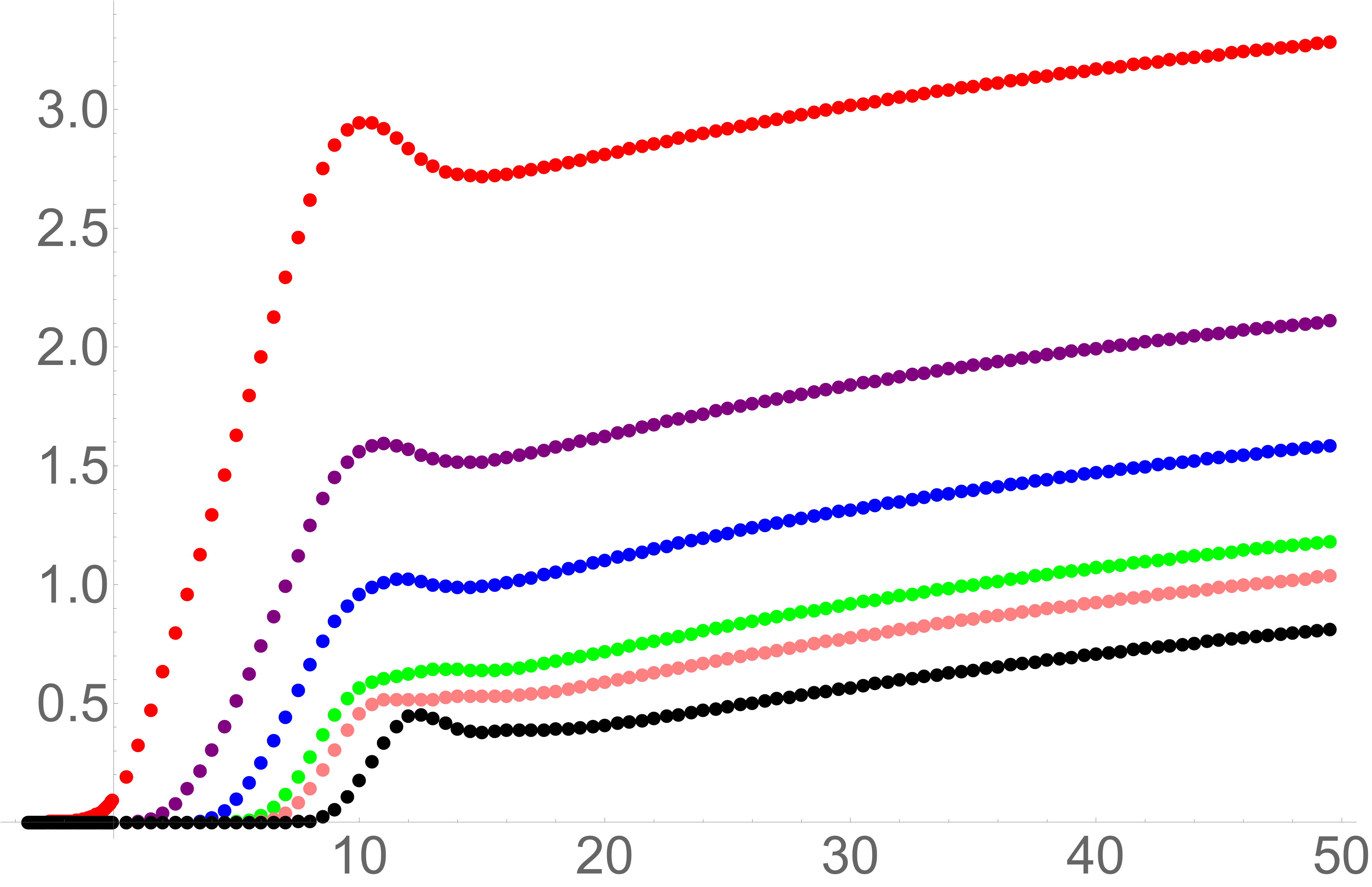}
 \end{center}
  \end{minipage}
     \put(-18,-38){$t \cdot E_{\text{kz}}$}
   \put(-160,52){$\Delta I_{A,B}$}
   \put(-180,65){(b)}
\vspace{3mm} \\ 
     \begin{minipage}{0.14\hsize}
  \begin{center}
   \includegraphics[width=45mm]{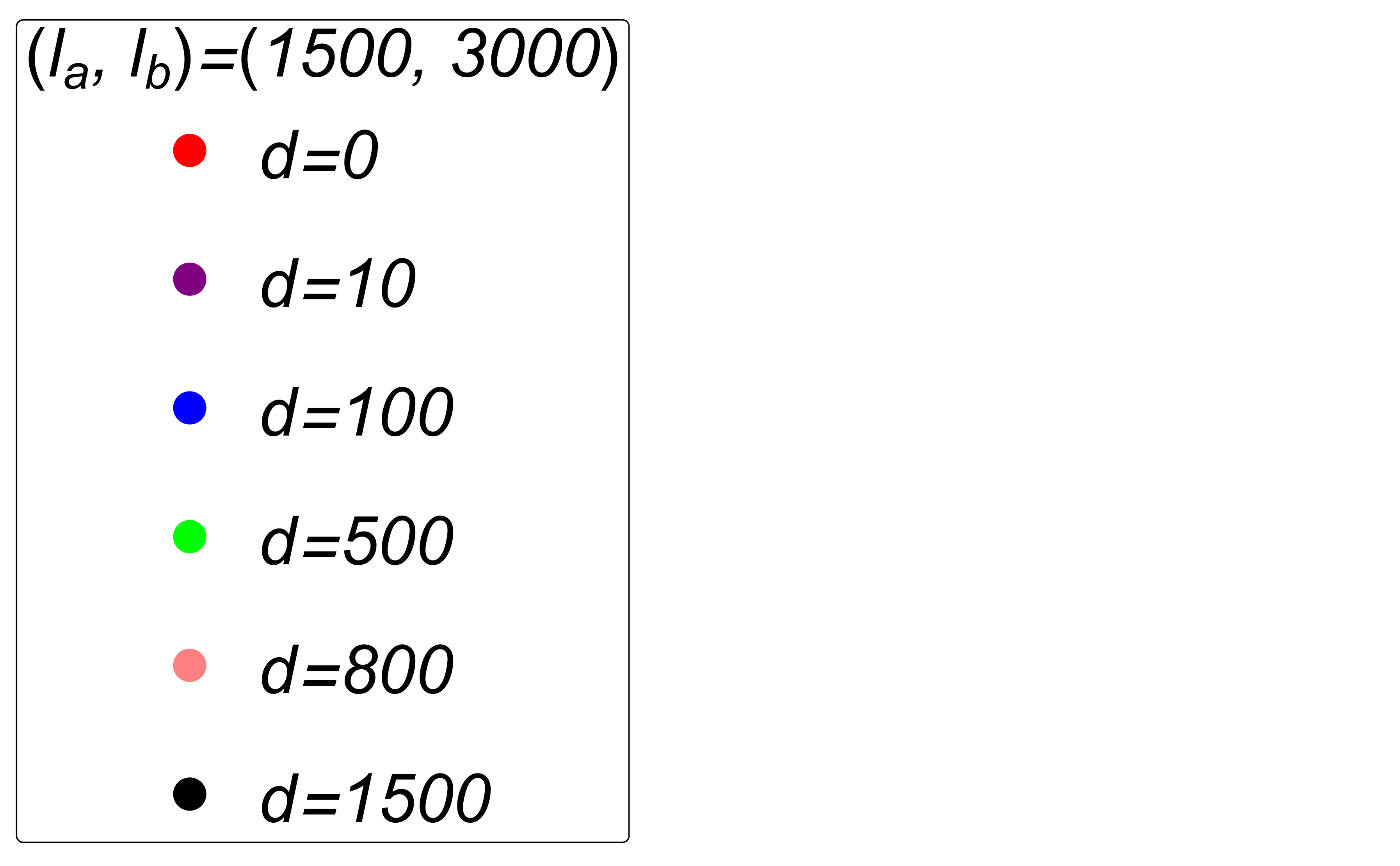}
  \end{center}
 \end{minipage}
 \begin{minipage}{0.4\hsize}
  \begin{center}
   \includegraphics[width=50mm]{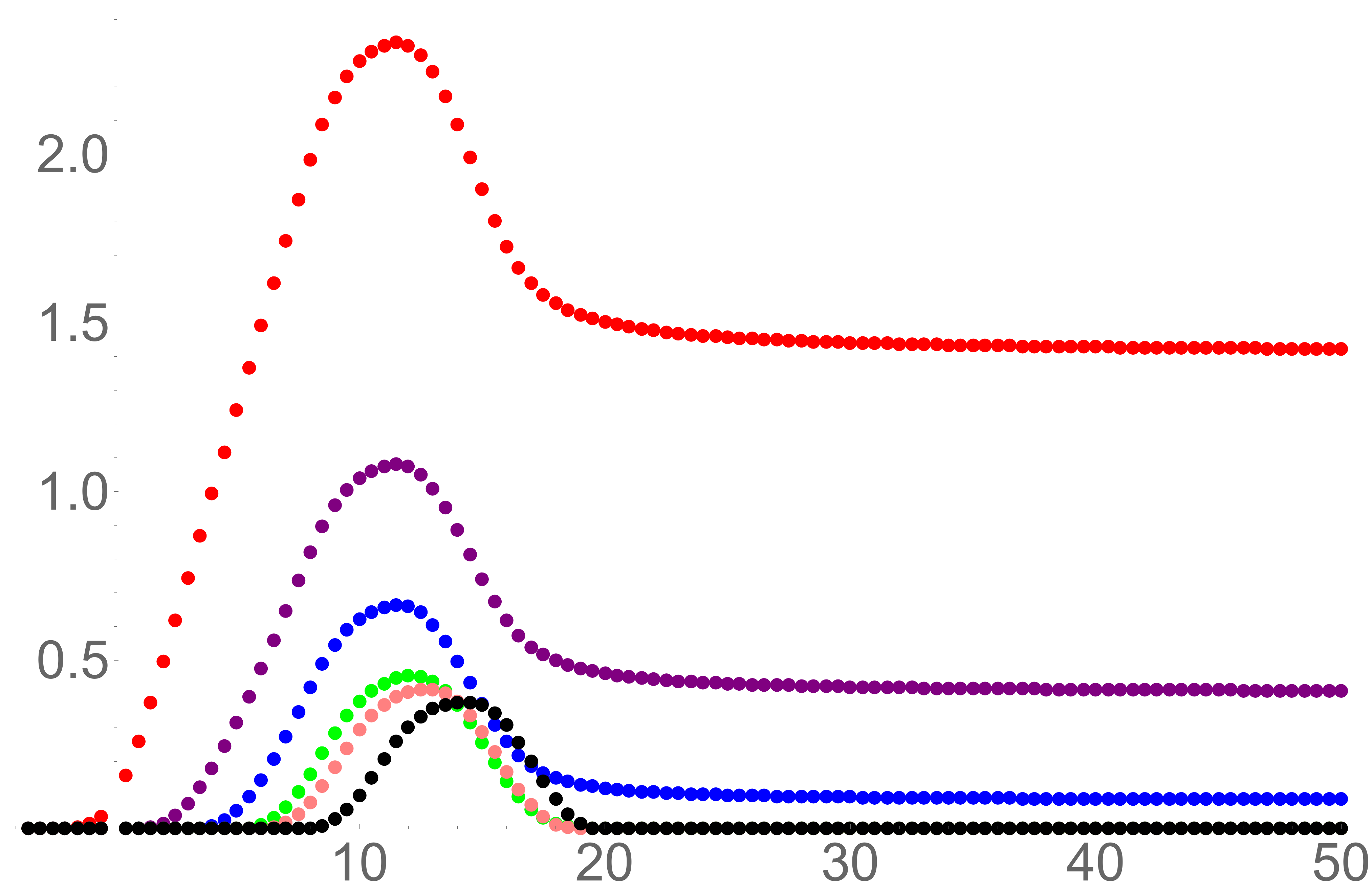}
  \end{center}
 \end{minipage}
   \put(-18,-38){$t \cdot E_{\text{kz}}$}
   \put(-160,52){$\Delta \mathcal{E}$}
      \put(-180,60){(c)}
         \hspace{5mm}
 \begin{minipage}{0.4\hsize}
 \begin{center}
  \includegraphics[width=50mm]{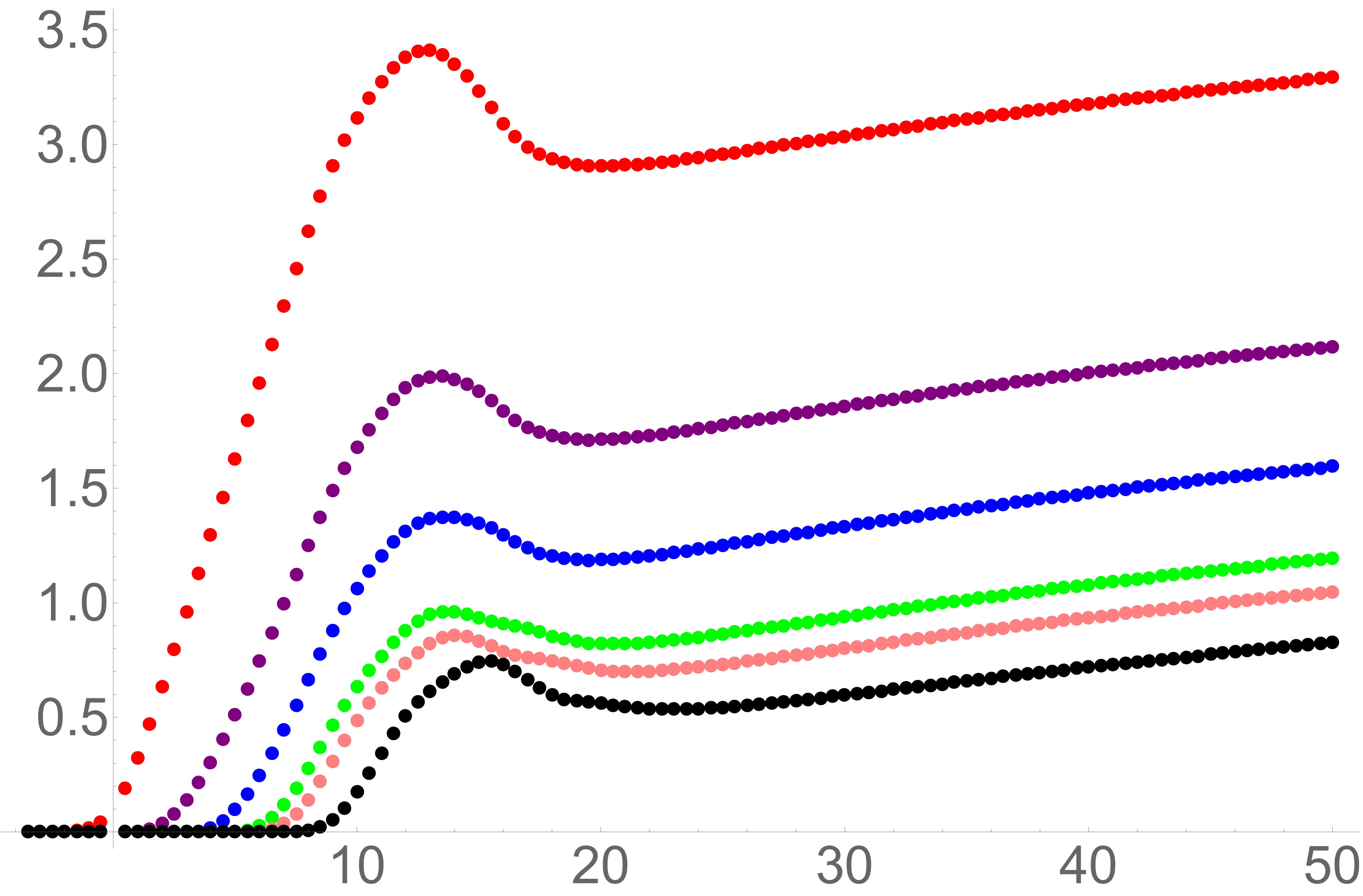}
 \end{center}
  \end{minipage}
     \put(-18,-38){$t \cdot E_{\text{kz}}$}
   \put(-170,55){$\Delta I_{A,B}$}
      \put(-190,60){(d)}
 \vspace{3mm} \\  
     \begin{minipage}{0.14\hsize}
  \begin{center}
   \includegraphics[width=45mm]{size_independent.pdf}
  \end{center}
 \end{minipage}
 \begin{minipage}{0.4\hsize}
  \begin{center}
   \includegraphics[width=50mm]{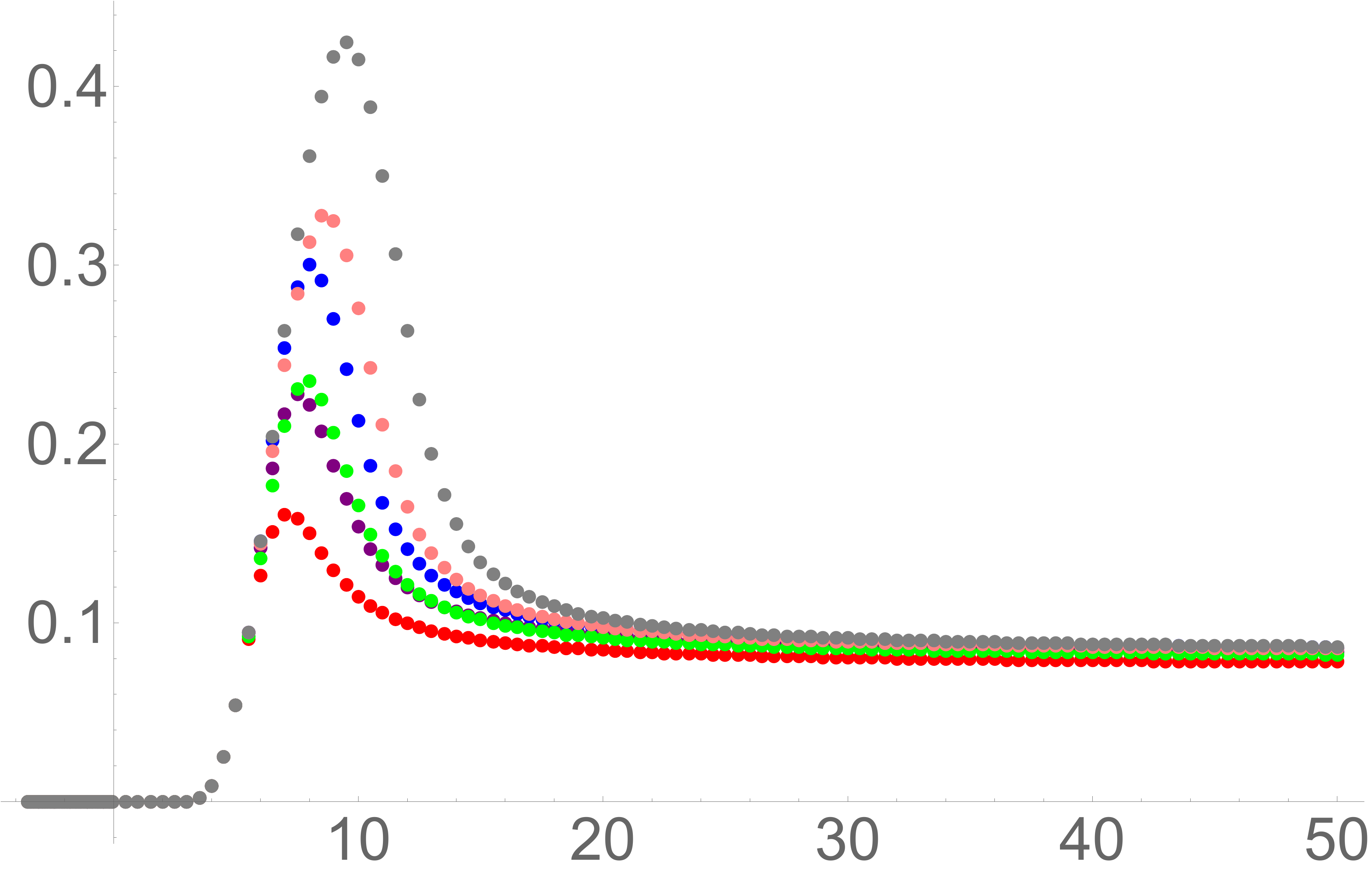}
  \end{center}
 \end{minipage}
   \put(-18,-38){$t \cdot E_{\text{kz}}$}
   \put(-160,55){$\Delta \mathcal{E}$}
      \put(-180,65){(e)}
     \hspace{5mm}
 \begin{minipage}{0.4\hsize}
 \begin{center}
  \includegraphics[width=50mm]{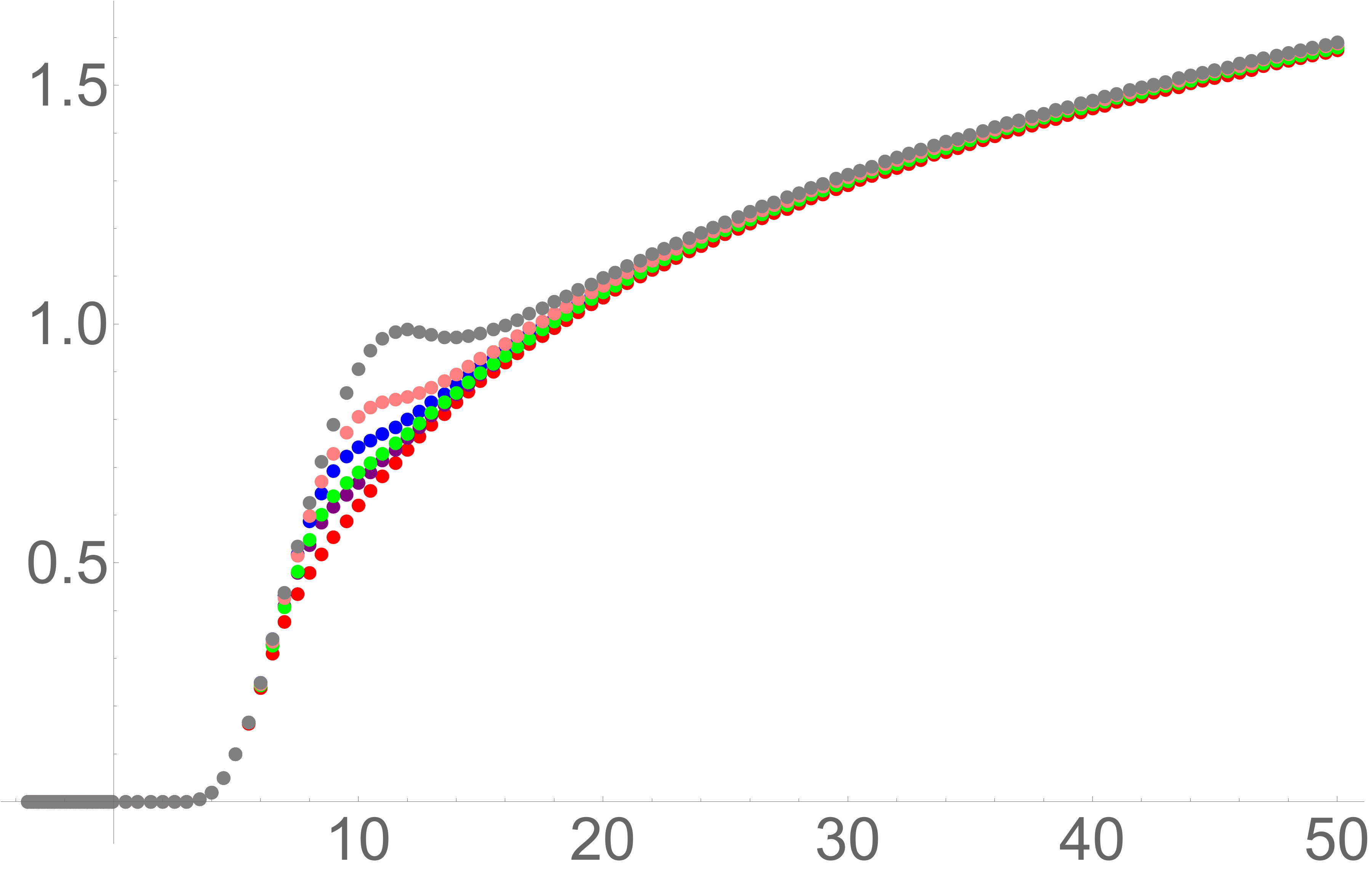}
 \end{center}
  \end{minipage}
     \put(-18,-38){$t \cdot E_{\text{kz}}$}
   \put(-160,55){$\Delta I_{A,B}$}
      \put(-180,65){(f)}
  \vspace{3mm} \\  
     \begin{minipage}{0.14\hsize}
  \begin{center}
   \includegraphics[width=45mm]{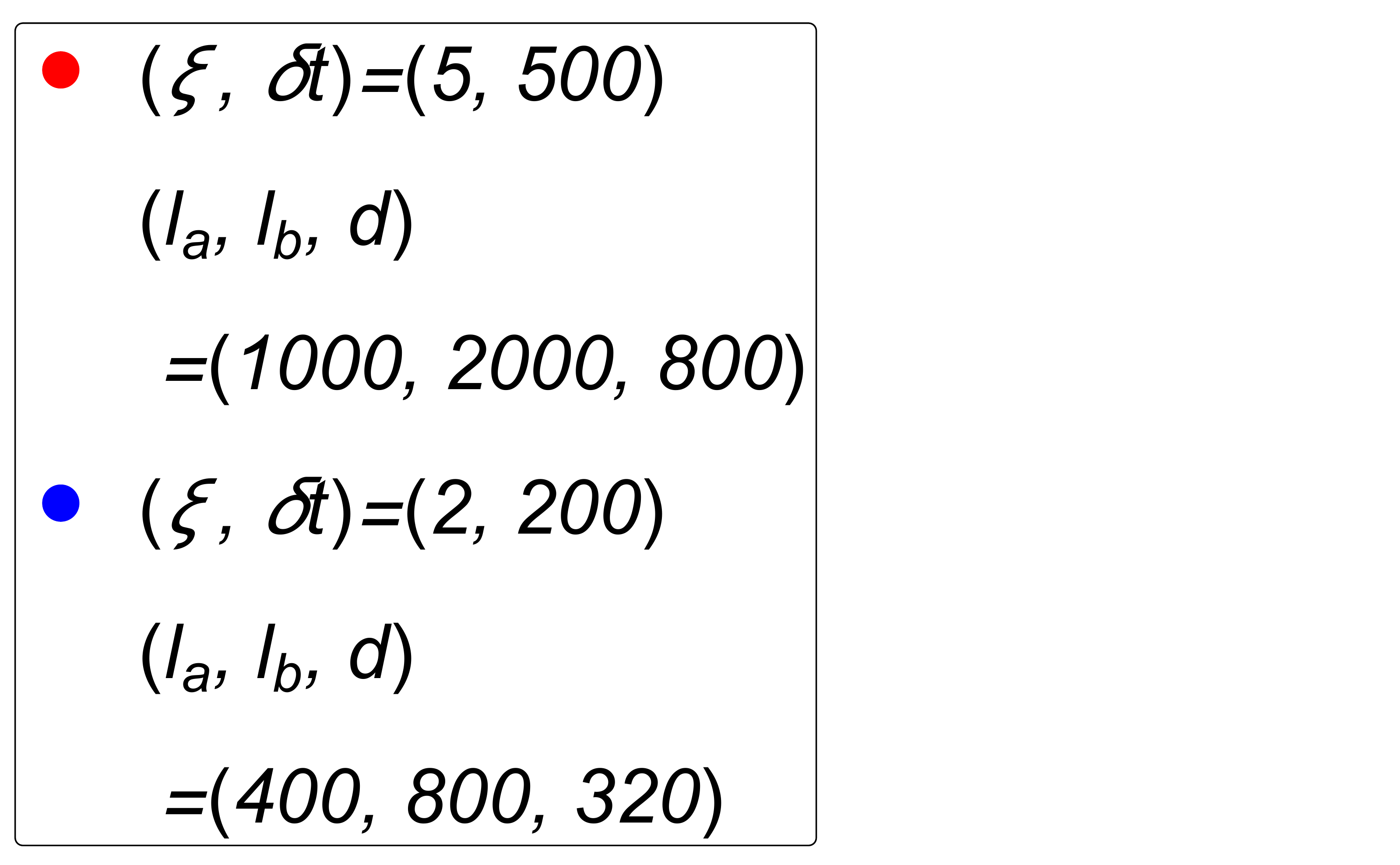}
  \end{center}
 \end{minipage}
 \begin{minipage}{0.4\hsize}
  \begin{center}
   \includegraphics[width=50mm]{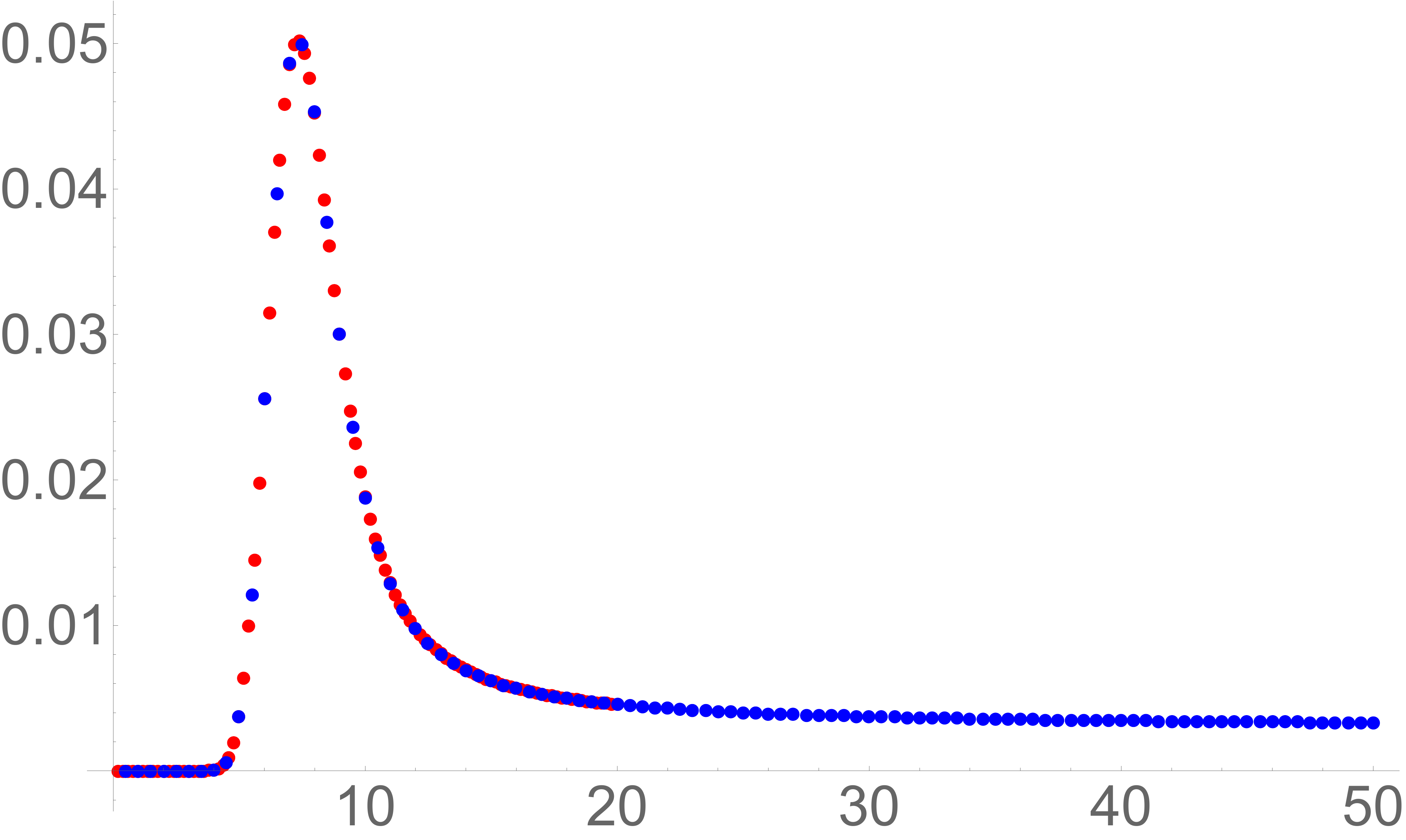}
  \end{center}
 \end{minipage}
   \put(-15,-35){$t \cdot E_{\text{kz}}$}
   \put(-170,50){$\Delta \mathcal{E}$}
      \put(-190,60){(g)}
     \hspace{5mm}
 \begin{minipage}{0.4\hsize}
 \begin{center}
  \includegraphics[width=50mm]{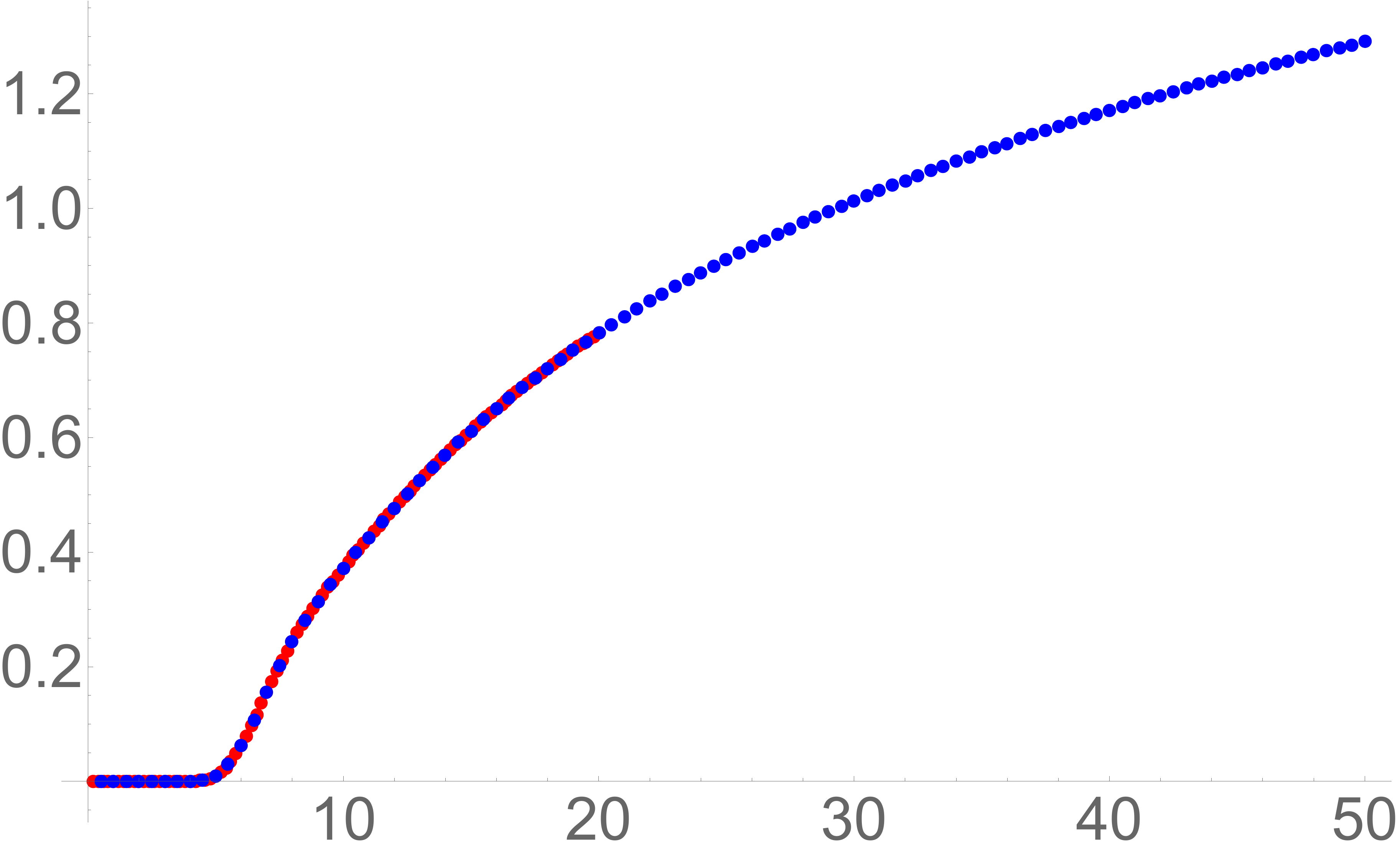}
 \end{center}
  \end{minipage}
     \put(-15,-35){$t \cdot E_{\text{kz}}$}
   \put(-170,50){$\Delta I_{A,B}$}
      \put(-190,60){(h)}
        \end{tabular}
  \caption{Time evolution of $\Delta \mathcal{E}$ and $\Delta I_{A,B}$ in the slow limit. The horizontal axis is labelled with $t \cdot E_{\text{kz}}$. The panels (a), (b), (c), and (d) show how $\Delta \mathcal{E}$ and $\Delta I_{A,B}$ time-evolve with fixed subsystem sizes, $(l_a, l_b)=(1500, 1500)$ and $(l_a, l_b)=(1500, 3000)$, in the protocol $(\xi, \delta t)=(2,200)$. Here we take the distance between $A$ and $B$ to $d=0, 10, 100, 500, 800$, and $1500$. The panels (e) and (f) show the time evolution of $\Delta \mathcal{E}$ and $\Delta I_{A,B}$ for various subsystem sizes with $d=100$. In the bottom panels (g) and (h), we plot $\Delta \mathcal{E}$ and $\Delta I_{A,B}$ with two protocols: one is $(\xi, \delta t)=(5,500)$ with $(l_a, l_b, d)=(1000, 2000, 800)$, the other is $(\xi, \delta t)=(2,200)$ with $(l_a, l_b, d)=(400, 800, 320)$.}
  \label{ecps}
\end{figure}

Next we show some plots of slow ECP in Figure \ref{ecps}. The contents of the figure are as follows: the panels (a) and (b) show how $\Delta \mathcal{E}$ and $\Delta I_{A,B}$ time-evolve, and how $\Delta \mathcal{E}$ and $\Delta I_{A,B}$ change for the distance between $A$ and $B$. Here, we take the subsystem sizes to $l_a=l_b=1500$, and a pair of parameters for the protocol to $(\xi, \delta t)= (2,200)$. 
The panels (c) and (d) are similar to the panels (a) and (b), but for different subsystem sizes, $l_b=2l_a=3000$. The panels (e) and (f) show the time-evolution of  $\Delta \mathcal{E}$ and $\Delta I_{A,B}$ for various subsystem sizes with $d=100$. 
In the bottom panels (g) and (h), we plot $\Delta \mathcal{E}$ and $\Delta I_{A,B}$ with two pairs of parameters, $(\xi, \delta t)=(5,200)$ and $(\xi, \delta t)=(2,200)$, to see the scaling law.

From these results, we find the following properties of $\Delta \mathcal{E}$ and $\Delta I_{A,B}$:

\begin{enumerate}[(1)]
\item Compared with the fast limit $t_s$, which is defined in the property (1) of the fast limit, is shifted by the Kibble--Zurek time $t_{\text{kz}}$ for $d\gg \f{1}{E_{\text{kz}}}$,
\be \label{ts1}
t_s \sim 0 ~~ \text{for } d \ll \f{1}{E_{\text{kz}}},~~~~ t_s \sim \f{d}{2}+t_{\text{kz}}~~ \text{for } d \gg \f{1}{E_{\text{kz}}}.
\ee 
For the slow ECP with $(\xi, \delta t) =(2,200)$, the Kibble--Zurek time defined in \eqref{ECPkz} $t_{\text{kz}}$ is $t_{\text{kz}} \sim 921$. Again, $t_s$ is independent of the subsystem sizes.
\item The quantities $\Delta \mathcal{E}$ and $\Delta I_{A,B}$ in the window $t_s \lessapprox t  \lessapprox t_s+\f{l_a}{2}$ are fitted by  following linear functions in $t$, $F_s(t,E_{\text{kz}})$ and $G_s(t, E_{\text{kz}})$;
\begin{subequations}
\begin{align}
&F_s(t, E_{\text{kz}}) \sim a'_1 t \cdot E_{\text{kz}}+a'_2,
\\
&G_s(t, E_{\text{kz}}) \sim b'_1 t \cdot E_{\text{kz}} +b'_2,
\end{align}
\end{subequations}
where $a'_1$ and $b'_1$ are the coefficients of $t \cdot E_{\text{kz}}$, and $a'_2$ and $b'_2$ are time-independent terms. The slopes of $F_s$ and $G_s$ are independent of the subsystem sizes as one can see in Figure \ref{ecps}, so that it is enough to evaluate $a'_1$ and $b'_1$ for various $d$ with a pair of the subsystem sizes $(l_a, l_b)$. The fitting results for $d=0, 10, 100, 500, 800, 1500$ with $l_a = l_b = 600$ are summarized in Table \ref{ta1}. The coefficients $a'_1$ and $b'_1$ appear to be independent of $d$ for $d\gg \f{1}{E_{\text{kz}}}$, and $a'_1$ is independent of $\xi$ and $\delta t$ for $d =0 $, as well as the property (2) in the fast limit. Its value is $a'_1\sim 0.25$.

\item The time $t_{M}$ depends on $l_a$ and $d$ as follows,
\be
\begin{split}
t_{M} > \f{l_a+d}{2}~ \text{for } d \ll \f{1}{E_{\text{kz}}}, ~~
t_M \sim \f{l_a+d}{2}+t_{\text{kz}} ~\text{for } d \gg \f{1}{E_{\text{kz}}}.
\end{split}
\ee

\item The width of the plateau $w$ in the slow limit is the same as the fast limit, $w \sim \frac{l_b-l_a}{2}$ for $d \gg \f{1}{E_{\text{kz}}}$, and the local maximum value in time appears to be independent of the distance.
\item The quantities $\Delta \mathcal{E}$ and $\Delta I_{A,B}$ in the window $t_s+\f{l_a}{2}+w \lessapprox t \lessapprox t_s+l_a+w$ monotonically decrease.
\item The quantities $\Delta \mathcal{E}$ and $\Delta I_{A,B}$ for $l_a,l_b, d, \xi \gg 1$ obey the following scaling law,
\begin{subequations}
\begin{align}
&\Delta \mathcal{E} = \Delta \mathcal{E} \biggl(l_a \cdot E_{\text{kz}},\hspace{0.3mm} l_b \cdot E_{\text{kz}},\hspace{0.3mm} d \cdot E_{\text{kz}},\hspace{0.3mm} t \cdot E_{\text{kz}},\hspace{0.3mm}\omega \biggr),
\\
&\Delta I_{A,B} = \Delta I_{A,B} \biggl(l_a \cdot E_{\text{kz}},\hspace{0.3mm} l_b \cdot E_{\text{kz}},\hspace{0.3mm} d \cdot E_{\text{kz}},\hspace{0.3mm} t \cdot E_{\text{kz}},\hspace{0.3mm}\omega \biggr).
\end{align}
\end{subequations}
\end{enumerate}
As well as the fast ECP, the properties (1), (2), (3), (4), and (5) can be seen from the panels (a), (b), (c), and (d) of Figure \ref{ecps} . We find the last property (6) from the panels (g) and (h) of Figures \ref{ecps}.

$\Delta \mathcal{E}$ in the window $t\gg l_b+t_{\text{kz}}$ approaches non-zero constant for $d \ll \f{1}{E_{\text{kz}}}$, whereas, $\Delta \mathcal{E}$ in the window $t\gg l_b+d+t_{\text{kz}}$ vanishes for $d \gg \f{1}{E_{\text{kz}}}$ as one can see in the panel (a) of Figure \ref{ecps}. On the other hand, in the panel (b) of Figure \ref{ecps}, $\Delta I_{A,B}$ grows logarithmically after $t\gg \frac{l_a + l_b + d}{2}+t_{\text{kz}}$. These properties are almost the same as (LN1), (MI1), and (MI2) in the fast limit, but the time is shifted by $t_{\text{kz}}$.

The main difference in time scale compared with the fast limit is that $t_s$ and $t_M$ are shifted by $t_{\text{kz}}$ in the slow limit. We will interpret the difference as our toy model in section \ref{ecpquasi}.

\begin{table}[htb]
  \begin{center}
 \caption{A fitting function $F_s(t,E_{\text{kz}})$ for $\Delta \mathcal{E}$. \label{ta1}}
 \begin{tabular}{|c|c|c|c|c|c|c|c|c|c|} \hline
    ($\xi$, $\delta t$) & ($l_a$, $l_b$, $d)$  & Fitting range & $F_s(t,E_{\text{kz}})$  \\ \hline \hline
    ($2$, $200$) & ($600$, $600$, $0$)      & $1    \le t\cdot E_{\text{kz}} \le \frac{11}{2}$ &$ 0.246 \hspace{1.5mm} t\cdot E_{\text{kz}}+0.00589 $  \\ 
    ($2$, $200$) & ($600$, $600$, $10$)    & $\f{7}{2} \le t\cdot E_{\text{kz}} \le 6   $ &$0.131 \hspace{1.5mm} t\cdot E_{\text{kz}}-0.343$ \\ 
    ($2$, $200$) & ($600$, $600$, $100$)  & $\f{9}{2} \le t\cdot E_{\text{kz}} \le 6   $ &$0.0681 \hspace{1.5mm} t\cdot E_{\text{kz}}-0.283$ \\ 
    ($2$, $200$) & ($600$, $600$, $500$)  & $6    \le t\cdot E_{\text{kz}} \le 7   $ &$0.0225 \hspace{1.5mm} t\cdot E_{\text{kz}}-0.125$  \\ 
    ($2$, $200$) & ($600$, $600$, $800$)  & $\f{13}{2} \le t\cdot E_{\text{kz}} \le \frac{15}{2}$ &$0.0152 \hspace{1.5mm} t\cdot E_{\text{kz}}-0.0949$ \\ 
    ($2$, $200$) & ($600$, $600$, $1500$) & $\f{17}{2} \le t\cdot E_{\text{kz}} \le 9   $ &$0.0117 \hspace{1.5mm} t\cdot E_{\text{kz}}-0.0938$  \\  \hline
   ($4$, $800$) & ($600$, $600$, $0$)      & $\f{3}{4}    \le t\cdot E_{\text{kz}} \le \frac{9}{2}$ &$ 0.245 \hspace{1.5mm} t\cdot E_{\text{kz}}+12.4 $  \\ 
    ($4$, $800$) & ($600$, $600$, $10$)    & $\f{25}{8} \le t\cdot E_{\text{kz}} \le \f{19}{4}   $ &$0.132 \hspace{1.5mm} t\cdot E_{\text{kz}}-205$ \\ 
    ($4$, $800$) & ($600$, $600$, $100$)  & $\f{31}{8} \le t\cdot E_{\text{kz}} \le 5   $ &$0.0584 \hspace{1.5mm} t\cdot E_{\text{kz}}-159$ \\ 
    ($4$, $800$) & ($600$, $600$, $500$)  & $5  \le t\cdot E_{\text{kz}} \le \f{23}{4}   $ &$0.00723 \hspace{1.5mm} t\cdot E_{\text{kz}}-25.2$  \\ 
    ($4$, $800$) & ($600$, $600$, $800$)  & $\f{41}{8} \le t\cdot E_{\text{kz}} \le 6$ &$0.00190 \hspace{1.5mm} t\cdot E_{\text{kz}}-7.20$ \\ 
    ($4$, $800$) & ($600$, $600$, $1500$) & $\f{23}{4} \le t\cdot E_{\text{kz}} \le \f{13}{2}   $ &$0.00108 \hspace{1.5mm} t\cdot E_{\text{kz}}-0.463$  \\ \hline
    \end{tabular}
  \end{center}
  \end{table}
  \begin{table}[htb]
  \begin{center}
 \caption{A fitting function $G_s(t,E_{\text{kz}})$ for $\Delta I_{A,B}$. \label{ta2}}
 \begin{tabular}{|c|c|c|c|c|c|c|c|c|} \hline
    ($\xi$, $\delta t$) & ($l_a$, $l_b$, $d$)  & Fitting range  &$G_s(t,E_{\text{kz}})$     \\ \hline \hline
    ($2$, $200$)& ($600$, $600$, $0$)       & $\f{1}{2} \le t\cdot E_{\text{kz}} \le 6   $ &$0.323 \hspace{1.5mm} t\cdot E_{\text{kz}}-0.00106$  \\ 
    ($2$, $200$)& ($600$, $600$, $10$)     & $\f{7}{2} \le t\cdot E_{\text{kz}} \le 7   $ &$0.201 \hspace{1.5mm} t\cdot E_{\text{kz}}-0.493$ \\ 
    ($2$, $200$)& ($600$, $600$, $100$)   & $5    \le t\cdot E_{\text{kz}} \le \f{15}{2}$ &$0.136 \hspace{1.5mm} t\cdot E_{\text{kz}}-0.583$ \\ 
    ($2$, $200$)& ($600$, $600$, $500$)   & $\f{13}{2} \le t\cdot E_{\text{kz}} \le 9   $ &$0.0698 \hspace{1.5mm} t\cdot E_{\text{kz}}-0.397$ \\ 
    ($2$, $200$)& ($600$, $600$, $800$)   & $9    \le t\cdot E_{\text{kz}} \le 14 $ &$0.0405 \hspace{1.5mm} t\cdot E_{\text{kz}}-0.223$ \\ 
    ($2$, $200$)& ($600$, $600$, $1500$) & $\f{17}{2} \le t\cdot E_{\text{kz}} \le 10 $ &$0.0377 \hspace{1.5mm} t\cdot E_{\text{kz}}-0.301$ \\  \hline
    ($4$, $800$)& ($600$, $600$, $0$)       & $\f{5}{4} \le t\cdot E_{\text{kz}} \le \f{15}{4}   $ &$0.329 \hspace{1.5mm} t\cdot E_{\text{kz}}+1.39$  \\ 
    ($4$, $800$)& ($600$, $600$, $10$)     & $\f{25}{8} \le t\cdot E_{\text{kz}} \le \f{35}{8}   $ &$0.204 \hspace{1.5mm} t\cdot E_{\text{kz}}-296$ \\ 
    ($4$, $800$)& ($600$, $600$, $100$)   & $5    \le t\cdot E_{\text{kz}} \le \f{25}{4}$ &$0.139 \hspace{1.5mm} t\cdot E_{\text{kz}}-403$ \\ 
    ($4$, $800$)& ($600$, $600$, $500$)   & $\f{55}{8} \le t\cdot E_{\text{kz}} \le \f{35}{4}   $ &$0.0889 \hspace{1.5mm} t\cdot E_{\text{kz}}-358$ \\ 
    ($4$, $800$)& ($600$, $600$, $800$)   & $\f{55}{8}    \le t\cdot E_{\text{kz}} \le \f{35}{4} $ &$0.0799 \hspace{1.5mm} t\cdot E_{\text{kz}}-360$ \\ 
    ($4$, $800$)& ($600$, $600$, $1500$) & $\f{15}{2} \le t\cdot E_{\text{kz}} \le 10 $ &$0.0638 \hspace{1.5mm} t\cdot E_{\text{kz}}-328$ \\ \hline
    \end{tabular}
  \end{center}
  \end{table}

\subsubsection{Quasi-particle interpretation}\label{ecpquasi}

The mass potential in the ECP changes as follows: The mass term in the very early time $t\ll -\delta t$ is constant, deceases monotonically after $t\sim -\delta t$, and vanishes asymptotically at late time $t\gg \delta t$. Entanglement structures of states in the early time therefore changes adiabatically. 

In the fast limit, $\Delta \mathcal{E}$ and $\Delta I_{A,B}$ monotonically increases from $t_s \sim d/2$ for $d\gg \xi$. On the other hand, in the slow limit,  $t_s$ is shifted by $t_{\text{kz}}$, $t_s \sim d/2 + t_{\text{kz}}$ for $d \gg \f{1}{E_{\text{kz}}}$. This means that the entangled pairs, which are composed of right- and left-moving particles $p_R$ and $p_L$, are created everywhere when the adiabaticity is broken. The left-moving particle $p_L$ is entangled with $p_R$. Entanglement between $p_L$ and $p_R$ contributes to $\Delta \mathcal{E}$ and $\Delta I_{A,B}$ only when $p_L$ is in $A$ (or $B$), and $p_R$ is in $B$ (or $A$). 
When $p_L$ or $p_R$ go away from $A$ or $B$, $\Delta \mathcal{E}$ and $\Delta I_{A,B}$ decrease. 
The entangled pair created on the middle of two regions reaches $A$ and $B$ at $t=t_s\sim d/2$, and goes away from there at $t=t_M \sim t_s + l_a/2$. When the subsystem sizes are the same $l_a=l_b$, $\Delta \mathcal{E}$ and $\Delta I_{A,B}$ decrease after $t_M$. Otherwise, $\Delta \mathcal{E}$ shows a plateau from $t_M$ to $t_M + l_b/2-l_a/2$ since the inflow and outflow of the entangled pairs into $A$ and $B$ balance each other. The similar behavior is observed in $\Delta I_{A,B}$.

In our computation, $\Delta \mathcal{E}$ does not vanish in the late time $t \gg \frac{l_b + d}{2}$ for $d \ll \xi$ in the fast limit or $d \ll 1/E_{\text{kz}}$ in the slow limit, whereas $\Delta \mathcal{E}$ vanishes in 2d CFT analysis. The mutual information $\Delta I_{A,B}$ in our computation increases logarithmically in the late time for arbitrary $l_a, l_b, d, \xi, \delta t$. However, this vanishes in the 2d CFT.

In the 2d CFT, only particle pairs with speed of light are created, and contribute to $\Delta \mathcal{E}$ and $\Delta I_{A,B}$. In our setup, there are entangled particles propagating with various velocities less than or equal to the speed of light as mentioned in the previous work \cite{Nishida:2017hqd}. The lattice artifact also contributes to them since we calculate on the lattice. Therefore, non-zero values of $\Delta \mathcal{E}$ and $\Delta I_{A,B}$ in the late time are due to the entangled pairs with the velocities less than the speed of light and the lattice artifact.

The quantity $\Delta \mathcal{E}$ after $t\gg \frac{l_b + d}{2}$ vanishes for $d \gg \xi$ in the fast limit or $d \gg 1/E_{\text{kz}}$ in the slow limit, which implies that the slow modes do not contribute to $\Delta \mathcal{E}$ for these configurations, and the lattice artifact is suppressed. As a result, the quantities in this configuration behaves in similar manner to those in the 2d CFT.

\begin{figure}[htb]
\centering
\begin{tabular}{c}
\includegraphics[width=6cm]{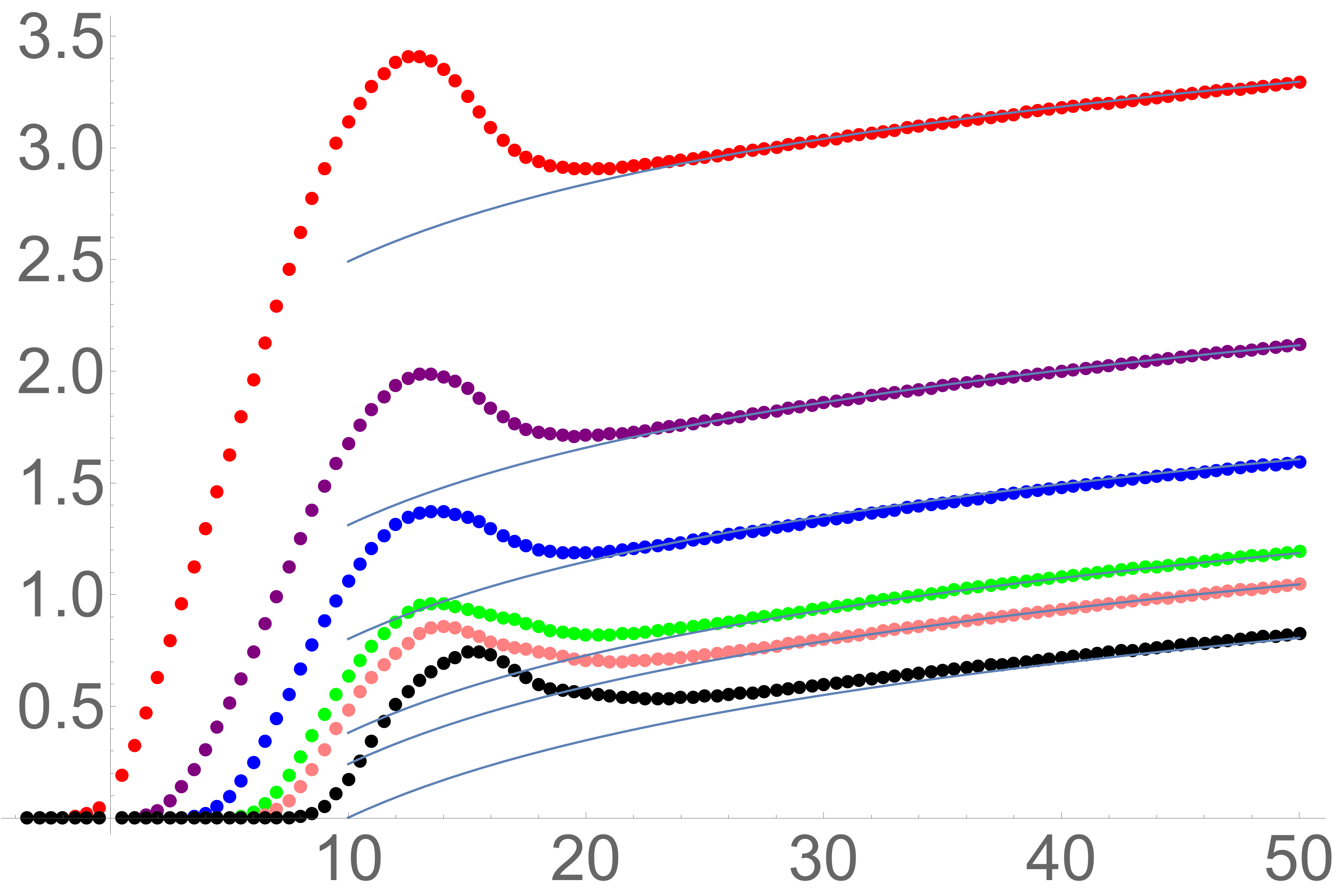}
\put(8,5){$t\cdot E_{\text{kz}}$}
\put(-175,120){$\Delta I_{A,B}$}
\end{tabular}
\caption{The fitting for $\Delta I_{A,B}$ by the logarithmic function in $t$. The numerical results are for $(l_a,l_b)=(1500,3000)$ with $d=0,10,100,500,800, 1500$ in the protocol $(\xi, \delta t)=(2,200)$. The blue curves are $1/2 \log t \cdot E_{\text{kz}} + c$, where $c$ is a size dependent constant. This confirms our interpretation.}
\label{log_fit}
\end{figure}

Since the late-time entanglement entropy grows as $1/2 \log t$ in the smooth quenches \cite{Nishida:2017hqd} and $\Delta I_{A,B}$ is given by the summation of entanglement entropies for $A$, $B$, and $A\cup B$ as defined in  \eqref{def MI}, $\Delta I_{A,B}$ at the late time increases as (see also Figure \ref{log_fit})
\begin{eqnarray}
\Delta I_{A,B} \sim 1/2 \log t + 1/2 \log t - 1/2 \log t = 1/2 \log t.
\end{eqnarray}

\subsection{CCP}
Here, we study the time evolution of $\Delta\mathcal{E}$ and $\Delta I_{A, B}$ in the fast and slow CCPs numerically\footnote{In the CCP, we  compute numerically $\Delta\mathcal{E}$ and $\Delta I_{A, B}$ with $\varepsilon=0.00001, 0.000001$, where $\varepsilon$ is defined in footnote \ref{f5}. In our numerical computations, $\Delta\mathcal{E}$ and $\Delta I_{A, B}$ do not depend on $\varepsilon$ so much.   }. There are several papers about the dynamics of quantum entanglement in the CCPs including, for example, the scaling property of entanglement entropy at $t=0$ \cite{Caputa:2017ixa}, the time evolution of entanglement entropy \cite{Nishida:2017hqd}, and complexity \cite{Camargo:2018eof}.  
The quantities $\Delta\mathcal{E}$ and $\Delta I_{A, B}$ in the CCPs oscillate in time. We will see that the amplitude of the oscillation in the fast CCP depends on the inter-subsystems distance $d$, and the behavior of $\Delta\mathcal{E}$ and $\Delta I_{A, B}$ with large $d$ in the fast CCP  is similar to that of $\Delta\mathcal{E}$ and $\Delta I_{A, B}$  in the fast ECP. We will also discuss the quasi-particle interpretation of the behavior of $\Delta\mathcal{E}$ and $\Delta I_{A, B}$ in the CCPs.

\subsubsection{Fast limit}
First, we study the time evolution of $\Delta\mathcal{E}$ and $\Delta I_{A, B}$ at the fast limit in the CCP. They oscillate in time.  We compute $\Delta\mathcal{E}$ and $\Delta I_{A, B}$ in the fast CCP with the parameters $(\xi=100, \delta t=5)$, $(\xi=100, \delta t=10)$, and $(\xi=200, \delta t=10)$. Here, we explain the time evolution of $\Delta I_{A,B}$ and $\Delta \mathcal{E}$ in the fast CCP with $(\xi=100, \delta t=5)$, which are computed numerically.

Figure \ref{ccpf} shows the time-dependence of the logarithmic negativity $\Delta\mathcal{E}$ and the mutual information $\Delta I_{A, B}$ in the fast CCP  with $l_a=l_b$ and $2l_a=l_b$. We find the following properties in the fast CCP limit:

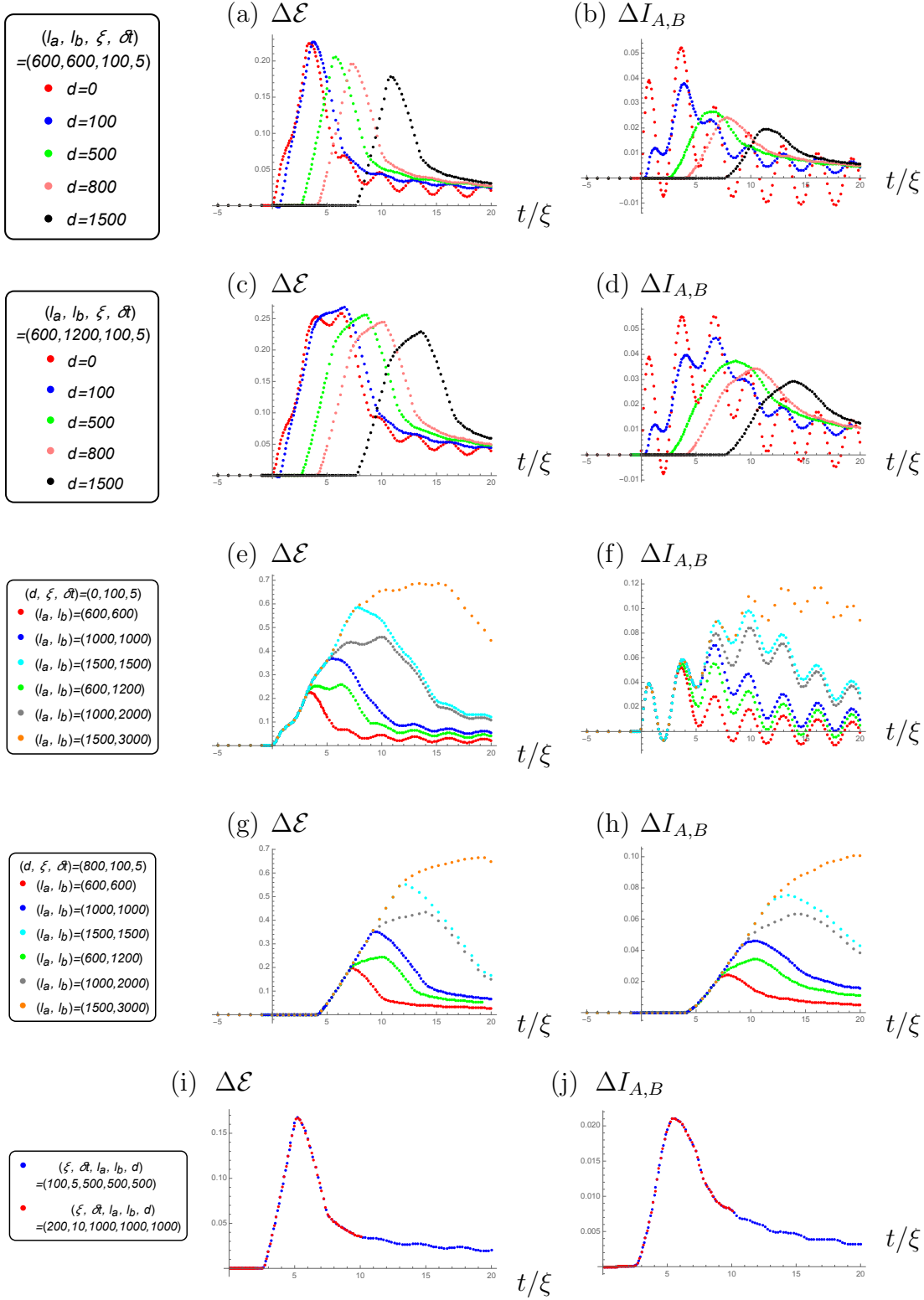
\begin{figure}[htbp]
\begin{tabular}{c}
 \begin{minipage}{0.15\hsize}
\begin{center}
\includegraphics[width=25mm]{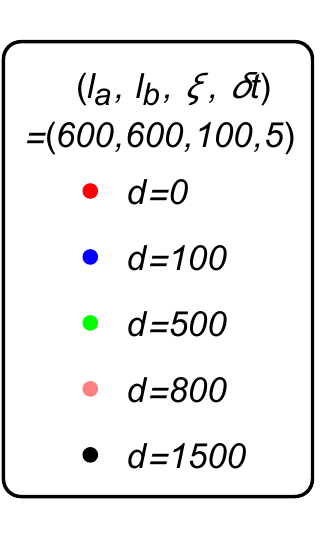}
 \end{center}
 \end{minipage}
 \begin{minipage}{0.35\hsize}
  \begin{center}
   \includegraphics[width=45mm]{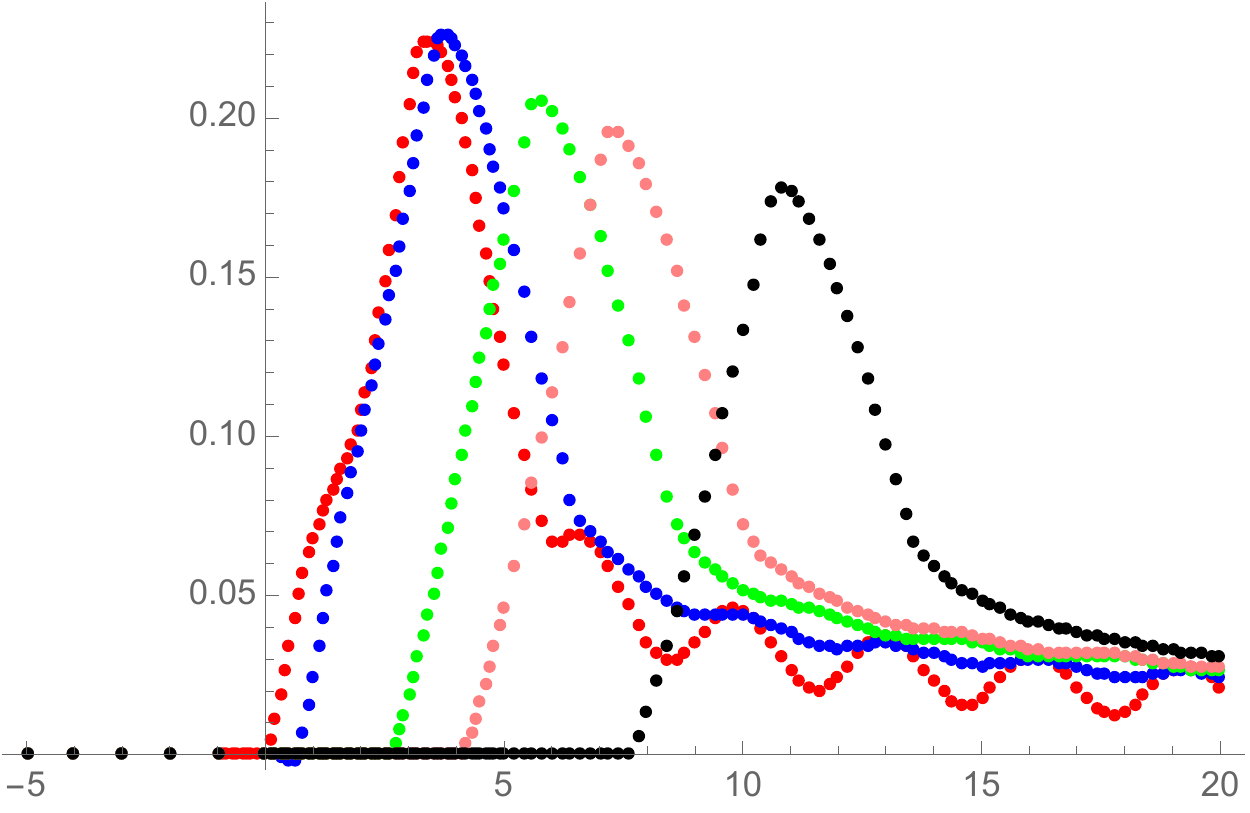}
  \end{center}
 \end{minipage}
   \put(-10,-40){$t/\xi$}
   \put(-120,50){$\Delta \mathcal{E}$}
   \put(-140,50){(a)}
 \begin{minipage}{0.35\hsize}
 \begin{center}
  \includegraphics[width=45mm]{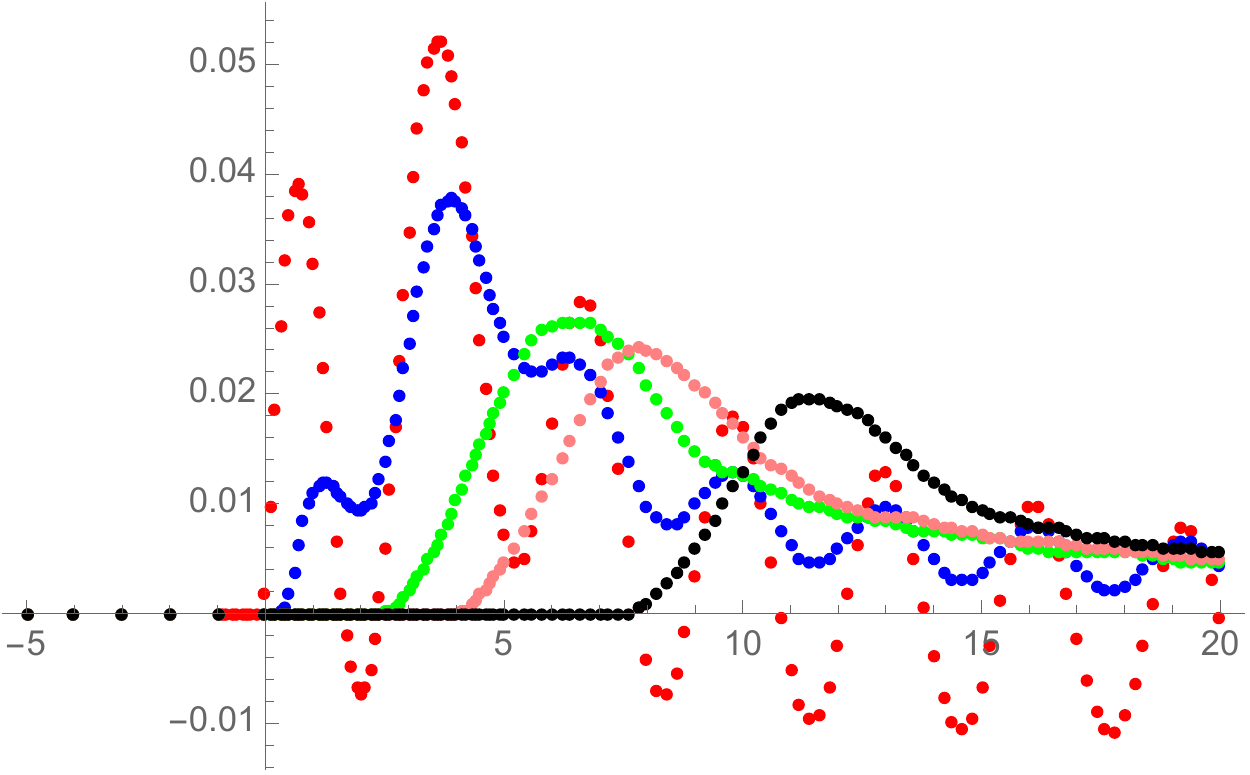}
 \end{center}
  \end{minipage}
     \put(-10,-25){$t/\xi$}
   \put(-130,50){$\Delta I_{A,B}$}
   \put(-150,50){(b)}
\vspace{0mm} \\
     \begin{minipage}{0.15\hsize}
  \begin{center}
   \includegraphics[width=25mm]{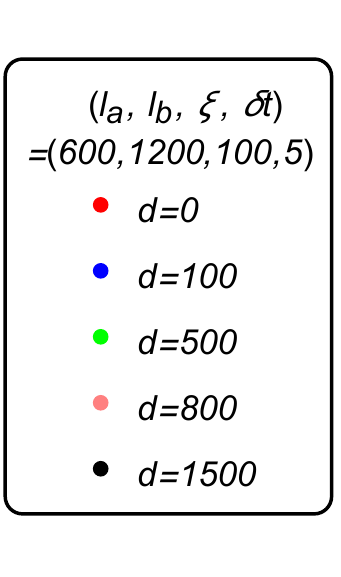}
  \end{center}
 \end{minipage}
 \begin{minipage}{0.35\hsize}
  \begin{center}
   \includegraphics[width=45mm]{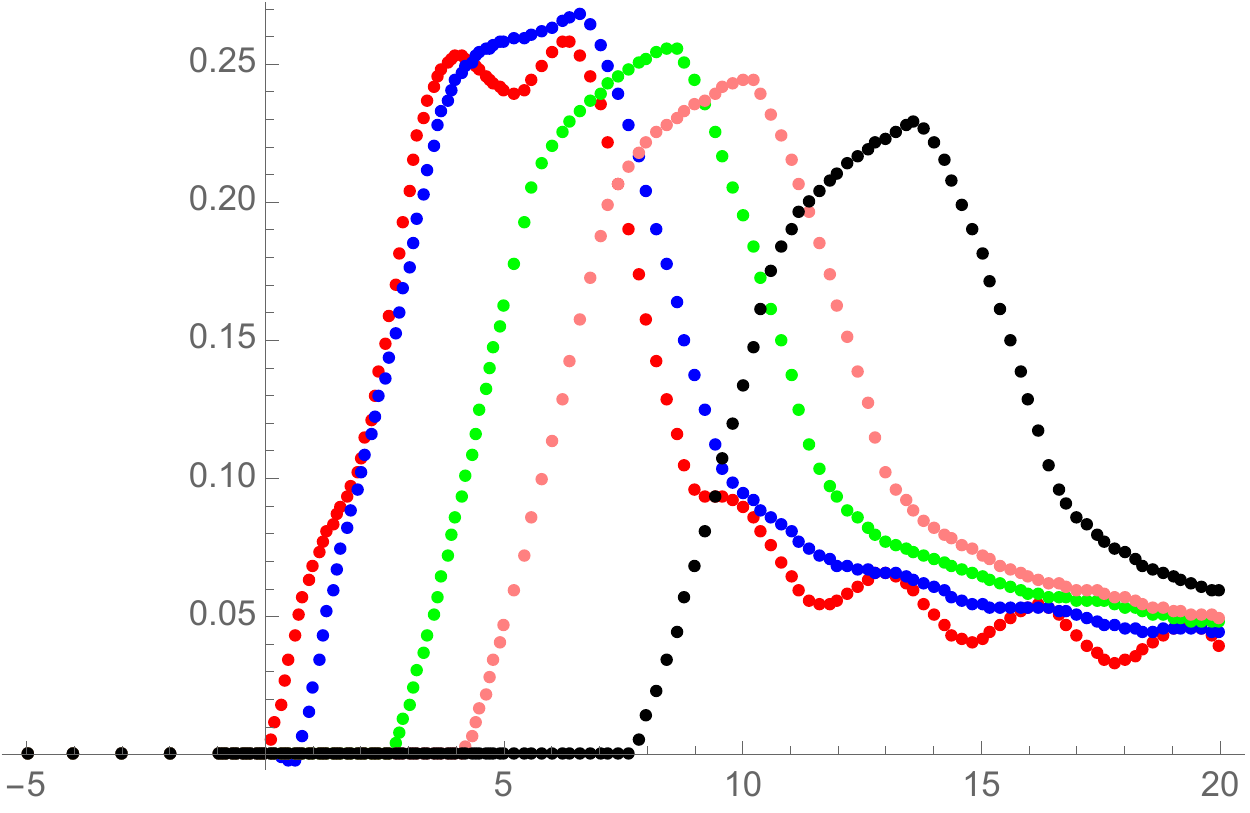}
  \end{center}
 \end{minipage}
   \put(-10,-30){$t/\xi$}
   \put(-120,50){$\Delta \mathcal{E}$}
      \put(-140,50){(c)}
 \begin{minipage}{0.35\hsize}
 \begin{center}
  \includegraphics[width=45mm]{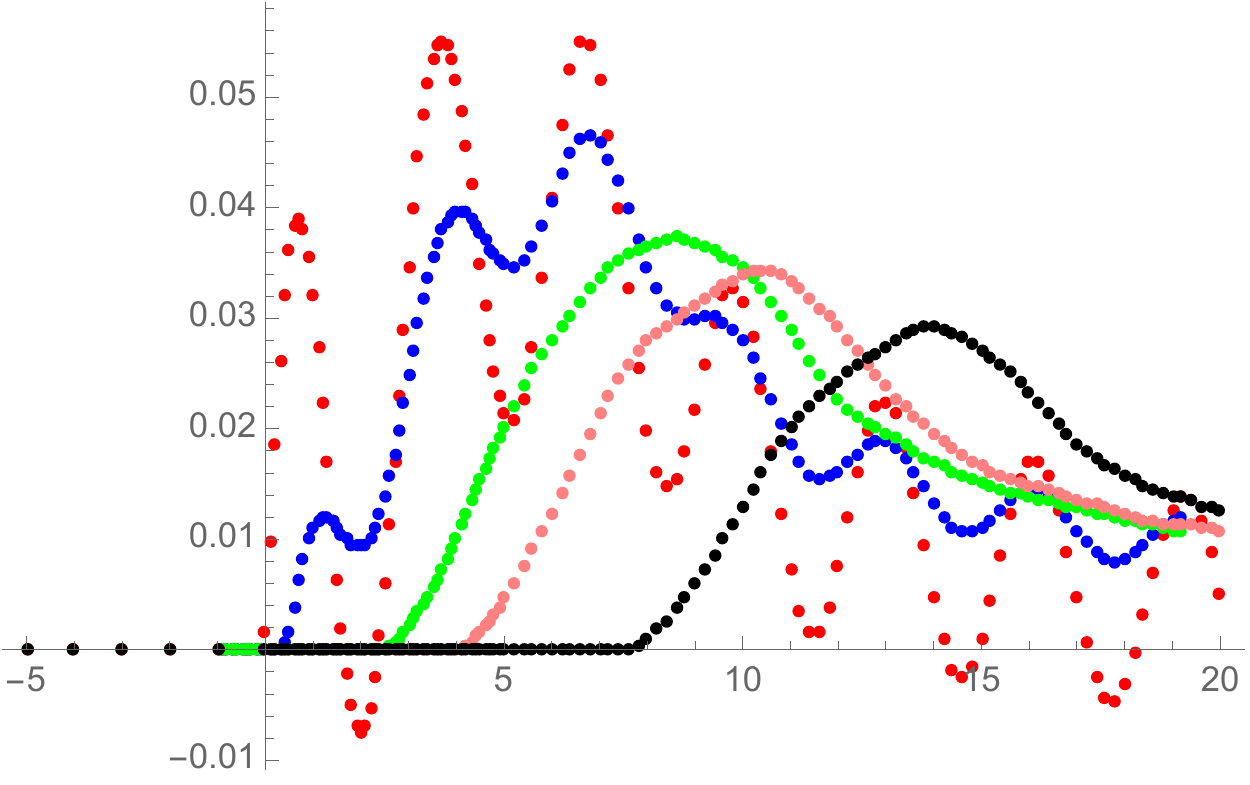}
 \end{center}
  \end{minipage}
     \put(-10,-30){$t/\xi$}
   \put(-120,50){$\Delta I_{A,B}$}
      \put(-140,50){(d)}
 \vspace{0mm} \\
     \begin{minipage}{0.15\hsize}
  \begin{center}
   \includegraphics[width=25mm]{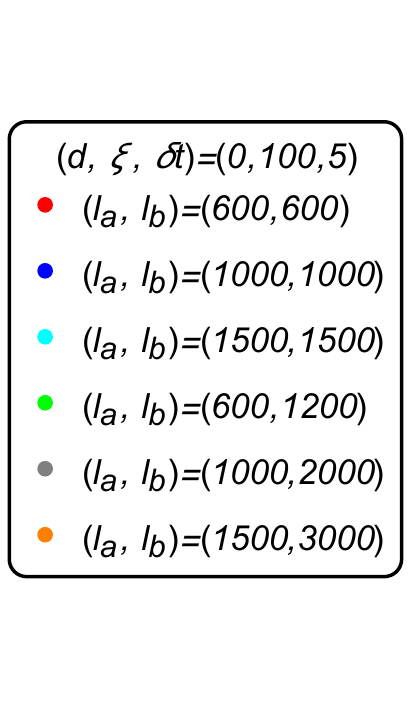}
  \end{center}
 \end{minipage}
 \begin{minipage}{0.35\hsize}
  \begin{center}
   \includegraphics[width=45mm]{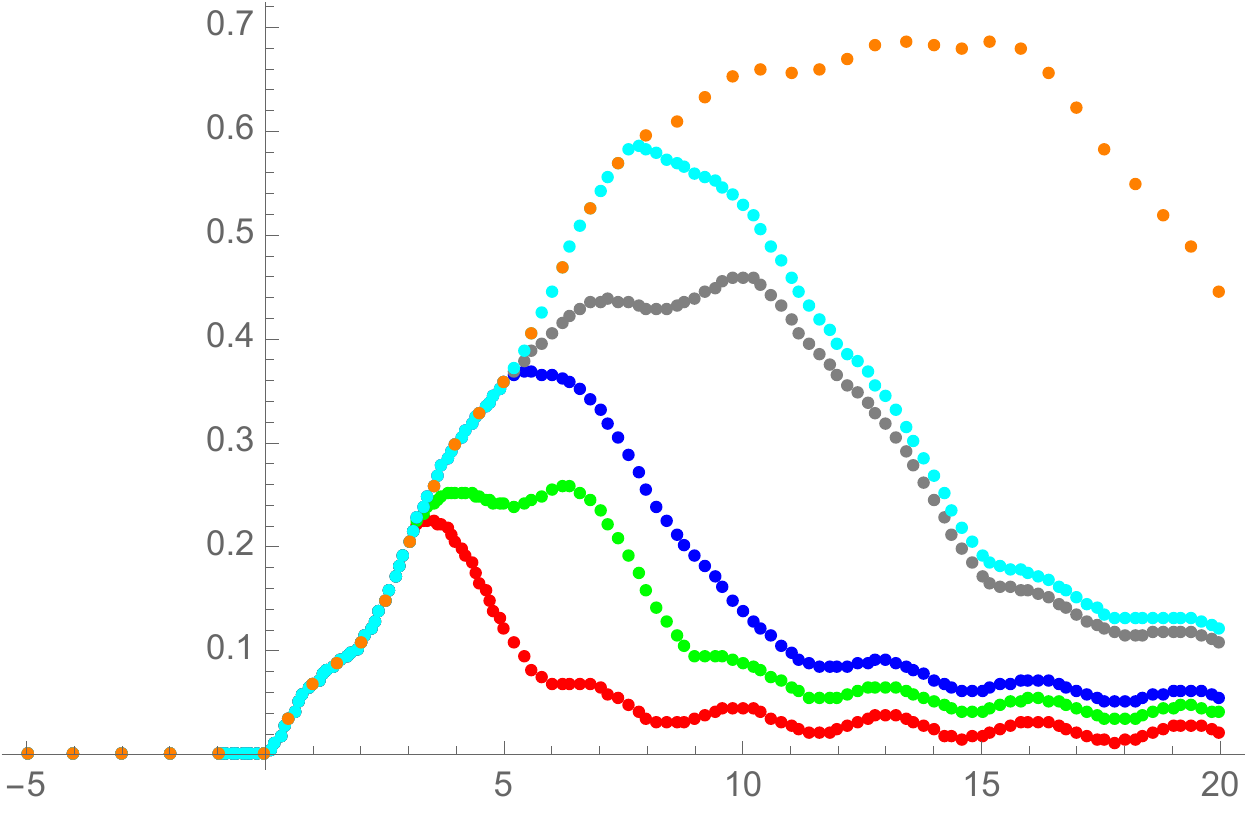}
  \end{center}
 \end{minipage}
   \put(-10,-40){$t/\xi$}
   \put(-120,50){$\Delta \mathcal{E}$}
      \put(-140,50){(e)}
 \begin{minipage}{0.35\hsize}
 \begin{center}
  \includegraphics[width=45mm]{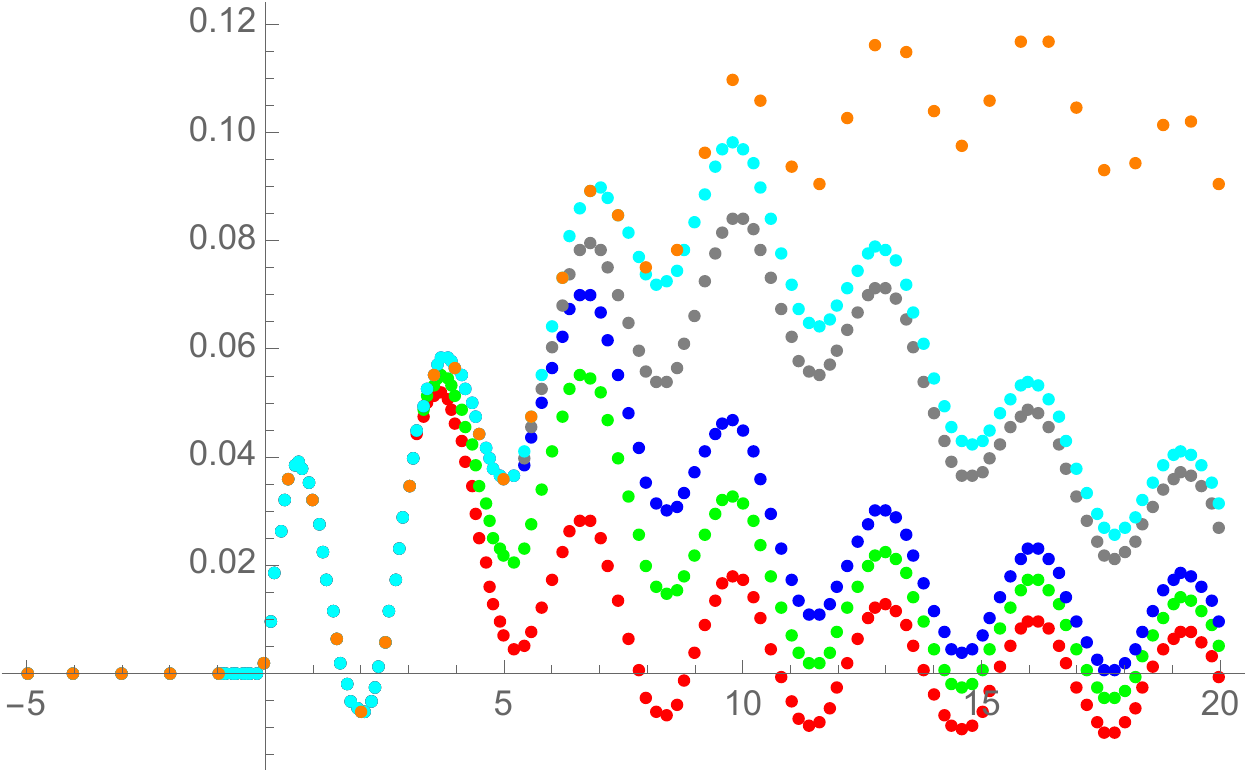}
 \end{center}
  \end{minipage}
     \put(-10,-30){$t/\xi$}
   \put(-120,50){$\Delta I_{A,B}$}
      \put(-140,50){(f)}
  \vspace{0mm} \\
       \begin{minipage}{0.15\hsize}
  \begin{center}
   \includegraphics[width=25mm]{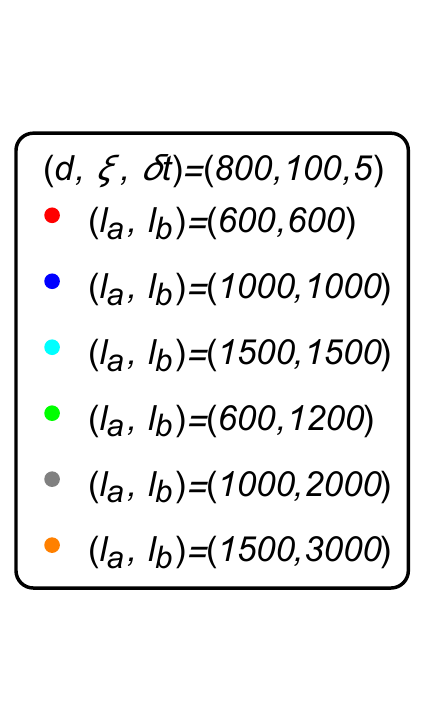}
  \end{center}
 \end{minipage}
 \begin{minipage}{0.35\hsize}
  \begin{center}
   \includegraphics[width=45mm]{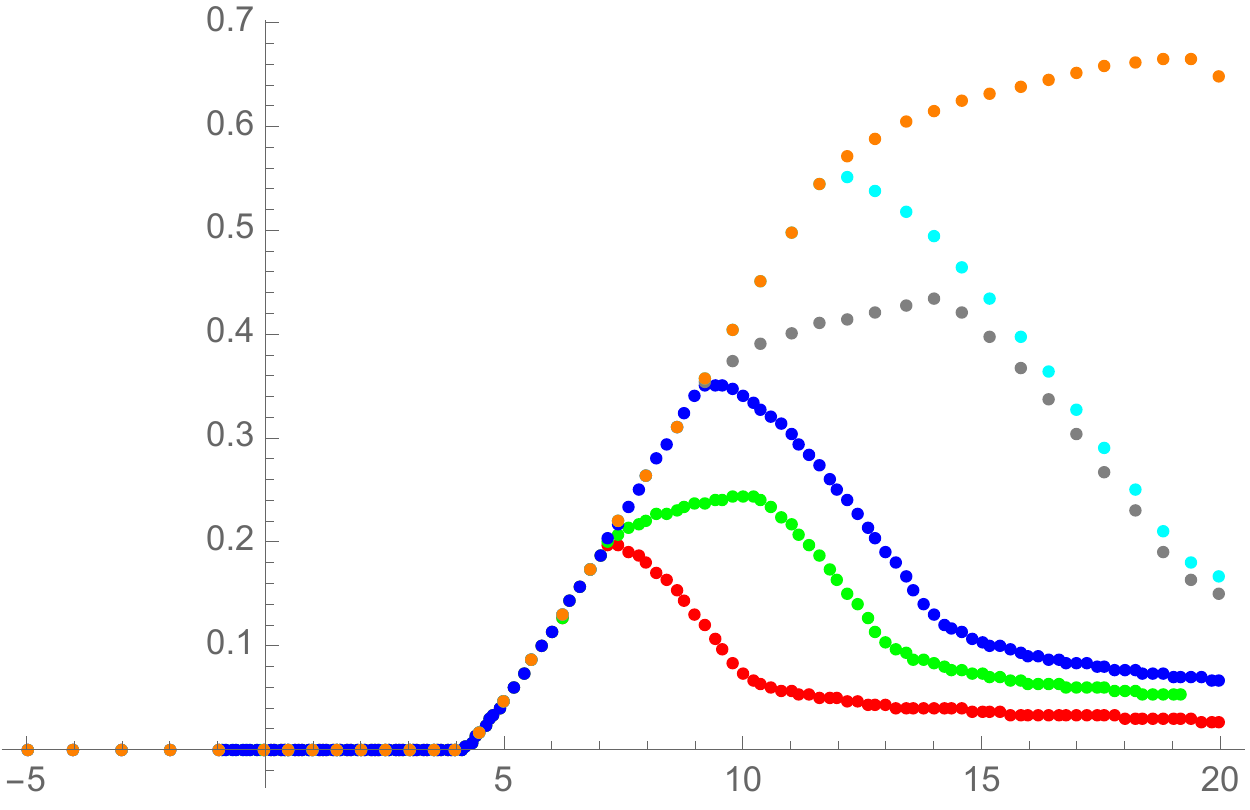}
  \end{center}
 \end{minipage}
   \put(-10,-40){$t/\xi$}
   \put(-120,50){$\Delta \mathcal{E}$}
      \put(-140,50){(g)}
 \begin{minipage}{0.35\hsize}
 \begin{center}
  \includegraphics[width=45mm]{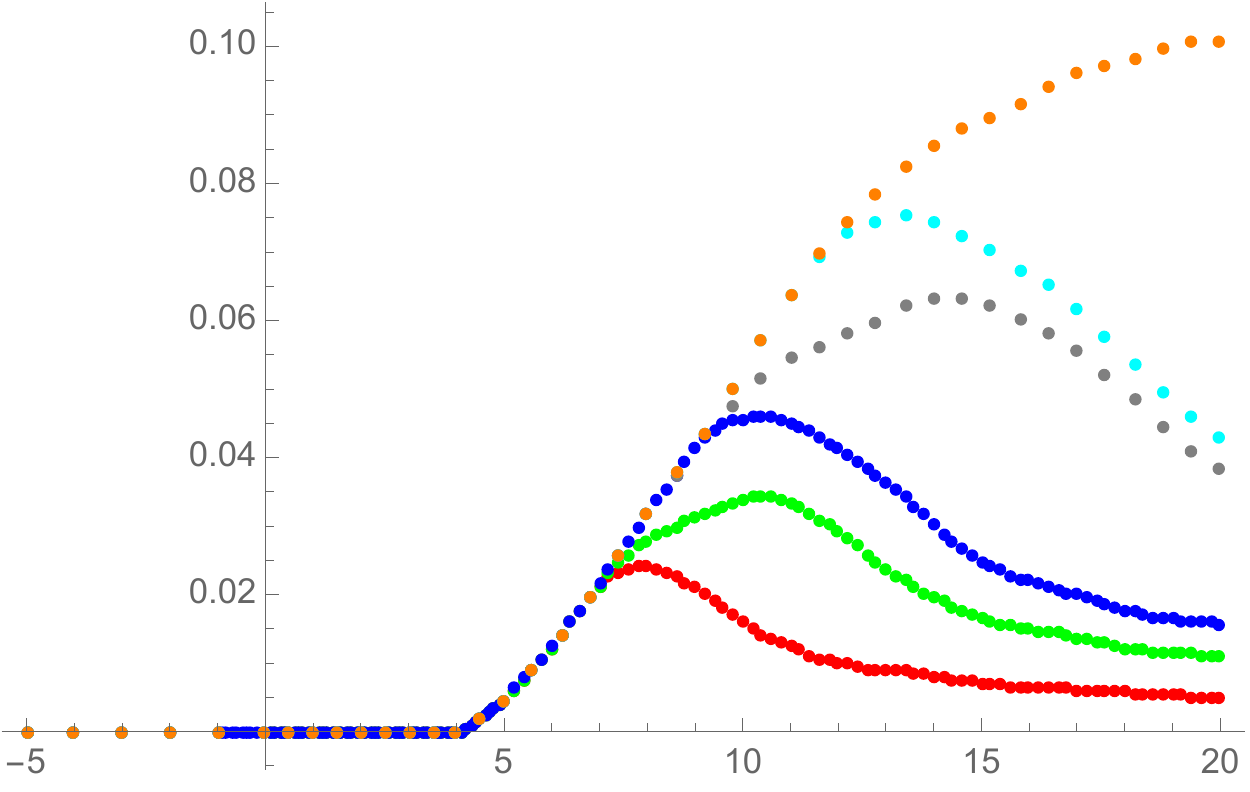}
 \end{center}
  \end{minipage}
     \put(-10,-40){$t/\xi$}
   \put(-120,50){$\Delta I_{A,B}$}
      \put(-140,50){(h)}
  \vspace{0mm} \\
     \begin{minipage}{0.15\hsize}
  \begin{center}
   \includegraphics[width=30mm]{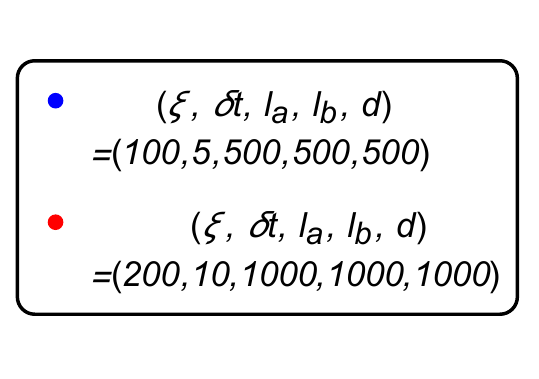}
  \end{center}
 \end{minipage}
 \begin{minipage}{0.35\hsize}
  \begin{center}
   \includegraphics[width=45mm]{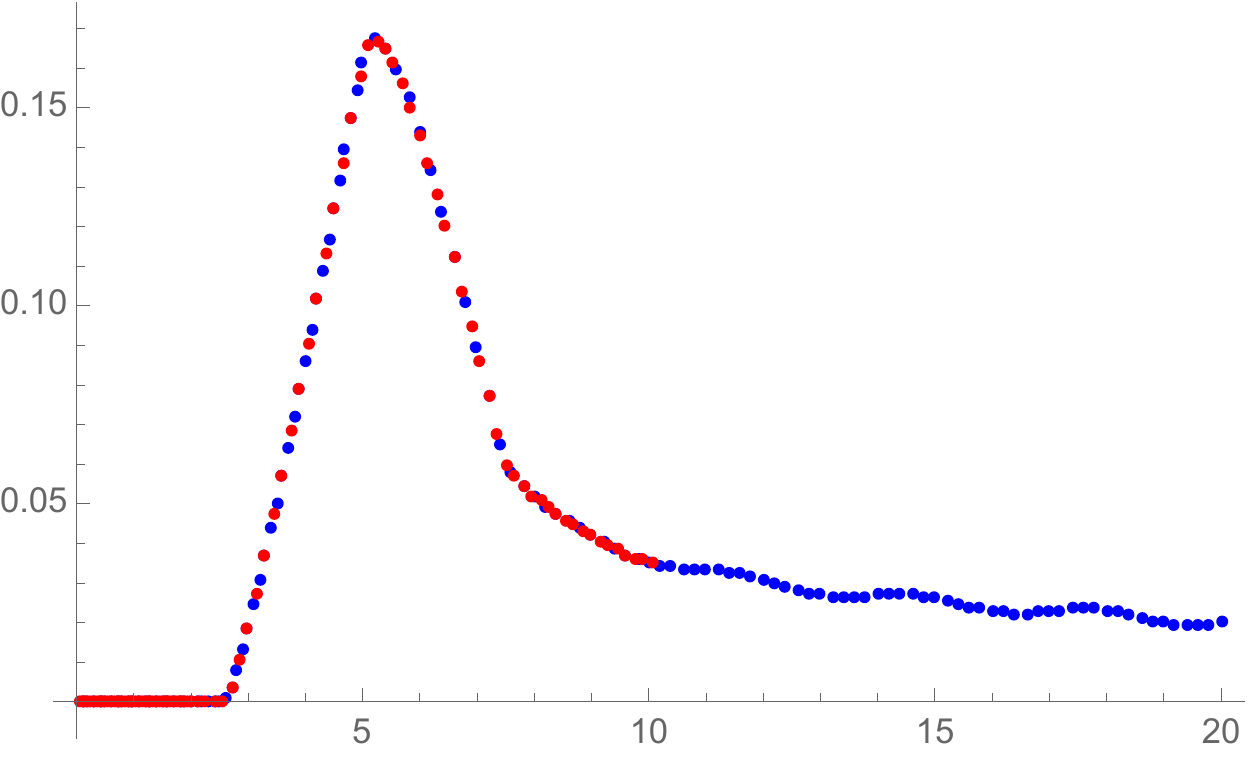}
  \end{center}
 \end{minipage}
   \put(-10,-40){$t/\xi$}
   \put(-145,50){$\Delta \mathcal{E}$}
      \put(-165,50){(i)}
 \begin{minipage}{0.35\hsize}
 \begin{center}
  \includegraphics[width=45mm]{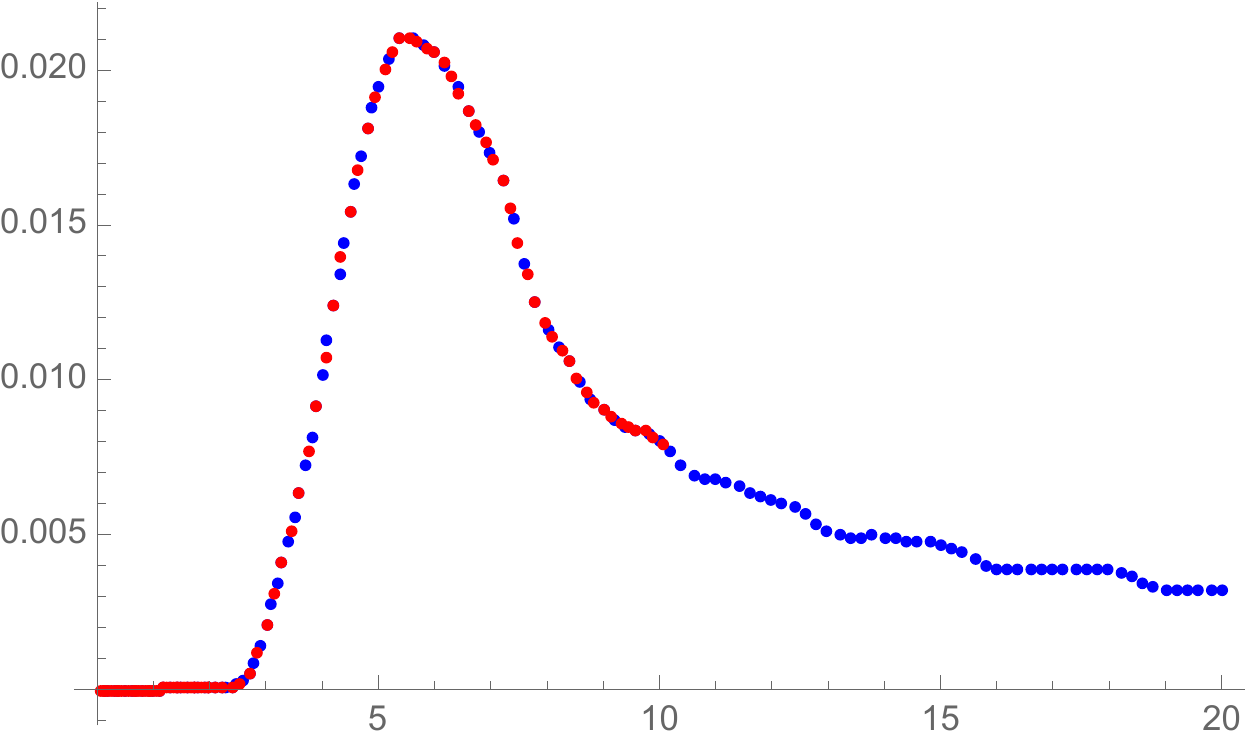}
 \end{center}
  \end{minipage}
     \put(-10,-20){$t/\xi$}
   \put(-140,50){$\Delta I_{A,B}$}
      \put(-160,50){(j)}
        \end{tabular}
  \caption{Time evolution of $\Delta \mathcal{E}$ (left side) and $\Delta I_{A,B}$ (right side) in the fast CCP. The panels (a), (b), (c), and (d) show $\Delta \mathcal{E}$ and $\Delta I_{A,B}$ with the  various distances between the two subsystems, and the panels (e), (f), (g), and (h) show $\Delta \mathcal{E}$ and $\Delta I_{A,B}$ with the various subsystem sizes $l_a$ and $l_b$. The panels (i) and (j) show $\Delta \mathcal{E}$ and $\Delta I_{A,B}$ have the scaling law.}
  \label{ccpf}
\end{figure}

\begin{enumerate}[(1)] 
\item 
 The quantities $\Delta \mathcal{E}$ and $\Delta I_{A,B}$ in the fast CCP oscillate.

\item 
 If $d \gg \xi$, the amplitude of oscillation of $\Delta\mathcal{E}$ and $\Delta I_{A, B}$  in the fast CCP seems to be small, and their time evolution becomes similar to that in the fast ECP.  The quantities $\Delta \mathcal{E}$ and $\Delta I_{A, B}$ with $d \gg \xi$ time-evolve as follows: They start to increase around $t\sim \frac{d}{2}$, and their slopes  with $l_a<l_b$  become small around $t\sim\frac{\l_a+d}{2}.$ Then, $\Delta \mathcal{E}$ and $\Delta I_{A, B}$ start to decrease around $t\sim\frac{\l_b+d}{2}$, and their slopes become small after $t\sim\frac{l_a+l_b+d}{2}$. 
 
 \item If $d \ll\xi$, the amplitude of the oscillation becomes large. A frequency $\omega$ of oscillation of $\Delta \mathcal{E}$ and $\Delta I_{A,B}$ at late time $t\gg\delta t$ is $\omega \sim \f{1}{\pi \xi}$.
The amplitude of the oscillation in $\Delta I_{A, B}$ is relatively larger than the one in $\Delta\mathcal{E}$. The mutual information $\Delta I_{A, B}$ with $d=0$ has a first dip at $t\sim2\xi$, and the logarithmic negativity $\Delta\mathcal{E}$ with $d=0$ also has a small dip around $t\sim2\xi$.

\item At early time $t< \frac{\l_a+d}{2}$, $\Delta\mathcal{E}$ and $\Delta I_{A, B}$ with fixed $d$ are independent of $l_a$ and $l_b$.
\item There is the scaling law for $\Delta\mathcal{E}$ and $\Delta I_{A, B}$ in the panels (i) and (j) of Figure \ref{ccpf}. If $l_a,l_b, \xi, d \gg 1$, we expect $\Delta\mathcal{E}$ and $\Delta I_{A, B}$ to have the following scaling law: 
\begin{align}
\Delta\mathcal{E}&\sim \Delta\mathcal{E}\left(\frac{l_a}{\xi}, \frac{l_b}{\xi}, \frac{d}{\xi}, \frac{t}{\xi}, \frac{\delta t}{\xi}\right),\notag\\
\Delta I_{A, B}&\sim \Delta I_{A, B}\left(\frac{l_a}{\xi}, \frac{l_b}{\xi}, \frac{d}{\xi}, \frac{t}{\xi}, \frac{\delta t}{\xi}\right).\label{scalinglawccpf}
\end{align}
\end{enumerate}

The panels of Figure \ref{ccpf} show that $\Delta \mathcal{E}$ and $\Delta I_{A,B}$ have the properties (1)-(5).
The panels (a) and (b) show the time evolution of $\Delta\mathcal{E}$ and $\Delta I_{A, B}$ for the subsystems with $l_a=l_b$, and the panels (c) and (d) show that with $l_a<l_b$. We plot the time evolution of $\Delta\mathcal{E}$ and $\Delta I_{A, B}$ with the various distance $d$ to study the $d$-dependence of $\Delta\mathcal{E}$ and $\Delta I_{A, B}$.  The panels of Figure \ref{ccpf} show that $\Delta \mathcal{E}$ and $\Delta I_{A,B}$ in the fast CCP oscillate (the property (1)).   The amplitude of their oscillation with $d \gg \xi$ is much smaller than the amplitude for $d \ll \xi$.
The time evolution of $\Delta \mathcal{E}$ and $\Delta I_{A, B}$ with $d \gg \xi$ is similar to their evolution in the fast ECP\footnote{In the panels (a) and (c) of Figure \ref{ccpf}, $\Delta \mathcal{E}$ with $d=100$ is negative around $t\sim50$. We expect this decrease to be related to $d\sim\xi$. Since we require more computations with $d\sim\xi$ for a better interpretation of  the decrease, we leave it for future work.}. In particular, $\Delta \mathcal{E}$ and $\Delta I_{A, B}$ in the fast CCP with $d \gg \xi$  start to increase at $t\sim \f{d}{2}$, and their slopes with $l_a<l_b$ becomes smaller around $t\sim \f{d+l_a}{2}$. They start to decrease at $t\sim \f{d+l_b}{2}$, and almost vanish after $t\sim\f{d+l_a+l_b}{2}$ (the property (2)).   The panels (e) and (f) show the evolution of $\Delta\mathcal{E}$ and $\Delta I_{A, B}$ with $d=0$, and the panels (g) and (h) show them with $d=800$. The panels (e) and (f)
show that $\Delta \mathcal{E}$ and $\Delta I_{A,B}$ oscillate with the frequency, which at late time $t\gg\delta t$ is expected to be $\omega \sim \f{1}{\pi \xi}$  as in \cite{Nishida:2017hqd}.
If we define the amplitude of oscillation of $\Delta \mathcal{E}$ and $\Delta I_{A,B}$ by $\mathcal{A}_{\mathcal{E}}$ and $\mathcal{A}_{I}$, the ratio of $\mathcal{A}_{\mathcal{E}}$ to $\left|\Delta\mathcal{E}(t) \right|$ is much smaller than the ratio of $\mathcal{A}_{I}$ to $\left|\Delta I_{A,B}(t) \right|$. The change $\Delta I_{A,B}$ with $d=0$ has a first dip at $t\sim 2\xi$, and $\Delta\mathcal{E}$ with $d=0$ also has a small dip around $t\sim2\xi$ (the property (3)).
At  early time  $t< \frac{\l_a+d}{2}$, all plots in the panels (e), (f), (g), and (h) lie on the same curve, and they therefore are independent of $l_a$ and $l_b$ at the early time (the property (4)). 
The panel (c) of Figure \ref{ecpf} shows that $\Delta \mathcal{E}$  in the fast ECP with $l_a<l_b$ has a plateau in the window  $\frac{l_a+d}{2}\lessapprox t\lessapprox\frac{l_b+d}{2}$, but $\Delta\mathcal{E}$ in the fast CCP with $l_a<l_b$ in the panel (g) of Figure \ref{ccpf} increases in this window. 

The panels (i) and (j) show the plots of $\Delta\mathcal{E}$ and $\Delta I_{A, B}$ with two different pairs of parameters $(\xi, \delta t)=(100,5)$ and $(200,10)$. 
These panels show that the plots for $\Delta \mathcal{E}$ and $\Delta I_{A,B}$ in the CCP with $(\xi, \delta t, l_a, l_b, d)=(100,5,500, 500,500)$ are on the plots with $(\xi, \delta t, l_a, l_b, d)=(200,10,1000, 1000,1000)$,  and we expect $\Delta\mathcal{E}$ and $\Delta I_{A, B}$ to have the scaling law in (\ref{scalinglawccpf}) if the parameters $\xi, l_a, l_b$, and $d$ are much larger than one (the property (5)).

\subsubsection{Slow limit}

Here, we study $\Delta\mathcal{E}$ and $\Delta I_{A, B}$ in the slow CCP. In the slow CCP, as our previous study of entanglement entropy \cite{Nishida:2017hqd}, it does not seem possible to use the quasi-particle interpretation with the speed of light for determining the time scales at which the behavior of $\Delta\mathcal{E}$ and $\Delta I_{A, B}$ changes.  We calculate $\Delta\mathcal{E}$ and $\Delta I_{A, B}$ in the slow CCP  with $(\xi=10, \delta t=1000)$, $(\xi=20, \delta t=2000)$, and $(\xi=10, \delta t=4000)$. Here, we mainly show $\Delta \mathcal{E}$ and $\Delta I_{A,B}$ for $(\xi=20, \delta t=2000, \xi_{\text{kz}}=200)$, which are computed numerically. Since it takes a long time to compute $X_\kappa(t), P_\kappa(t)$, and $D_\kappa(t)$ in the slow CCP numerically, we use the adiabatic approximation of them for large $|\kappa|$ as well as the computation of entanglement entropy in the slow CCP \cite{Nishida:2017hqd} \footnote{In the numerical computations for the slow CCP, we use the adiabatic approximation of $f_\kappa(t)$ with $|\kappa|>\kappa_*$. We  check the $\kappa_*$-dependence of $\Delta \mathcal{E}$ and $\Delta I_{A,B}$ with $(\xi=10, \delta t=1000)$ by computing them  with $\kappa_*=0.05, 0.3, 0.5$, and the computation results  show no significant difference between them.   }. See the appendix of \cite{Nishida:2017hqd} for more details of the approximation.
Figure \ref{ccps} shows the time evolution of $\Delta\mathcal{E}$ and $\Delta I_{A, B}$ in the slow CCP. 
We are not able to find any significant differences between $\Delta \mathcal{E}$ and $\Delta I_{A,B}$.
We find the following properties of $\Delta\mathcal{E}$ and $\Delta I_{A,B}$ in the slow CCP:

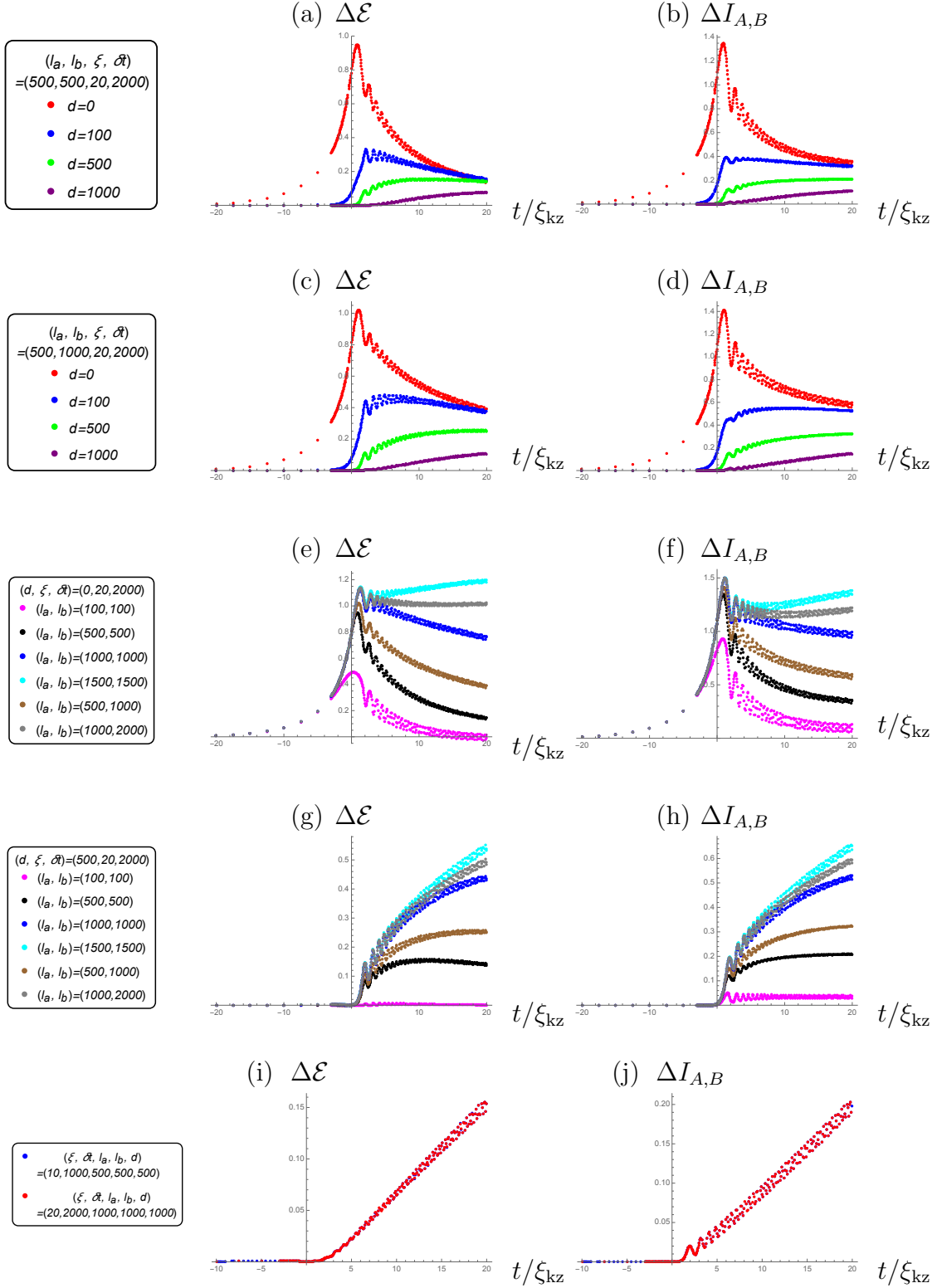
\begin{figure}[htbp]
\begin{tabular}{c}
 \begin{minipage}{0.15\hsize}
\begin{center}
\includegraphics[width=25mm]{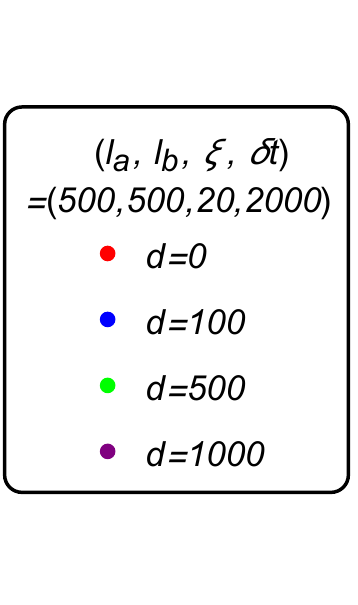}
 \end{center}
 \end{minipage}
 \begin{minipage}{0.35\hsize}
  \begin{center}
   \includegraphics[width=45mm]{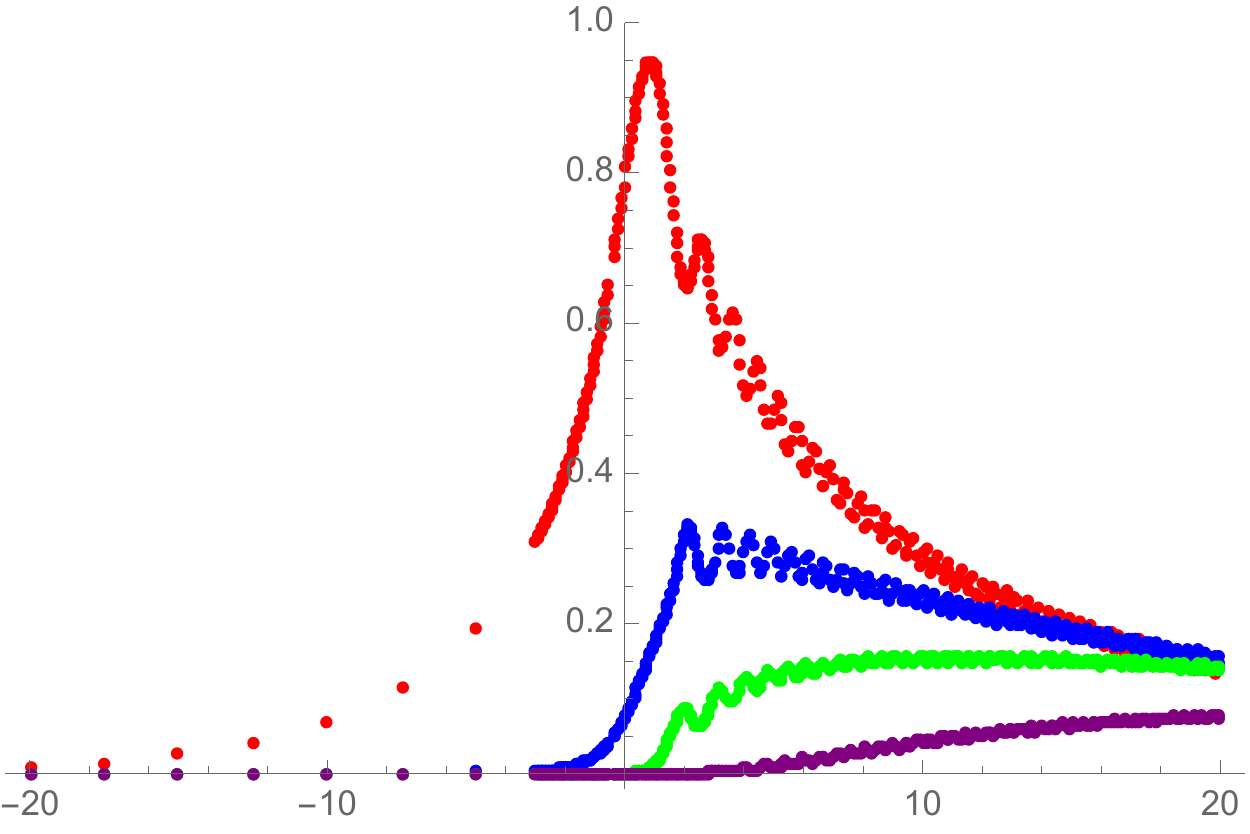}
  \end{center}
 \end{minipage}
   \put(-10,-40){$t/\xi_{\text{kz}}$}
   \put(-90,50){$\Delta \mathcal{E}$}
   \put(-110,50){(a)}
 \begin{minipage}{0.35\hsize}
 \begin{center}
  \includegraphics[width=45mm]{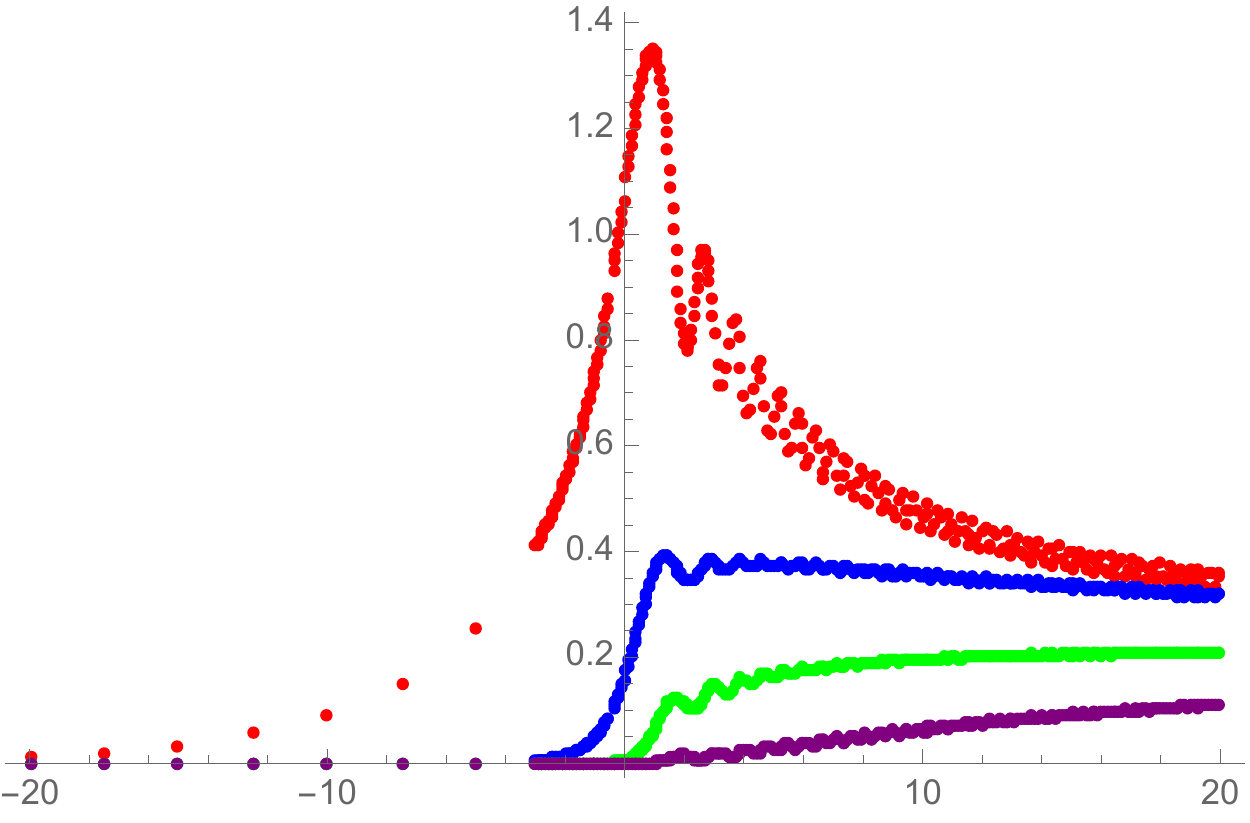}
 \end{center}
  \end{minipage}
     \put(-10,-40){$t/\xi_{\text{kz}}$}
   \put(-90,50){$\Delta I_{A,B}$}
   \put(-110,50){(b)}
\vspace{0mm} \\
     \begin{minipage}{0.15\hsize}
  \begin{center}
   \includegraphics[width=25mm]{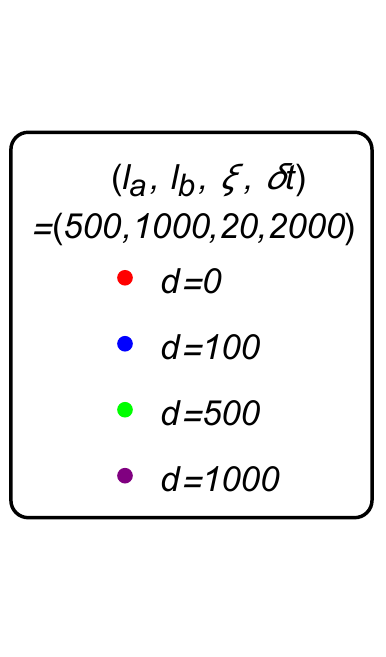}
  \end{center}
 \end{minipage}
 \begin{minipage}{0.35\hsize}
  \begin{center}
   \includegraphics[width=45mm]{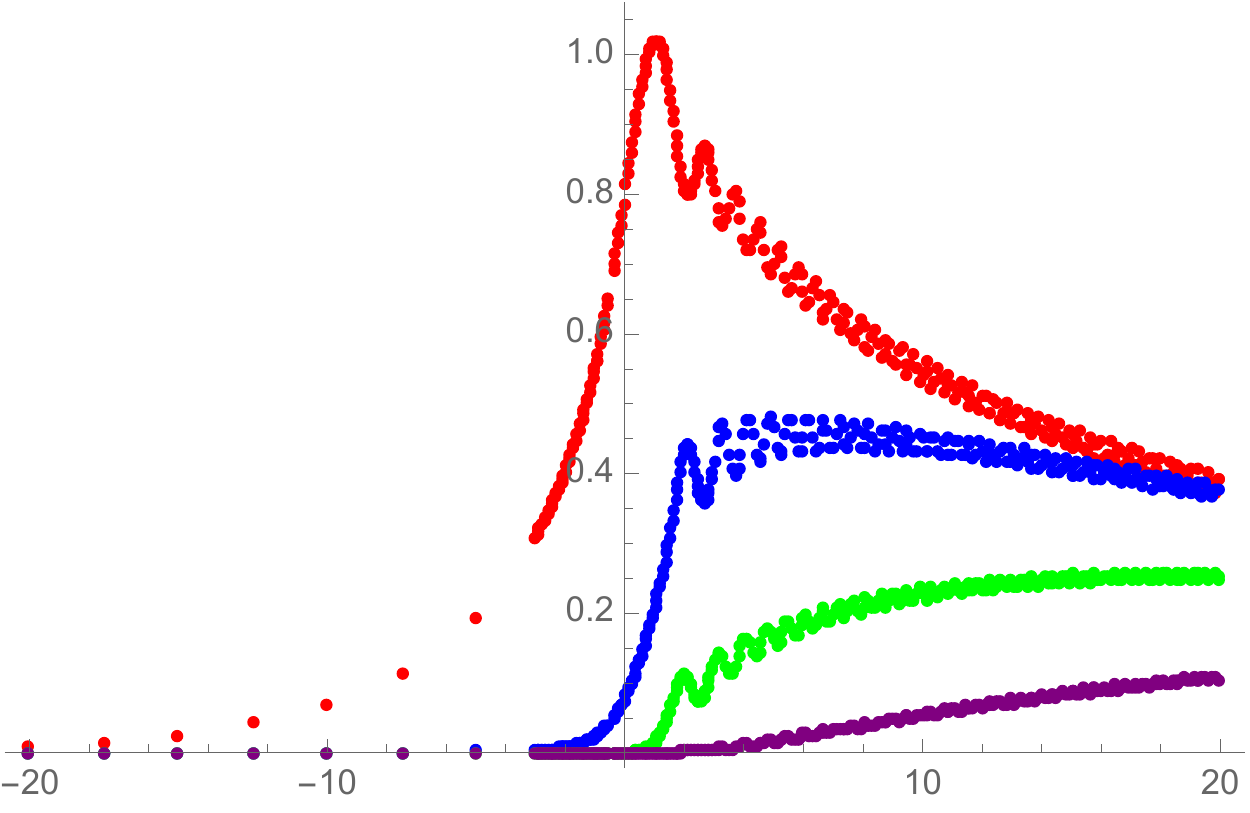}
  \end{center}
 \end{minipage}
   \put(-10,-30){$t/\xi_{\text{kz}}$}
   \put(-90,50){$\Delta \mathcal{E}$}
      \put(-110,50){(c)}
 \begin{minipage}{0.35\hsize}
 \begin{center}
  \includegraphics[width=45mm]{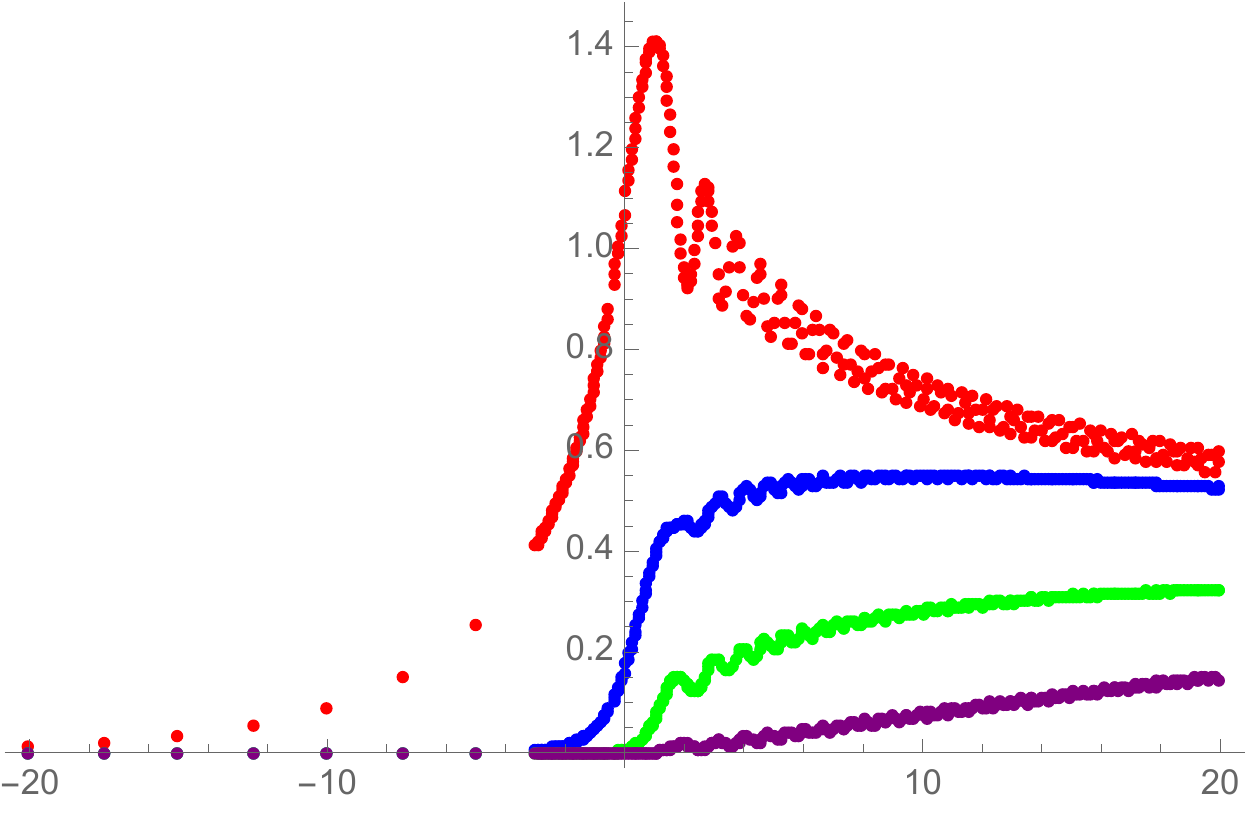}
 \end{center}
  \end{minipage}
     \put(-10,-30){$t/\xi_{\text{kz}}$}
   \put(-90,50){$\Delta I_{A,B}$}
      \put(-110,50){(d)}
 \vspace{0mm} \\
     \begin{minipage}{0.15\hsize}
  \begin{center}
   \includegraphics[width=25mm]{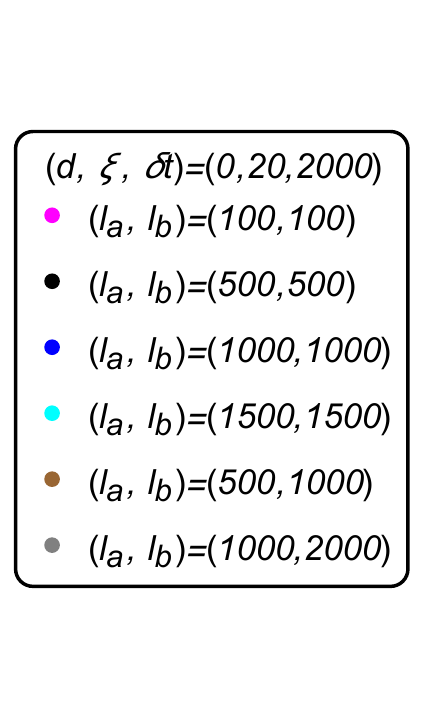}
  \end{center}
 \end{minipage}
 \begin{minipage}{0.35\hsize}
  \begin{center}
   \includegraphics[width=45mm]{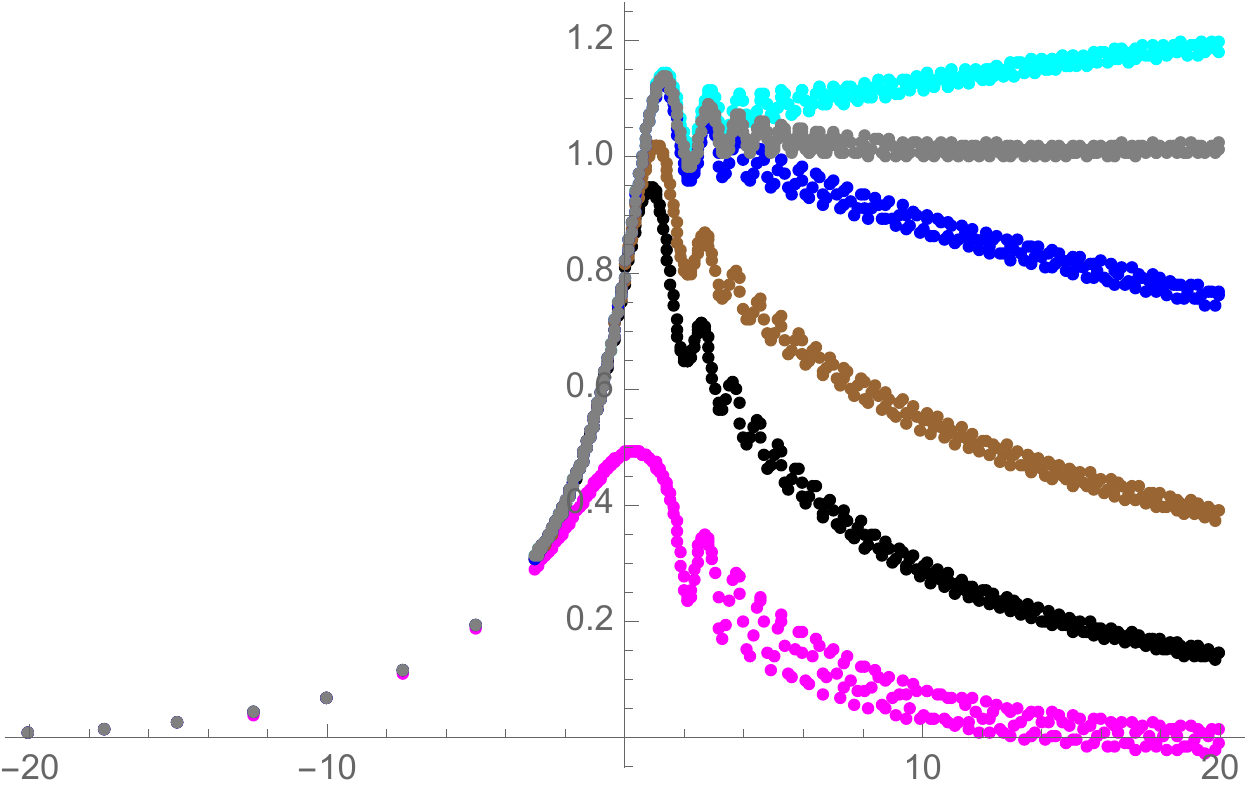}
  \end{center}
 \end{minipage}
   \put(-10,-40){$t/\xi_{\text{kz}}$}
   \put(-90,50){$\Delta \mathcal{E}$}
      \put(-110,50){(e)}
 \begin{minipage}{0.35\hsize}
 \begin{center}
  \includegraphics[width=45mm]{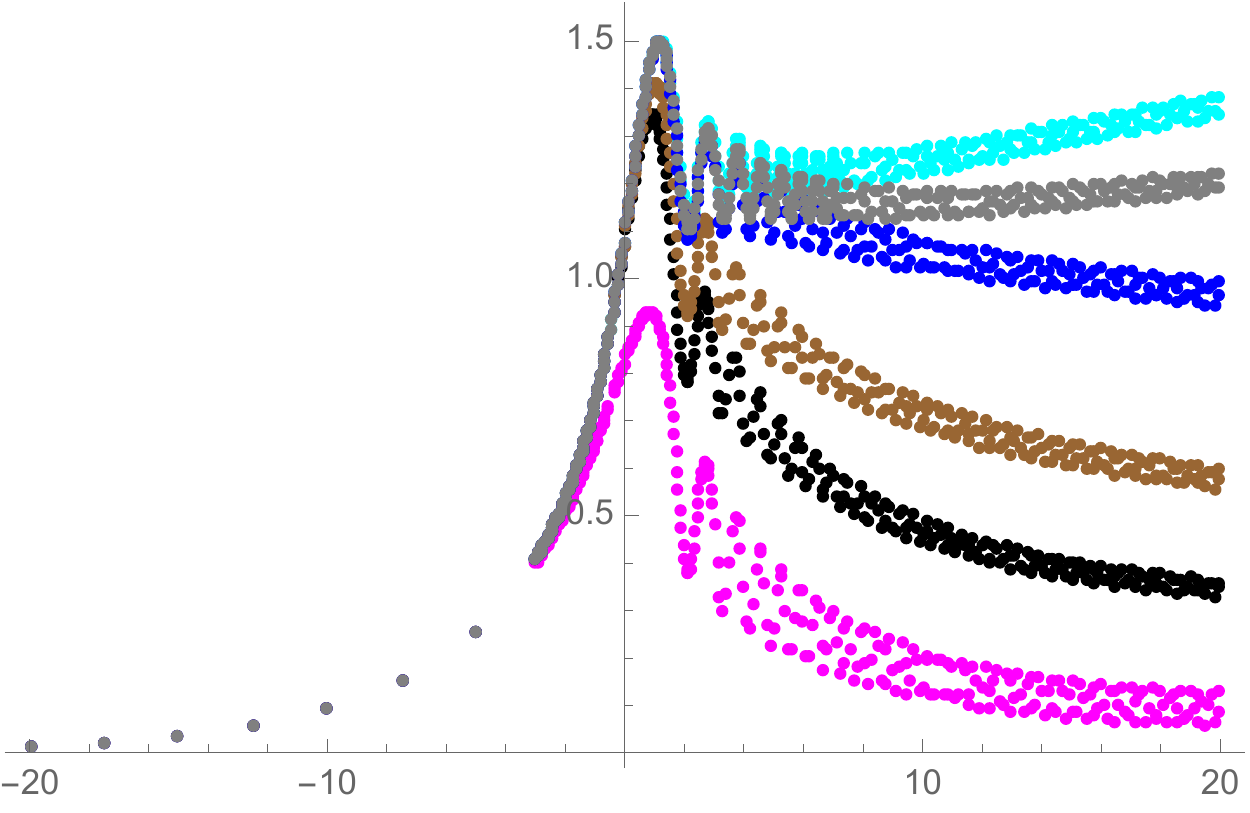}
 \end{center}
  \end{minipage}
     \put(-10,-30){$t/\xi_{\text{kz}}$}
   \put(-90,50){$\Delta I_{A,B}$}
      \put(-110,50){(f)}
  \vspace{0mm} \\
       \begin{minipage}{0.15\hsize}
  \begin{center}
   \includegraphics[width=25mm]{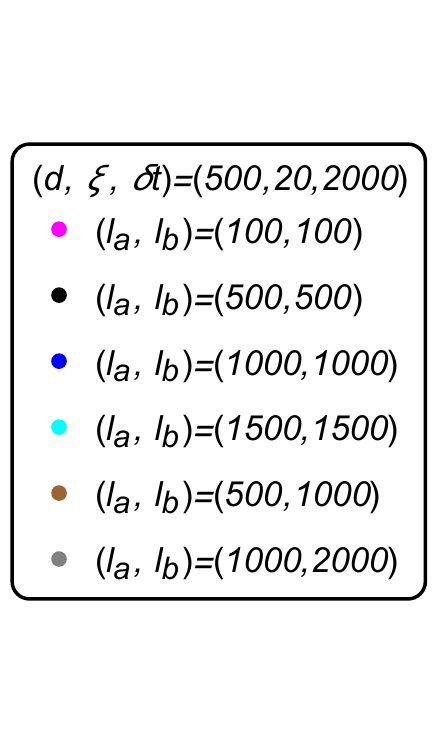}
  \end{center}
 \end{minipage}
 \begin{minipage}{0.35\hsize}
  \begin{center}
   \includegraphics[width=45mm]{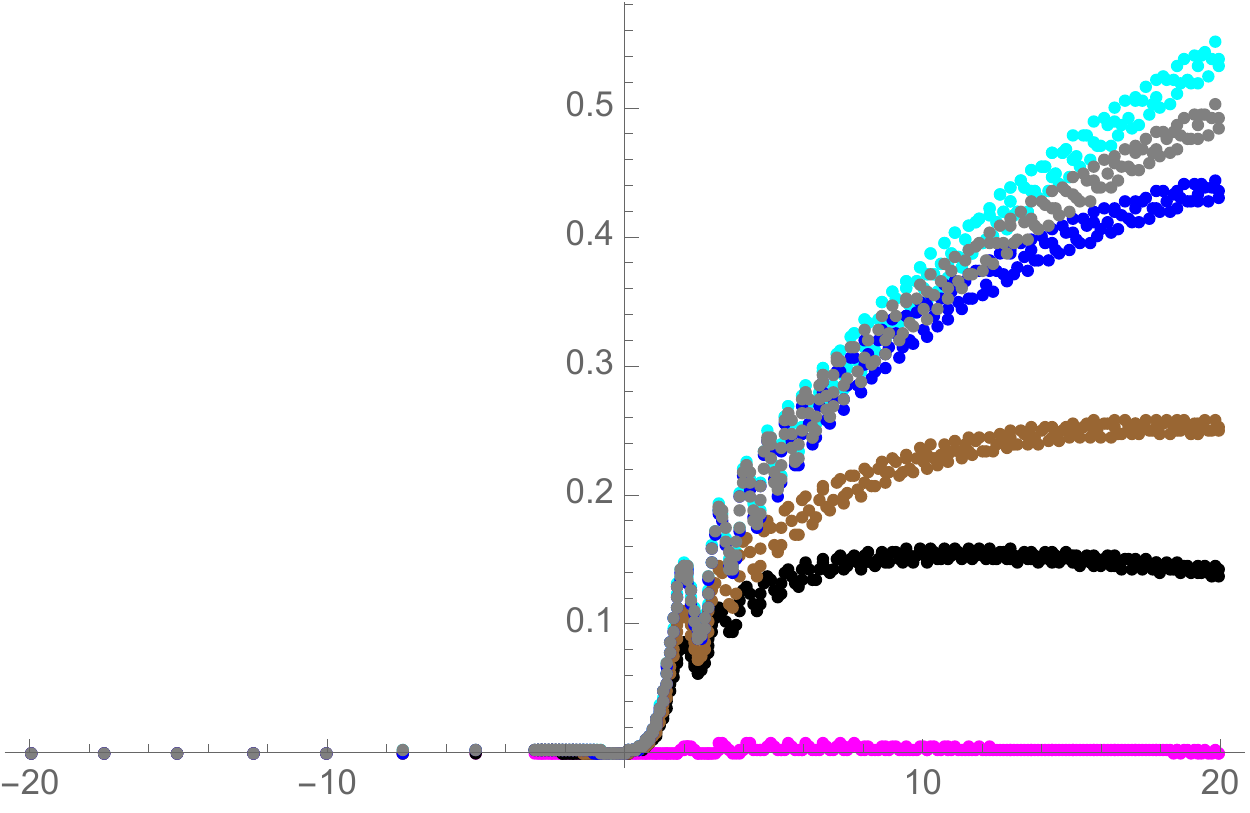}
  \end{center}
 \end{minipage}
   \put(-10,-40){$t/\xi_{\text{kz}}$}
   \put(-90,50){$\Delta \mathcal{E}$}
      \put(-110,50){(g)}
 \begin{minipage}{0.35\hsize}
 \begin{center}
  \includegraphics[width=45mm]{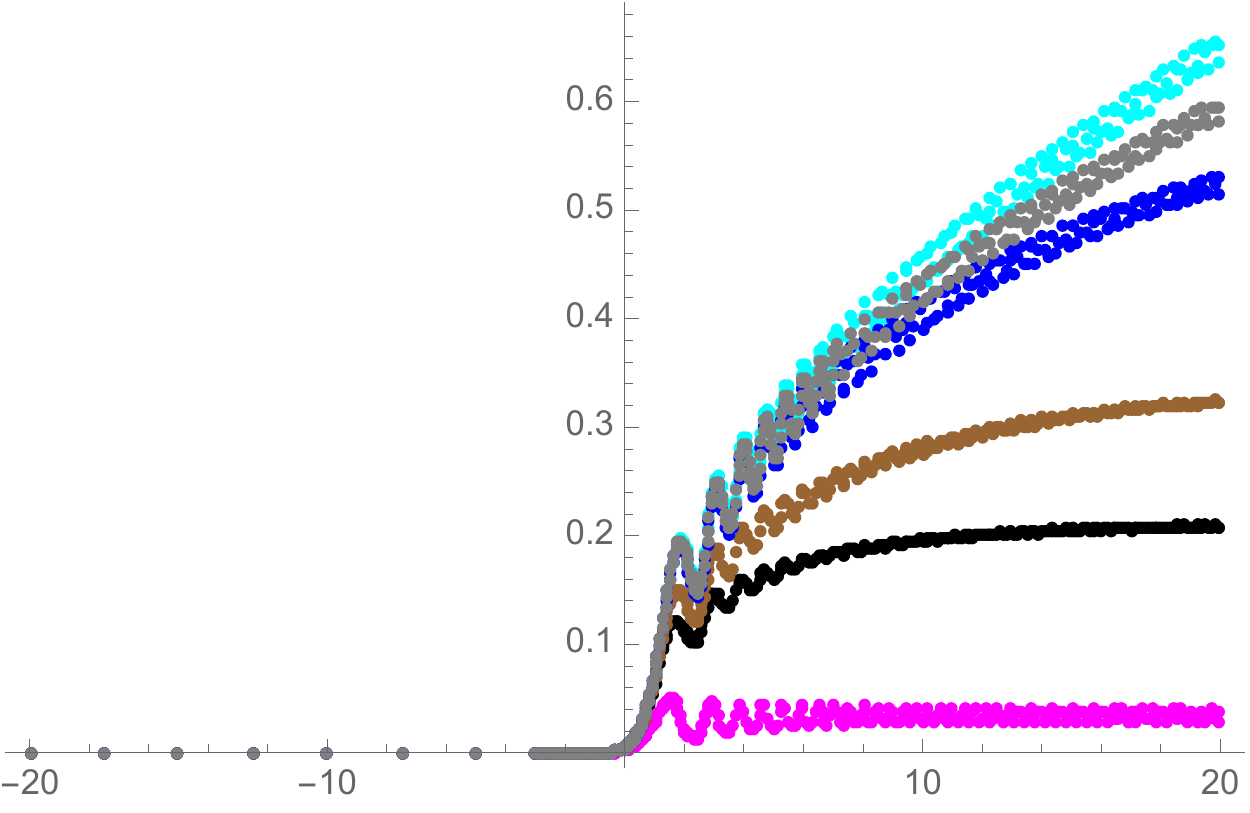}
 \end{center}
  \end{minipage}
     \put(-10,-40){$t/\xi_{\text{kz}}$}
   \put(-90,50){$\Delta I_{A,B}$}
      \put(-110,50){(h)}
  \vspace{0mm} \\
     \begin{minipage}{0.15\hsize}
  \begin{center}
   \includegraphics[width=30mm]{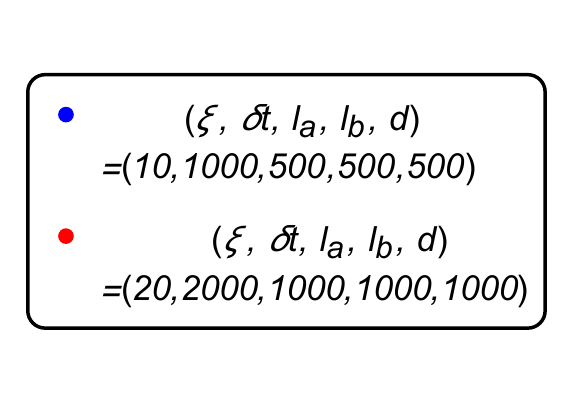}
  \end{center}
 \end{minipage}
 \begin{minipage}{0.35\hsize}
  \begin{center}
   \includegraphics[width=45mm]{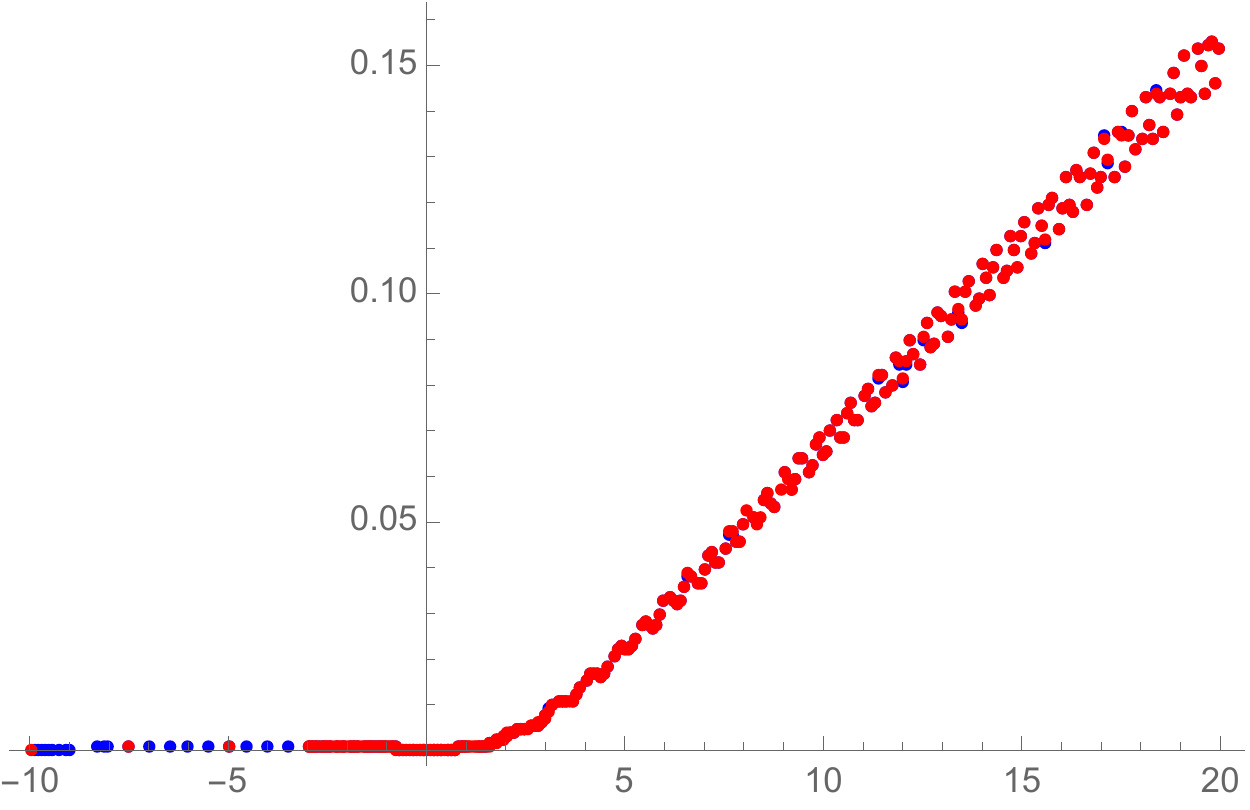}
  \end{center}
 \end{minipage}
   \put(-10,-40){$t/\xi_{\text{kz}}$}
   \put(-110,50){$\Delta \mathcal{E}$}
      \put(-130,50){(i)}
 \begin{minipage}{0.35\hsize}
 \begin{center}
  \includegraphics[width=45mm]{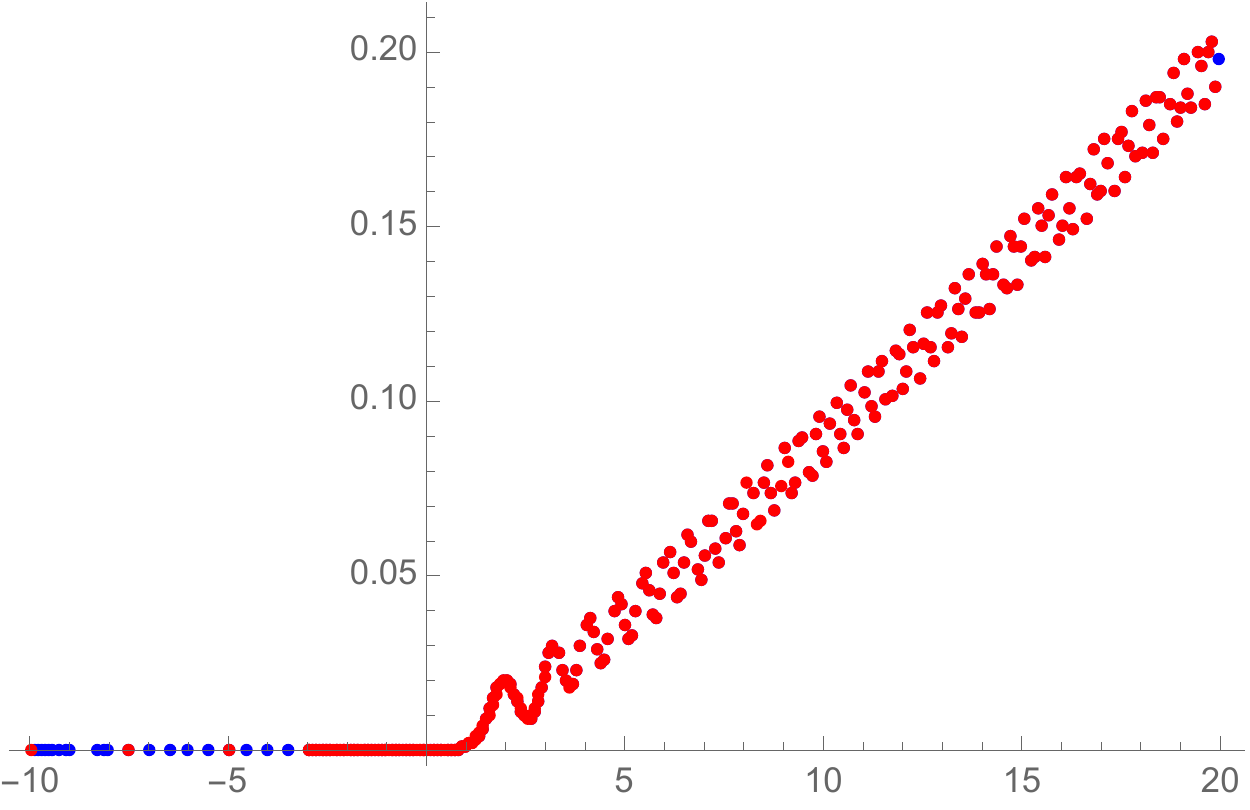}
 \end{center}
  \end{minipage}
     \put(-10,-40){$t/\xi_{\text{kz}}$}
   \put(-110,50){$\Delta I_{A,B}$}
      \put(-130,50){(j)}
        \end{tabular}
  \caption{Time evolution of $\Delta \mathcal{E}$ (left side) and $\Delta I_{A,B}$ (right side) in the slow CCP. In the panels (a), (b), (c), and (d), we plot $\Delta \mathcal{E}$ and $\Delta I_{A,B}$ with   various distances. In the panels (e), (f), (g), and (h), we plot the time evolution of $\Delta \mathcal{E}$ and $\Delta I_{A,B}$ with fixed $d$ by changing $l_a$ and $l_b$. The panels (i) and (j) show $\Delta \mathcal{E}$ and $\Delta I_{A,B}$ in the slow CCP have a scaling law.}
  \label{ccps}
\end{figure}

\begin{enumerate}[(1)]
\item  The time, when the time evolution of $\Delta\mathcal{E}$ and $\Delta I_{A, B}$ starts to increase from zero, becomes later as $d$ increases. \item The measures $\Delta \mathcal{E}$ and $\Delta I_{A,B}$ oscillate, and their frequency at the late time $t\gg\delta t$ is expected to be $\f{1}{\pi \xi}$. These quantities  $\Delta \mathcal{E}$ and $\Delta I_{A,B}$ with $d=0$ have a first dip around $t\sim2\xi_{\text{kz}}$. 

\item 
The larger the subsystem sizes $l_a$ and $l_b$ are, the larger the late-time $\Delta\mathcal{E}$ and $\Delta I_{A, B}$ appear to be.
\item 
The early-time  $\Delta\mathcal{E}$ and $\Delta I_{A, B}$ are independent of $l_a$ and $l_b$. 
\item 
The panels (i) and (j) of Figure \ref{ccps} show that $\Delta \mathcal{E}$ and $\Delta I_{A,B}$  for $l_a,l_b, \xi, d\gg 1$ are expected to have the following scaling law, 
\begin{align}
\Delta \mathcal{E}&\sim \Delta \mathcal{E}\left(\frac{l_a}{\xi_{\text{kz}}}, \frac{l_b}{\xi_{\text{kz}}}, \frac{d}{\xi_{\text{kz}}}, \frac{t}{\xi_{\text{kz}}}, \frac{\delta t}{\xi_{\text{kz}}}\right),\notag\\
\Delta I_{A, B}&\sim \Delta I_{A, B}\left(\frac{l_a}{\xi_{\text{kz}}}, \frac{l_b}{\xi_{\text{kz}}}, \frac{d}{\xi_{\text{kz}}}, \frac{t}{\xi_{\text{kz}}}, \frac{\delta t}{\xi_{\text{kz}}}\right).\label{slccps}
\end{align}

\end{enumerate}

These properties can be seen from Figure \ref{ccps} as follows. 
The panels (a) and (b) of Figure \ref{ccps} show the time evolution of $\Delta\mathcal{E}$ and $\Delta I_{A, B}$ with $l_a=l_b$ in the slow CCP, respectively, and (c) and (d) show the evolution of $\Delta\mathcal{E}$ and $\Delta I_{A, B}$ with $l_a<l_b$, respectively. The larger $d$ is, the later the time when $\Delta I_{A,B}$ and $\Delta \mathcal{E}$ start to increase from zero  becomes (the property (1)).   As the time evolution of $\Delta \mathcal{E}$ and $\Delta I_{A,B}$ in the fast CCP, that in the slow CCP oscillate with the period which at late time is well-approximated by $\pi \xi$. In the slow CCP, $\Delta \mathcal{E}$ and $\Delta I_{A,B}$ with $d=0$ have a first dip around $t\sim2\xi_{\text{kz}}$ (the property (2)).

In the panels (e), (f), (g), and (h), we plot the time evolution of $\Delta\mathcal{E}$ and $\Delta I_{A, B}$ with various $l_a$ and $l_b$. We can see that $\Delta\mathcal{E}$ and $\Delta I_{A, B}$ at late time become larger as $l_a$ and $l_b$ increase (the property (3)). As in the fast CCP, all plots at early time in the panels (e), (f), (g), and (h) lie on the same curve, and those at the early time are independent of the subsystem sizes $l_a$ and $l_b$ (the property (4)). The panels (i) and (j) show $\Delta \mathcal{E}$ and $\Delta I_{A,B}$ have the scaling law (\ref{slccps})  as the changes in the other protocols in this paper(the property (5)).

\subsubsection{Quasi-particle interpretation}

The measures $\Delta \mathcal{E}$ and $\Delta I_{A,B}$ in the fast CCP oscillate, and the amplitude of their oscillation for $d\gg \xi$ is smaller than that for $d \ll \xi$. The evolution of $\Delta \mathcal{E}$ and $\Delta I_{A,B}$ with $d \gg \xi$ in the fast CCP is similar to that in the 2D CFT (\ref{2dCFT1}) and the fast ECP because of the small amplitude of their oscillation, and they can be explained by the propagation of quasi-particles with the speed of light as in the sudden quench \cite{CM4} and the ECP.

We suppose entangled pairs are produced at around $t\sim0$ since the adiabaticity in the fast CCP is broken at that time. 
We assume that the created particles move with the group velocity $v_\kappa=\frac{d \omega_\kappa}{dk}|_{t \sim 0}$, where $\omega_\kappa=\sqrt{4\sin^2{\left(\f{\kappa}{2}\right)}+m^2(t)}$. Since $m^2(t)$ at $t\sim 0$ is very small, the fastest-moving particles move with  the maximum group velocity $|v_{max}| \sim1$ \footnote{The quasi-particles propagate with the momentum-dependent velocities, and the particles with $\kappa\sim 0, \pm \pi$ propagate slowly \cite{ CM4, CM5, Alba:2017pnas, Alba:2017lvc, Alba:2018hie}}.  Therefore, the time evolution of $\Delta \mathcal{E}$ and $\Delta I_{A,B}$ in the fast CCP is well-interpreted in terms of the relative propagation of quasi-particles. Since the amplitude of their oscillation with large $d$ is small, their time evolution is well-described in the same manner as in the fast ECP.  Therefore, their late-time evolution is similar although their late-time protocols are different.

The logarithmic negativity $\Delta\mathcal{E}$  with $l_a<l_b$ and $d\gg\xi$ in the fast ECP shows a plateau from $t\sim\frac{\l_a+d}{2}$ to $t\sim\frac{\l_b+d}{2}$, and the negativity  becomes zero after $t\sim\frac{\l_a+\l_b+d}{2}$. However, $\Delta\mathcal{E}$ for this configuration  in the fast CCP  increases from $t\sim\frac{\l_a+d}{2}$ to $t\sim\frac{\l_b+d}{2}$ and is nonzero after $t\sim\frac{\l_a+\l_b+d}{2}$ (See, for example, the panel (g)  of Figure \ref{ccpf}). This is the difference between $\Delta\mathcal{E}$ in the fast ECP and CCP with $d\gg\xi$, and we expect this difference to come from the slow $\kappa$-modes of the quasi-particles.

The quantum measures at both limits in CCP $\Delta \mathcal{E}$ and $\Delta I_{A,B}$ oscillate in time, and their period at late time is well approximated by $\f{\pi}{m}  $, where $m$ is the late-time mass. Since the late-time period of their oscillation is the same as that of entanglement entropy, we expect the period to be interpreted as in  \cite{Nishida:2017hqd}. The time evolution of two point function $X_{ab}$, $P_{ab}$ and $D_{ab}$ determines the time evolution of $\Delta \mathcal{E}$ and $\Delta I_{A,B}$. Fourier modes of $X_{ab}$, $P_{ab}$ and $D_{ab}$ are given by (\ref{kspect}).  Each $\kappa$-mode  at late time is well approximated by
\begin{align}
X_{\kappa}&\sim \mathcal{C}^x_{\kappa}+\mathcal{D}^x_{\kappa}\cos(2\omega_{\kappa}(\infty)t+\Theta^x_{\kappa}),\notag\\
P_{\kappa}&\sim \mathcal{C}^p_{\kappa}+\mathcal{D}^p_{\kappa}\cos(2\omega_{\kappa}(\infty)t+\Theta^p_{\kappa}),\label{kspect}\\
D_{\kappa}&\sim \mathcal{D}^d_{\kappa}\cos(2\omega_{\kappa}(\infty)t+\Theta^d_{\kappa}),\notag
\end{align}
where $\mathcal{C}^i_{\kappa}$, $\mathcal{D}^i_{\kappa}$, and $\Theta^i_{\kappa}$ are independent of $t$. 
We expect the time evolution of $\Delta \mathcal{E}$ and $\Delta I_{A,B}$ at late time to be determined by the slowly-moving particles. Since the late-time velocities for quasi-particles are determined by $\omega_{\kappa}(\infty)$, and the particles with $\kappa\sim 0$ propagate very slowly, then they contribute dominantly to the time evolution of $\Delta \mathcal{E}$ and $\Delta I_{A,B}$ at late time.
At ${\kappa}=0$, the period of (\ref{kspect}) is $\frac{\pi}{\omega_0(\infty)}=\pi\xi$ and consistent with the period of $\Delta\mathcal{E}$ and $\Delta I_{A, B}$. 
\section{Discussion and future directions}
In this paper, we have studied the quantum correlation between the subsystems $A$ and $B$ in 2D free scalar field theory with the time-dependent mass $m(t)$. 
The protocols where we have studied the correlation are ECP and CCP which depend on a pair of parameter $(\xi, \delta t)$: The mass potential $m^2(t)$ in ECP is almost constant at early time $t \ll -\delta t$, starts to decrease sharply around $t\sim -\delta t$, and vanishes asymptotically in the region $t \gg \delta t$. On the other hand, $m^2(t)$ in CCP is almost constant at early time $t \ll -\delta t$, starts to decrease sharply around $t\sim -\delta t$, vanishes at $t=0$, increases monotonically after $t=0$, and becomes constant asymptotically in the region $t \gg \delta t$. 

The initial state, for which we have studied the  correlation $A$ and $B$ in the time evolution, is the ground state in the massive free scalar field theory. Since the Hamiltonian, which drives the system, is time-dependent, the state is no longer the ground state except for the region $t \ll - \delta t$, and the correlation between $A$ and $B$ changes in time.
We have studied how the time evolution of quantum correlation depends on $(\xi, \delta t)$ in the two limits: the fast limit $\delta t \ll \xi$ and the slow limit $\delta t \gg \xi$.

We have studied the changes of mutual information and logarithmic negativity, which are defined by subtracting the information and negativity for the initial state from those for the excited state at $t$, in order to study how the quantum correlation between $A$ and $B$ changes in time. We have called the changes of mutual information and logarithmic negativity $\Delta I_{A,B}$ and $\Delta \mathcal{E}$. 
We have found the time evolution of $\Delta \mathcal{E}$ with large $d$ in the fast and slow ECPs  is similar to that in the sudden quench \cite{CM4}. We have also found its evolution in the ECPs is well-described in terms of entangled pairs:  The pairs in the fast limit are created everywhere around $t\sim 0$, and that in the slow limit are created at $t \sim t_{\text{kz}}$, or the effective time when the adiabaticity is broken. 
The pair is composed of left- and right-moving particles with the speed of light which entangle with each other. 
Only when the left moving particle is in $A$(or $B$), and the right one is in $B$(or $A$), quantum entanglement between them contributes to the quantum correlation between $A$ and $B$. 

The relativistic propagation of entangled pairs well-describes the time evolution of $\Delta I_{A,B}$ in the fast and slow ECPs except for its logarithmic growth in time.
 The time evolution of $\Delta I_{A,B}$ is well-interpreted in terms of the ``generalized quasi-particle picture": Each pair created at $t \sim 0$ ($t\sim t_{\text{kz}}$) are composed of left- and right-moving particles with various values of velocities. The change $\Delta I_{A,B}$ at late time grows logarithmically in time, which comes from entanglement between slow-moving quasi-particles.
 
We have also studied the time evolution of $\Delta I_{A,B}$ and $\Delta \mathcal{E}$ in the fast and slow CCPs, and found that those in both limits oscillate with the frequency which, at late time, is well-approximated by $\f{m}{\pi}$. The longer the distance $d$ becomes, the smaller the amplitude of oscillation of $\Delta I_{A,B}$ and $\Delta \mathcal{E}$ in the fast ECP do. Since the evolution of $\Delta \mathcal{E}$ and $\Delta I_{A,B}$ in the fast CCP for $d\gg \xi$ is similar to that of  $\Delta \mathcal{E}$ and $\Delta I_{A,B}$ in the fast ECP, their evolution in the fast CCP is well-described in terms of the generalized quasi-particle picture which explains that in the fast ECP. We found the late-time period of their oscillation is interpreted in terms of quasi-particles propagating with the group velocity $v_\kappa$.
The late-time time evolution of $\Delta I_{A,B}$ and $\Delta \mathcal{E}$ is expected to come from the entanglement between slow-moving particles, and the late time velocity $v_\kappa$ is small around $\kappa=0$. Here we neglect the modes around $\kappa\sim \pm \pi$ though their velocities are small. This is because these modes are expected to suffer from the lattice artifacts. Also, we have found the time evolution of $\Delta \mathcal{E}$ and $\Delta I_{A,B}$ for the adjacent interval in the fast CCP has a first dip at $t\sim2\xi$. On the other hands, that of $\Delta \mathcal{E}$ and $\Delta I_{A,B}$   for the adjacent interval in the slow one has the first dip at $t\sim2 \xi_{\text{kz}}$. Thus, the time interval from $t=0$ to the time when first dip appears is given by twice as the effective correlation length: in the fast limit the initial correlation length $\xi$, in the slow limit the Kibble-Zurek length, $\xi_{\text{kz}}$.

\subsubsection*{Future directions}

\begin{itemize}
\item
In the fast CCP with large $d$, $\Delta \mathcal{E}$ and $\Delta I_{A,B}$ have a similar behavior in our numerical computation. It is interesting to study whether the time evolution of $\Delta \mathcal{E}$ is different from that of $\Delta I_{A,B}$ in the very-late-time regime where we have not computed them. 
\item
Another future direction  is to understand why $\Delta \mathcal{E}$ in the fast CCP with large $d$ does not have the plateau in the window  $\frac{l_a+d}{2}\lessapprox t\lessapprox\frac{l_b+d}{2}$ although $\Delta \mathcal{E}$ in the fast ECP has the plateau in this window.
\item
In the fast ECP with large $d$, we can explain the time evolution of $\Delta \mathcal{E}$ by using the quasi-particles which move at the speed of light. However, we need to utilize  not only quasi-particles at the speed of light, but also the slowly-moving particles to explain the time evolution of $\Delta I_{A,B}$ because of the logarithmic increase at the late time. It is important to investigate why we need such a slowly-moving particle picture in only $\Delta I_{A,B}$.   \end{itemize}

\section*{Acknowledgement}
We would like to thank Keun-Young Kim, Min-Sik Seo, Tadashi Takayanagi, and Akio Tomiya for comments and useful discussions. Y. Sugimoto thanks Interdisciplinary Center for Theoretical Study, University of Science and Technology of China for hospitality during my visit. The work of M.~Nishida was supported by Basic Science Research Program through the National Research Foundation of Korea (NRF) funded by the Ministry of Science, ICT \& Future Planning (NRF-2017R1A2B4004810) and GIST Research Institute (GRI) grant funded by the GIST in 2018. The work of Y. Sugimoto was supported in part by the JSPS Research Fellowship for Young Scientists (No. JP17J00828). M.Nozaki is also partially supported by RIKEN iTHEMS Program.  The work by H.~Fujita was supported in part by the JSPS Research Fellowship for Young Scientists (No.JP16J04752).

\appendix 

\end{document}